\documentclass[11pt,a4paper]{article}
\pdfoutput=1
\usepackage{jstyle}
\usepackage{amsmath, amsfonts, amssymb}
\usepackage{graphicx, hyperref, color, verbatim}
\usepackage[dvipsnames]{xcolor}
\usepackage{float}
\usepackage{caption}
\usepackage{subcaption}
\usepackage[bb=stix]{mathalpha}

\usepackage[LGR,T1]{fontenc}
\DeclareSymbolFont{ttgreek}{LGR}{cmtt}{m}{n}
\DeclareMathSymbol{\ttalpha}{\mathord}{ttgreek}{`a}

\makeatletter
\DeclareFontFamily{OMX}{MnSymbolE}{}
\DeclareSymbolFont{MnLargeSymbols}{OMX}{MnSymbolE}{m}{n}
\SetSymbolFont{MnLargeSymbols}{bold}{OMX}{MnSymbolE}{b}{n}
\DeclareFontShape{OMX}{MnSymbolE}{m}{n}{
    <-6>  MnSymbolE5
   <6-7>  MnSymbolE6
   <7-8>  MnSymbolE7
   <8-9>  MnSymbolE8
   <9-10> MnSymbolE9
  <10-12> MnSymbolE10
  <12->   MnSymbolE12
}{}
\DeclareFontShape{OMX}{MnSymbolE}{b}{n}{
    <-6>  MnSymbolE-Bold5
   <6-7>  MnSymbolE-Bold6
   <7-8>  MnSymbolE-Bold7
   <8-9>  MnSymbolE-Bold8
   <9-10> MnSymbolE-Bold9
  <10-12> MnSymbolE-Bold10
  <12->   MnSymbolE-Bold12
}{}

\let\llangle\@undefined
\let\rrangle\@undefined
\DeclareMathDelimiter{\llangle}{\mathopen}%
                     {MnLargeSymbols}{'164}{MnLargeSymbols}{'164}
\DeclareMathDelimiter{\rrangle}{\mathclose}%
                     {MnLargeSymbols}{'171}{MnLargeSymbols}{'171}
\makeatother

\makeatletter
\def\@fpheader{\hfill USTC-ICTS/PCFT-26-14}
\makeatother

\title{Defect Approach to Giant Graviton Dynamics}

\author[a,b]{Junding Chen}
\author[c,d]{Yunfeng Jiang}
\author[b,d]{Xinan Zhou}

\affiliation[a]{School of Physical Sciences, University of Chinese Academy of Sciences, No.19A Yuquan Road, Beijing 100049, China}
\affiliation[b]{Kavli Institute for Theoretical Sciences, University of Chinese Academy of Sciences, Beijing 100190, China}
\affiliation[c]{School of physics \& Shing-Tung Yau Center, Southeast University}
\affiliation[d]{Peng Huanwu Center for Fundamental Theory, Hefei, Anhui 230026, China}

\abstract{We develop a framework of zero dimensional defects for analyzing light-light-heavy-heavy (LLHH) correlators in conformal field theories. We specifically apply this formalism to correlators of giant gravitons in $\mathcal{N}=4$ super Yang-Mills to probe the nontrivial physics beyond planarity. By combining this framework with bootstrap techniques, we compute all four-point functions at strong coupling involving two maximal giant gravitons and two supergravitons of arbitrary dimensions. We identify a partially broken, higher-dimensional hidden symmetry -- a defect extension of 10d hidden conformal symmetry -- present at both strong and weak coupling, which allows these correlators to be packaged into a single generating function. Furthermore, we perform a systematic OPE analysis of the strong-coupling correlators, extracting the complete spectrum of anomalous dimensions for the defect-channel double-particle operators. Finally, we argue that the defect perspective provides the natural nonperturbative description for any LLHH correlator by showing that four-point conformal blocks reduce to defect two-point blocks in the heavy limit.}

\begin{document}
\maketitle
\tableofcontents

\newpage

\section{Introduction}
The large $N$ expansion of 't Hooft \cite{tHooft:1973alw} provides a topological reorganization of the gauge theory, where the $1/N$ expansion mirrors the genus expansion of a dual string theory. For 4d $\mathcal{N}=4$ super Yang-Mills (SYM), this qualitative picture finds a concrete realization via the AdS/CFT correspondence \cite{Maldacena:1997re,Witten:1998qj,Gubser:1998bc}, with  dual strings propagating in AdS$_5\times$S$^5$. Following the original vision of 't Hooft, the theory is most tractable in the planar limit where $N\to\infty$. In the dual AdS description, the planar limit suppresses string loop corrections in the bulk, significantly simplifying the calculation of observables. This can be further augmented by the strong-weak duality of the correspondence which allows us to analytically study the infinite 't Hooft coupling limit in terms of weakly coupled supergravity. Significant progress has been made on this front by adopting the bootstrap strategy, as was initiated in \cite{Rastelli:2016nze,Rastelli:2017udc}, and general efficient computational frameworks have been established (see \cite{Bissi:2022mrs}  for a recent summary). Parallel to AdS/CFT, another remarkable simplification in the planar limit is the emergence of integrability \cite{Minahan:2002ve}. Powerful tools such as the Quantum Spectral Curve \cite{Gromov:2013pga} and the hexagon formalism \cite{Basso:2015zoa} now allow for the study of $\mathcal{N}=4$ SYM at finite coupling, offering a path toward an exact solution of the planar theory. 

While the planar limit is in relatively good control, non-planarity remains largely terra incognita. Perturbatively in $1/N$, the planar techniques can be recycled. In AdS, unitarity methods similar to those in flat space allow one to glue together tree-level correlators into loop-level corrections \cite{Aharony:2016dwx,Alday:2017xua,Aprile:2017bgs}. Using integrability, one can also tessellate hexagons around handles to account for higher genus \cite{Eden:2017ozn,Bargheer:2017nne}. However, at the truly non-planar level -- where we go beyond just $1/N$ corrections and start to see the discrete granular nature of color -- both results and techniques are currently lacking. In $\mathcal{N}=4$ SYM, the giant graviton operators \cite{McGreevy:2000cw,Balasubramanian:2001nh,Hashimoto:2000zp} serve as an ideal probe of this regime. Maximal giant gravitons, for instance, take the form of the determinant in color space with dimension $\Delta=N$. The determinants invalidate the naive ribbon graph counting and make the large $N$ limit much more nontrivial. The large conformal dimension also signals the onset of nonperturbative physics. Concretely, one can study giant gravitons and the associated non-planar effects by computing their correlators. However, it turns out the study of these correlators is extremely challenging with very limited results  beyond free theory\footnote{In the free theory,  the problem reduces to the Wick contraction of operators. This is a combinatorial task for which several methods have been developed to manage the complexity, see for example \cite{Corley:2001zk,deMelloKoch:2019dda,deMelloKoch:2004crq,Berenstein:2013md,Lin:2022gbu,Holguin:2022drf,Holguin:2023naq}.}. 
\begin{itemize}
    \item Most studies focus on three-point functions, in particular of the type $\langle \mathcal{O}\mathcal{D}\mathcal{D}\rangle$ where one has two heavy giant graviton operators $\mathcal{D}$ and one light operator $\mathcal{O}$ \cite{Bissi:2011dc,Caputa:2012yj,Jiang:2019xdz,Jiang:2019zig,Holguin:2022zii}. However, such correlators are in a sense not sufficiently nontrivial because symmetries fix them up to an overall coefficient which depends on the dynamics. 
    \item The next simplest but nontrivial case is four-point functions where correlators develop interesting dependence on their insertion locations. Giant graviton correlators of the form $\langle\mathcal{O}\mathcal{O}\mathcal{D}\mathcal{D}\rangle$ have been studied, but only in the weak coupling limit and up to three loops \cite{Jiang:2019xdz,Jiang:2019zig,Jiang:2023uut}. No results are available at strong coupling, except for predictions for certain integrated correlators using supersymmetric localization \cite{Brown:2024tru,Brown:2024yvt,Brown:2025huy}.
\end{itemize}
One of the goals of this work is to bridge this gap by establishing a new powerful computational framework. We obtain general results for giant gravitons at both strong and weak coupling, a subset of which has been reported in the short paper \cite{Chen:2025yxg}. In this paper, we give a more detailed account and continue to develop the results which we will show to be useful also in broader contexts out of giant gravitons, even in the absence of supersymmetry. Let us give a summary of our main findings below.

\vspace{0.5cm}

\noindent{\bf Unconventional defects.} The most important conceptual ingredient of our framework is the shift of perspective towards defects. Conventionally, the giant graviton correlators are viewed as correlation functions of local operators. We will however argue that it is more advantageous to view them as the correlation functions of light operators in the presence of a ``zero dimensional'' defect, defined by
\begin{equation}
    \llangle \mathcal{O}\mathcal{O}\rrangle=\frac{\langle \mathcal{O}\mathcal{O}\mathcal{D}\mathcal{D}\rangle}{\langle\mathcal{D}\mathcal{D}\rangle}\;.
\end{equation}
The unconventional defect dimension can be more naturally understood in terms of holography, where objects generally gain an extra dimension by moving into the bulk. The zero dimensional defect in the CFT then corresponds to a line in AdS. This is in fact consistent with our expectation from string theory  where the giant gravitons corresponds to a D3 brane wrapping an S$^3$ inside of S$^5$. In AdS$_5$ the giant graviton behaves like a heavy particle traveling along a 1d geodesic which we interpret as the dual line defect. 

In fact, the unusual defect dimension can also be understood without the use of AdS/CFT, but in terms of conformal symmetry breaking in embedding space. Any CFT$_d$ can be uplifted into the $d+2$ dimensional embedding space where conformal symmetry is linearized as the $SO(d+1,1)$ rotations. The insertion of a $p$ dimensional defect partially breaks the symmetry by separating the dimensions into two groups of $p+2$ and $d-p$ dimensions. The rotations within each group $SO(p+1,1)\times SO(d-p)$ constitute the remaining unbroken conformal symmetry. The picture remains valid as long as the first group has at least one dimension, i.e., $p+2\geq 1$. Therefore, the notion of defects can be extended  in CFTs to include more generally the values $p=-1,0$. The case of $p=-1$ has been considered from this defect perspective in \cite{Zhou:2024ekb} where the defect setup is realized by placing $\mathcal{N}=4$ SYM on four dimensional real projective space.

\vspace{0.5cm}

\noindent{\bf Bootstrap.} The benefit of such a defect perspective is that the computation of giant graviton correlators can now be placed under the umbrella of the recent extension of bootstrap techniques for holographic correlators of defect CFTs \cite{Gimenez-Grau:2023fcy,Chen:2023yvw,Zhou:2024ekb}. This gives us a lot of technical control at strong coupling and allows us to use existing techniques (which we will also significantly extend in the current context) to study an otherwise difficult problem in gauge theory. Using bootstrap techniques, we will show that all infinitely many defect two-point functions $\llangle \mathcal{O}_{k_1}\mathcal{O}_{k_2}\rrangle$, where $\mathcal{O}_k$ is a light supergraviton operator of Kaluza Klein (KK) level $k$ and the defect is maximal giant gravitons, can be uniquely fixed at strong coupling using symmetries and consistency conditions.

\vspace{0.5cm}

\noindent{\bf CFT data and hidden symmetry.} The defect two-point functions contains a wealth of CFT data which encode the dynamics of the giant gravitons. We systematically extract this dynamical information from our strong coupling results. In particular, we obtain the complete AdS tree-level anomalous dimensions of certain ``double-particle'' operators which can be interpreted as the ``binding energies'' of a pair of light supergraviton and giant graviton. We should emphasize that this analysis is greatly facilitated by the zero dimensional defect formalism which provides the kinematic basis for decomposing the correlators. 

Moreover, it turns out that the defect perspective is also crucial in revealing hidden structures in the giant graviton correlators. We show that there is a defect extension of the 10d hidden conformal symmetry discovered in \cite{Caron-Huot:2018kta}, where the 10d symmetry is partially broken by a 4d defect. This is essentially the Weyl uplifting of the zero dimensional defect in the 4d CFT into the higher dimensional spacetime. This hidden structure allows for all infinitely many strong coupling two-point functions to be packaged into a single generating function. We further point out the hidden structure persists at weak coupling where it allows us to write down generating functions for loop correction integrands. We also find another manifestation of hidden symmetry at the level of CFT data. We show that the use of the hidden symmetry provides the automatic diagonalization of a mixing problem of which the solution is the anomalous dimension spectrum. The hidden symmetry also leads to a certain fourth-order differential operator which simplifies the computation of AdS loop corrections to the giant graviton correlators.

\vspace{0.5cm}

\noindent{\bf Defect description for LLHH.} The usefulness of the defect formalism extends well beyond supersymmetry and giant gravitons. We argue that the defect perspective provides a natural description for any Light-Light-Heavy-Heavy (LLHH) correlators by showing the four-point function conformal blocks reduce to defect two-point function conformal blocks in the heavy limit. Since conformal blocks form a basis for decomposing any correlator, this proves our claim at the nonperturbative level. We also complement the analysis with a study of Witten diagrams in the heavy limit where we find similar results. These observations naturally point to a nonperturbative approach of analytic functionals of which the main ideas we briefly outline. 

\vspace{0.5cm}

The rest of this paper is organized as follows. In Section \ref{Sec:corrkinematics}, we perform a detailed analysis of the kinematics of giant gravitons, from both the four-point function perspective and the defect two-point function perspective. We also present the results of conformal blocks, R-symmetry blocks and superconformal blocks which form the basis for extracting CFT data later. In Section \ref{Sec:sugraanalysis}, we review the supergravity description of giant gravitons. Results from this section will be used as the input and also checks of the bootstrap approach. We also clarify the difference between classical and quantum averaging over the moduli. In Section \ref{Sec:treelevelWittendiagrams}, we present a systematic study of defect Witten diagrams which provide the necessary ingredients for the bootstrap. The bootstrap analysis for giant graviton correlators is carried out in Section \ref{Sec:2ptfunbootstrap} where we present the result general defect two-point functions. We also point out a partially broken higher dimensional hidden conformal symmetry which is present at both strong and weak couplings. The dynamical information encoded in these giant graviton correlators is carefully examined in Section \ref{Sec:extractingCFTdata}. We show that an OPE analysis together with an unmixing procedure allow us to extract the complete anomalous dimension spectrum. We also explain the implication of hidden symmetry in the CFT data and how it helps to compute loop corrections in AdS. In Section \ref{Sec:defectforLLHH}, we extend the scope of our defect approach by considering general LLHH correlators without supersymmetry. We explain how the defect picture naturally emerges in the heavy limit. We conclude with a discussion of future research directions in Section \ref{Sec:outlook}. This paper is also accompanied by several appendices where we relegate some technical details.

\section{Kinematics of giant graviton correlators}\label{Sec:corrkinematics}
\subsection{Giant gravitons and supergravitons}
In this paper, we will consider two types of $\frac{1}{2}$-BPS operators, namely, supergravitons and giant gravitons. The supergravitons are light single-trace operators\footnote{To be more precise, we should define single-particle supergraviton operators. They contain additional multi-trace terms with $1/N$ suppressed coefficients to make operators with different particle numbers orthogonal \cite{Arutyunov:1999en,Arutyunov:2000ima,Aprile:2019rep,Alday:2019nin,Aprile:2020uxk}.}
\begin{equation}
    \mathcal{O}_k(x,t)={\rm tr}\left(X^it^i\right)^k\;,
\end{equation}
where $X^i$ with $i=1,2,\ldots,6$ are the six scalars of $\mathcal{N}=4$ SYM. The vectors $t^i$ are null, i.e., satisfying $t\cdot t=0$. This property ensures  the R-symmetry representation is projected to the rank-$k$ symmetric traceless representation of $SO(6)$. These operators have protected conformal dimensions $\Delta=k$. In the dual gravity theory in AdS$_5\times$S$^5$, they correspond to the KK modes of the 10d supergravity field. 

By contrast, the giant graviton operators are constructed in $\mathcal{N}=4$ SYM as heavy sub-determinants
\begin{equation}\label{defggop}
    \mathcal{D}_M(x,t)=\frac{1}{M!}\sum_{\sigma\in S_M}(-1)^{|\sigma|}\delta_{a_{\sigma_1}}^{b_1}\ldots \delta_{a_{\sigma_M}}^{b_M} (X\cdot t)^{a_1}_{b_1}\ldots (X\cdot t)^{a_M}_{b_M}\;,
\end{equation}
where $M\leq N$ with $N$ being the rank of the gauge group $SU(N)$. These operators have dimensions $\Delta=M$ and are in the rank-$M$ symmetric traceless representation of the $SO(6)$ R-symmetry. The giant graviton operators are called maximal if they saturate the bound $M=N$ (full determinant), and non-maximal otherwise. Such operators are also called sphere giant gravitons because they are dual to a D3 brane with one dimension in AdS$_5$ and three dimensions in the compact internal S$^5$ as an orbiting S$^3$. There  also exist dual giant graviton operators which correspond to a D3 brane wrapping an S$^3$ in AdS$_5$. For these operators, instead of the antisymmetric Schur polynomials in (\ref{defggop}), they are defined by symmetric Schur polynomials. However, the bootstrap analysis of this paper will only focus on the sphere giant gravitons.  

Because both supergravitons and giant gravitons are $\frac{1}{2}$-BPS, their three-point functions (e.g., two giant gravitons and a single supergraviton) are trivial in the sense that they are protected by non-renormalization theorems. In this paper, we focus on the simplest nontrivial case, namely four-point functions
\begin{equation}
    F_{k_1k_2}(x_i,t_i)=\langle \mathcal{O}_{k_1}(P_1,t_1)\mathcal{O}_{k_2}(P_2,t_2) \mathcal{D}_M(P_3,t_3)\mathcal{D}_M(P_4,t_4)\rangle\;.
\end{equation}
In the next two subsections, we will discuss the superconformal kinematics of these objects from two complementary perspectives, both as four-point functions and as defect two-point functions.

\subsection{Implications of superconformal symmetry on four-point functions}
The symmetry implications of four-point functions of $\frac{1}{2}$-BPS operators have been discussed in many papers and here we just summarize the relevant results.\footnote{We have set $M=N$ here for convenience since later we will mostly consider the maximal case. But it is easy to see that the results for the kinematics are the same for the non-maximal case (especially from the defect perspective).} Let us first look at the consequence of the bosonic symmetries, namely $SO(6)_R$ R-symmetry and conformal symmetry. The first thing to note is that the four-point function can only be nonzero when $k_1-k_2$ is an even integer. This can be easily seen by considering the R-symmetry tensor products $[0,N,0]\times [0,N,0]$ and $[0,k_1,0]\times [0,k_2,0]$. The irreducible representations appearing in these two tensor products can have an overlap only when $k_2-k_1$ is even. We can  extract a kinematic factor
\begin{equation}\label{FtomathcalF}
    F_{k_1k_2}(x_i,t_i)=\left(\frac{t_{34}}{x_{34}^2}\right)^{\frac{2N-k_1-k_2}{2}}\left(\frac{t_{23}t_{24}}{x_{23}^2x_{24}^2}\right)^{\frac{k_2-k_1}{2}}\left(\frac{t_{12}t_{34}}{x_{12}^2x_{34}^2}\right)^{k_1}\mathcal{F}_{k_1k_2}(U,V;\sigma,\tau)\;,
\end{equation}
where $t_{ij}=t_i\cdot t_j$ and $x_{ij}=x_i-x_j$. Without loss of generality, we have assumed $k_1\leq k_2$. Invariance under conformal symmetry and R-symmetry implies that the four-point function becomes a function of the conformal and R-symmetry cross ratios
\begin{equation}\label{crossratios4pt}
    U=\frac{x_{12}^2x_{34}^2}{x_{13}^2x_{24}^2}\;,\quad V=\frac{x_{14}^2x_{23}^2}{x_{13}^2x_{24}^2}\;,\quad \sigma=\frac{t_{13}t_{24}}{t_{12}t_{34}}\;,\quad \tau=\frac{t_{14}t_{23}}{t_{12}t_{34}}\;.
\end{equation}
In particular, $\mathcal{F}_{k_1k_2}$ is a degree $k_1$ polynomial in $\sigma$ and $\tau$. To consider the constraints from the fermionic generators, it is useful to make a change of variables
\begin{equation}\label{crossratiostozzbalphaalphabar}
    U=z\bar{z}\;,\quad V=(1-z)(1-\bar{z})\;,\quad \sigma=\alpha\bar{\alpha}\;,\quad \tau=(1-\alpha)(1-\bar{\alpha})\;.
\end{equation}
The constraints are known as the superconformal Ward identities \cite{Eden:2000bk,Nirschl:2004pa}
\begin{equation}\label{scfWardid}
    (z\partial_z-\alpha\partial_\alpha)\mathcal{F}(z,\bar{z};\alpha,\bar{\alpha})\big|_{\alpha=\frac{1}{z}}=0\;,
\end{equation}
and similarly with $z\leftrightarrow\bar{z}$, $\alpha\leftrightarrow\bar{\alpha}$. These constraints can be solved and the solution takes the form 
\begin{equation}\label{solscfWardid4pt}
   F_{k_1k_2}(x_i,t_i)=F_{k_1k_2}^{\rm free}(x_i,t_i)+t_{12}^2t_{34}^2x_{13}^4x_{24}^4 R\, K_{k_1k_2}(x_i,t_i)\;.
\end{equation}
Here $F_{k_1k_2}^{\rm free}$ is the four-point function in the free theory and $R$ is a factor determined by superconformal symmetry
\begin{equation}
R=(1-z\alpha)(1-z\bar{\alpha})(1-\bar{z}\alpha)(1-\bar{z}\bar{\alpha})\;.
\end{equation}
This also defines a new object $K_{k_1k_2}$ which is the reduced four-point function. Note that this reduced correlator has shifted R-symmetry charges $\{k_1-2,k_2-2,N-2,N-2\}$ and conformal dimensions $\{k_1+2,k_2+2,N+2,N+2\}$. Upon setting $\bar{\alpha}=1/\bar{z}$, the second term with the reduced correlator vanishes and the first term becomes a meromorphic function. This meromorphic function has the interpretation as the four-point correlator of a 2d chiral CFT and is independent of the coupling \cite{bllprv13}.  

\subsection{Giant gravitons as defects and defect two-point functions}\label{Subsec:4ptas2pt}
As we mentioned in the introduction, another useful perspective is to view these four-point functions as two-point functions in the presence of a defect created by the two giant gravitons. To explain how this works in detail, it is convenient to use the embedding space where the action of conformal symmetry is linearized. For each point $x^\mu\in\mathbb{R}^4$, we introduce a six dimensional null vector in $\mathbb{R}^{1,5}$
\begin{equation}
    P_A=\left(\frac{1+x^2}{2},\frac{1-x^2}{2},x^\mu\right)\;,
\end{equation}
where the first component is time-like. We will treat $P_3$ and $P_4$ as the defect. Usually defects are characterized by the directions in which they extend. But this can also be achieved by defining projectors. Let us define
\begin{equation}\label{defN}
    \mathbb{N}_{AB}=\frac{P_{3,A}P_{4,B}+P_{4,A}P_{3,B}}{P_3\cdot P_4}\;,
\end{equation}
which obviously satisfies the property
\begin{equation}
    \mathbb{N}_{AB}\mathbb{N}^B{}_C=\mathbb{N}_{AC}\;.
\end{equation}
Note that
\begin{equation}
    \mathbb{N}\cdot P_3=P_3\;,\quad \mathbb{N}\cdot P_4=P_4\;.
\end{equation}
Therefore, the projector $\mathbb{N}$ projects the embedding space $\mathbb{R}^{1,5}$ to the subspace spanned by $P_3$ and $P_4$. We can similarly define projectors for R-symmetry. From $t_3$ and $t_4$, we define 
\begin{equation}\label{defM}
\mathbb{M}_{IJ}=\delta_{IJ}-\frac{t_{3,I}t_{4,J}+t_{4,I}t_{3,J}}{t_3\cdot t_4}\;,
\end{equation}
which satisfies 
\begin{equation}
    \mathbb{M}_{IJ}\mathbb{M}^J{}_K=\mathbb{M}_{IK}\;.
\end{equation}
However, this R-symmetry projector has the property 
\begin{equation}
    \mathbb{M}\cdot t_3=\mathbb{M}\cdot t_4=0\;.
\end{equation}
Therefore, the R-symmetry projects to the space orthogonal to $t_3$ and $t_4$. 

For a purely bosonic defect without internal R-symmetry, the projector $\mathbb{N}$ would be sufficient for describing all structures showing up in correlators (and similarly $\mathbb{M}$ would be enough for describing a defect in the internal space without a spacetime part). To describe the giant gravitons, we also need the ``square roots'' of these projectors. These are the antisymmetric combinations
\begin{equation}\label{nmoperators}
   \mathbb{n}_{AB}=\frac{P_{3,A}P_{4,B}-P_{4,A}P_{3,B}}{P_3\cdot P_4}\;, \quad  \mathbb{m}_{IJ}=\frac{t_{3,I}t_{4,J}-t_{4,I}t_{3,J}}{t_3\cdot t_4}\;.
\end{equation}
It is easy to verify
\begin{equation}
    \mathbb{n}_{AB}\mathbb{n}^B{}_C=\mathbb{N}_{AC}\;,\quad \mathbb{m}_{IJ}\mathbb{m}^J{}_K=\delta_{IK}-\mathbb{M}_{IK}\;.
\end{equation}
The need for these additional tensors can be understood from the invariance of the correlator under $3\leftrightarrow 4$ exchange, which is most evident from the four-point function perspective. However, the invariance does not need to hold separately for $P_3\leftrightarrow P_4$ and $t_3\leftrightarrow t_4$. Only the invariance under their simultaneous actions is required. The square root tensors, which are odd under $3\leftrightarrow 4$ exchange and therefore cannot appear individually, can pair up to achieve overall even parity. We will also see the need for these additional objects in a slightly different way later in this subsection.

Let us also mention that from these projectors (and their square roots) it is clear to see what subgroups of symmetries are preserved. The $SO(5,1)$ conformal group is broken to $SO(4)\times SO(1,1)$. Similarly, the $SO(6)$ R-symmetry is broken to $SO(4)\times SO(2)$. In fact, the two giant gravitons also preserve half of the supercharges and the system has a remaining supersymmetry $SU(2|2)\times SU(2|2)\subset PSU(2,2|4)$ (see, e.g., \cite{Imamura:2021ytr}).
Here we would like to emphasize in particular the $SO(1,1)$ and $SO(2)$ factors. Naively, they are broken by the fixed vectors $P_3$, $P_4$ and $t_3$, $t_4$. However, the point of the defect picture is that we cannot talk about individual vectors of $P_3$, $P_4$ and $t_3$, $t_4$. Instead, we have only access to the projectors (and their square roots) formed by these vectors. While the $SO(1,1)$ and $SO(2)$ symmetries are broken by the vectors, they are preserved by the projectors and therefore by the giant graviton defect. 

Let us now consider correlators. In embedding space, operators are required to have the following scaling property
\begin{equation}
    \mathcal{O}_k(\lambda P,\bar{\lambda} t)=\left(\frac{\bar{\lambda}}{\lambda}\right)^k\mathcal{O}_k( P, t)\;.
\end{equation}
This property is carried over to  the defect correlators 
\begin{equation}
    \llangle \mathcal{O}_{k_1}(\lambda_1 P_1,\bar{\lambda}_1 t_1)\ldots \mathcal{O}_{k_n}(\lambda_n P_n,\bar{\lambda}_n t_n)  \rrangle =\prod_{i=1}^n \left(\frac{\bar{\lambda}_i}{\lambda_i}\right)^{k_i}  \llangle \mathcal{O}_{k_1}(P_1,t_1)\ldots \mathcal{O}_{k_n}(P_n,t_n) \rrangle\;.
\end{equation}
We first consider one-point functions. The only possibility which we can construct from $P$, $t$ and $\mathbb{N}$, $\mathbb{M}$ is\footnote{Note invariants constructed from $\mathbb{n}$ and $\mathbb{m}$ vanish automatically.}
\begin{equation}
    \llangle \mathcal{O}_k(P_i,t_i) \rrangle=a_k \frac{(\frac{1}{2}t_i\cdot\mathbb{M}\cdot t_i)^{\frac{k}{2}}}{(P_i\cdot \mathbb{N}\cdot P_i)^{\frac{k}{2}}}=a_k\left(\frac{t_{i3}t_{i4}x_{34}^2}{t_{34}x_{i3}^2x_{i4}^2}\right)^{\frac{k}{2}}\;.
\end{equation}
Here the one-point function coefficient $a_k$ is a physical data. Moreover, $k$ needs to be an even integer so that the R-symmetry representation $[0,k,0]$ has an identity representation when branched into representations of the subgroup $SO(4)=SU(2)\times SU(2)$  preserved by the defect. That defect one-point functions are fixed by symmetries can also be understood from the defect-free CFT perspective where we treat the giant gravitons as local operators. The defect one-point function is actually the ratio
\begin{equation}\label{1ptand3pt}
\llangle \mathcal{O}_k(P_i,t_i) \rrangle=\frac{\langle \mathcal{O}_k(P_i,t_i) \mathcal{D}(P_3,t_3)\mathcal{D}(P_4,t_4)\rangle }{\langle \mathcal{D}(P_3,t_3)\mathcal{D}(P_4,t_4)\rangle}=C_{\mathcal{D}\mathcal{D}\mathcal{O}_k}\left(\frac{t_{i3}t_{i4}x_{34}^2}{t_{34}x_{i3}^2x_{i4}^2}\right)^{\frac{k}{2}}\;.
\end{equation}
Here we divide by the giant graviton two-point function so that there are no conformal dimensions and R-symmetry charges with respect to 3 and 4, as it should be for the case of a defect one-point function. We have also used the fact that defect-free two- and three-point functions are fixed up to a coefficient in the latter. These two perspectives agree with each other and we have
\begin{equation}
  a_k= C_{\mathcal{D}\mathcal{D}\mathcal{O}_k}\;.
\end{equation}
Because three-point functions of $\frac{1}{2}$-BPS operators are protected by supersymmetry, it follows that the defect one-point functions are also protected and $a_k$ is independent of the coupling. 

Similarly, the defect two-point function should be defined from the four-point function as the ratio
\begin{equation}
    G_{k_1k_2}(x_i,t_i)=\llangle \mathcal{O}_{k_1}(x_1,t_1) \mathcal{O}_{k_2}(x_2,t_2) \rrangle=\frac{\langle \mathcal{O}_{k_1}(P_1,t_1)\mathcal{O}_{k_2}(P_2,t_2) \mathcal{D}(P_3,t_3)\mathcal{D}(P_4,t_4)\rangle}{\langle \mathcal{D}(P_3,t_3)\mathcal{D}(P_4,t_4)\rangle}\;.
\end{equation}
In this case, the unbroken conformal symmetry and R-symmetry are no longer constraining enough to fix the correlator and there are cross ratios. From the defect perspective, the natural definitions of conformal cross ratios are \cite{Billo:2016cpy}
\begin{equation}\label{defxichi}
\xi=\frac{-2P_1\cdot P_2}{\big( \prod_{i=1}^2 P_i\cdot\mathbb{N}\cdot P_i\big)^{\frac{1}{2}}}=\frac{U}{\sqrt{V}}\;,\quad 
\chi=\frac{2P_1\cdot P_2-2P_1\cdot \mathbb{N} \cdot P_2}{\big( \prod_{i=1}^2 P_i\cdot\mathbb{N}\cdot P_i\big)^{\frac{1}{2}}}=\frac{1-U+V}{\sqrt{V}}\;,
\end{equation}
and similarly for R-symmetry cross ratios we have
\begin{equation}
\label{defSigmaSigmabar}
    \Sigma=\frac{t_{12}}{\big(\prod_{i=1}^{2}\frac{1}{2} t_i \cdot \mathbb{M} \cdot t_i\big)^{\frac{1}{2}}}= \frac{1}{\sqrt{\sigma \tau}}\;, \quad  \bar{\Sigma}= \frac{t_{12}-t_1 \cdot \mathbb{M} \cdot t_2}{\big(\prod_{i=1}^{2}\frac{1}{2} t_i \cdot \mathbb{M} \cdot t_i\big)^{\frac{1}{2}}}= \frac{\sigma+\tau}{\sqrt{\sigma \tau}}\;.
\end{equation}
In analogy with higher dimensional defects, the numerators $-2P_1\cdot P_2$,  $2P_1\cdot P_2-2P_1\cdot\mathbb{N}\cdot P_2$ in the definitions of $\xi$, $\chi$ are correspondingly the total squared distance $x_{12}^2$ and the inner product in the transverse space $2x_1^jx_2^j$. After extracting a factor of one-point functions, we can write the two-point function as a function of the cross ratios 
\begin{equation}\label{defcalG}
    G_{k_1k_2}(x_i,t_i)=\prod_{i=1,2}\left(\frac{t_{i3}t_{i4}x_{34}^2}{t_{34}x_{i3}^2x_{i4}^2}\right)^{\frac{k_i}{2}}\mathcal{G}_{k_1k_2}(\xi,\chi;\Sigma,\bar{\Sigma})\;.
\end{equation}
But here comes an important subtlety. In defining the cross ratios (\ref{defxichi}) and (\ref{defSigmaSigmabar}), we have only used the symmetric projectors. Under the $3\leftrightarrow 4$ exchange, we find 
\begin{equation}\label{simulZ2}
U\to \frac{U}{V}\;,\quad V\to\frac{1}{V}\;,\quad \sigma\to\tau\;,\quad \tau\to\sigma\;.
\end{equation}
But this is a redundancy in the parameterization of the cross ratios $\xi$, $\chi$ and $\Sigma$, $\bar{\Sigma}$ where $\{U,V\}\to\{U/V,1/V\}$ leaves $\xi$, $\chi$ invariant and $\{\sigma,\tau\}\to\{\tau,\sigma\}$ leaves $\Sigma$, $\bar{\Sigma}$ invariant. Consequently, the two-point defect correlator appears to be {\it separately} invariant under these two $\mathbb{Z}_2$ actions. However, this is an unnecessarily constraining (and in general incorrect) constraint\footnote{It is obvious from the four-point function perspective that we have no such constraints.} because the correlator is only required to be invariant under the {\it simultaneous} $\mathbb{Z}_2$ action (\ref{simulZ2}). The way to resolve this issue is to also use the ``square root'' operators (\ref{nmoperators}). We can form new cross ratios 
\begin{equation}
 \frac{P_1\cdot \mathbb{n}\cdot P_2}{P_1\cdot P_2}=\frac{1-V}{U}\;,\quad \frac{t_1\cdot\mathbb{m}\cdot t_2}{t_1\cdot t_2}=\sigma-\tau\;,
\end{equation}
which have the opposite parity. A more economical statement is that we should just use $U$, $V$, $\sigma$, $\tau$ as the basic variables, and require invariance under the total action of $U\to U/V$, $V\to 1/V$, $\sigma\leftrightarrow \tau$.\footnote{A good analogy is four-point functions of $\frac{1}{2}$-BPS operators in 2d SCFT with small $\mathcal{N}=4$ superconformal symmetry \cite{Rastelli:2019gtj}. There the four-point correlators have two conformal cross ratios $\mathtt{z}$, $\mathtt{\bar{z}}$ and two R-symmetry cross ratios $\ttalpha$, $\bar{\ttalpha}$, with $\{\mathtt{z},\ttalpha\}$ and $\{\bar{\mathtt{z}},\bar{\ttalpha}\}$ associated with the left- and right-moving factors of the $PSU(1,1|2)\times PSU(1,1|2)$ subalgebra respectively. The analogue of (\ref{defxichi}) and (\ref{defSigmaSigmabar}) are the cross ratios $\mathtt{U}=\mathtt{z}\bar{\mathtt{z}}$, $\mathtt{V}=(1-\mathtt{z})(1-\bar{\mathtt{z}})$ and $\mathtt{\sigma}=\ttalpha\bar{\ttalpha}$, $\mathtt{\tau}=(1-\ttalpha)(1-\bar{\ttalpha})$. These are the only cross ratios which can be defined in higher dimensions and are separately invariant under $\mathtt{z}\leftrightarrow\bar{\mathtt{z}}$ and $\ttalpha\leftrightarrow\bar{\ttalpha}$. But the correlator is only required to be invariant under the simultaneous action $\{\mathtt{z},\ttalpha\}\leftrightarrow \{\bar{\mathtt{z}},\bar{\ttalpha}\}$ which exchanges the left- and right-movers.} We should point out that this subtlety does not arise for higher dimensional defects (such as surfaces defects in 6d $(2,0)$ theories considered in \cite{Chen:2023yvw}) where using $\xi$, $\chi$ is sufficient.\footnote{For defects with dimension $1\leq p\leq d-2$, we can use a conformal transformation to map the defect and correlators to a special 2d configuration where the defect two-point function also effectively looks like a four-point function. The defect intersects the complex plane at $0$ and $\infty$. One operator is inserted at $1$, and the other operator has complex coordinates $z$, $\bar{z}$. It seems that one could also define $U$, $V$ in terms of $z$, $\bar{z}$, and exchanging the defect intersections leads to the same transformations for $U$ and $V$. However, the important point is that this only affects $U$, $V$ and does not affect R-symmetry cross ratios, which implies $U\to U/V$, $V\to 1/V$ becomes a parameterization degeneracy. \label{footnotehigherp}} 

Viewed as defect correlators, it is also easy to see that $k_2-k_1$ must be an even integer in order for the two-point function to be non-vanishing. We can for example consider the $[0,k,0]$ representation which appears in the tensor product of $[0,k_1,0]\times [0,k_2,0]$ of the bulk channel OPE. The $SO(6)_R$ representation theory requires $k$ to have the same parity as $k_2-k_1$. As we have seen from the analysis of one-point function, $k$ must be even for the one-point function to exist and this shows $k_2-k_1$ must be even. 

Finally, the solution to the superconformal Ward identities (\ref{solscfWardid4pt}) implies that the defect two-point function can be written as
\begin{equation}\label{defGandH}
    G_{k_1k_2}=\frac{x_{34}^{2N}}{t_{34}^N}\left(F_{k_1k_2}^{\rm free}+t_{12}^2t_{34}^2x_{13}^4x_{24}^4 R\, K_{k_1k_2}\right)=G_{k_1k_2,{\rm free}}+\frac{t_{12}^2x_{13}^4x_{24}^4}{x_{34}^4}R\, H_{k_1k_2}\;.
\end{equation}
Here $G_{k_1k_2,{\rm free}}$ is the defect two-point function in the free theory and $H_{k_1k_2}$ is defined to be the reduced two-point function. The reduced correlator has shifted R-symmetry charges $\{k_1-2,k_2-2\}$ and conformal dimensions $\{k_1+2,k_2+2\}$. For convenience, we also write the reduced correlator as a function of cross ratios by extracting one-point functions according to these weights
\begin{equation}
\label{defcalH}
    H_{k_1k_2}=\prod_{i=1,2}\left(\frac{t_{i3}t_{i4}x_{34}^2}{t_{34}x_{i3}^2x_{i4}^2}\right)^{\frac{k_i+2}{2}} \frac{t_{34}^4}{t_{13}^2 t_{14}^2 t_{23}^2 t_{24}^2} \mathcal{H}_{k_1 k_2}(U,V,\sigma,\tau)\;.
\end{equation}
This implies
\begin{equation}
    \label{add}
    \mathcal{G}_{k_1 k_2} = \mathcal{G}_{k_1 k_2,\rm free} + (V \sigma \tau)^{-1} R \, \mathcal{H}_{k_1 k_2}\;.
\end{equation}

\subsection{Conformal blocks and R-symmetry blocks}
\label{subsec:Superconformal blocks}
To efficiently extract data from the giant graviton correlators we need to decompose them into basic kinematic objects which are the conformal blocks and R-symmetry blocks. These are further grouped together by supersymmetry into superconformal blocks. In this subsection, we will discuss these objects in turn and point out a few important properties. While our analysis will mostly take the defect perspective, we also highlight the connections of these objects with their counterparts in four-point functions in defect-free CFTs.

\vspace{0.5cm}

\noindent{\bf Conformal blocks.} Let us start with conformal blocks. We will see that the R-symmetry blocks are very similar, essentially because the symmetries are same (up to a Wick rotation). In the bulk channel, the defect correlator conformal blocks are the same as the four-point function conformal blocks. This is easiest to see from the OPE perspective. Let us consider the exchange of an operator $\mathcal{O}$ in the bulk channel. The corresponding conformal block (without stripping off the kinematic factors) can be written as 
\begin{equation}
    \mathbb{C}_{\mathcal{O}}(x_{12},\partial_{x_2})\llangle \mathcal{O}(x_2)\rrangle\;,
\end{equation}
where $\mathbb{C}_{\mathcal{O}}(x_{12},\partial_{x_2})$ is the differential operator in the $\mathcal{O}_1(x_1)\times \mathcal{O}_2(x)_2$ OPE for the appearance of the operator $\mathcal{O}$. As we pointed out in (\ref{1ptand3pt}), the defect one-point function $\llangle \mathcal{O}(x_2)\rrangle$ is nothing but the ratio of the three-point function $\langle \mathcal{O}(x_2)\mathcal{D}(x_3)\mathcal{D}(x_4) \rangle$ and the two-point function $\langle \mathcal{D}(x_3)\mathcal{D}(x_4) \rangle$. It is clear that the bulk-channel defect conformal block is proportional to 
\begin{equation}
    \mathbb{C}_{\mathcal{O}}(x_{12},\partial_{x_2})\langle \mathcal{O}(x_2) \mathcal{D}(x_3)\mathcal{D}(x_4)\rangle\;,
\end{equation}
which is the conformal block in the four-point function $\langle \mathcal{O}_1(x_1)\mathcal{O}_2(x_2) \mathcal{D}(x_3)\mathcal{D}(x_4)\rangle$ for exchanging the operator $\mathcal{O}$. As a consistency check, it is useful to notice that the four-point conformal blocks depend on the external dimensions only via the combinations $\Delta_2-\Delta_1$ and $\Delta_3-\Delta_4$. Since the two heavy operators have the same dimension, the dependence on their dimensions drop out in the conformal block and this nicely agrees with the defect picture. More precisely, we can decompose the defect two-point function $\mathcal{G}_{k_1k_2}$ into the following bulk channel conformal blocks where the exchanged operator has conformal dimension $\Delta$ and spin $\ell$ 
\begin{equation}
\label{bosoniccfblock}
    g_{\Delta,\ell}(z,\bar{z})=(-1)^{\ell} \frac{((1-z)(1-\bar z))^{\frac{k_2}{2}}}{(z-\bar{z})(z \bar z)^{\frac{k_1+k_2-2}{2}}} \left(p_{\Delta+\ell}(z) p_{\Delta-\ell-2}(\bar z)-(z\leftrightarrow \bar{z})\right)\;.
\end{equation}
Here the prefactor is extracted according to \eqref{FtomathcalF},  $k_{ij}=k_i-k_j$ and
\begin{equation}
    p_{h}(z)=z^{\frac{h}{2}}{}_2 F_1\left( \frac{h-k_{12}}{2},\frac{h}{2};h;z\right)\;.
\end{equation}

The discussion of conformal blocks in the defect channel is facilitated by the use of quadratic Casimir operators. As we pointed out, the giant graviton defect breaks the conformal group $SO(5,1)$ into the product $SO(1,1)\times SO(4)$. As in \cite{Billo:2016cpy}, we have a set of two decoupled Casimir equations, one for each subgroup factor
\begin{equation}
\begin{aligned}
\label{defectCasimireqs}
 \left(\widehat{{\rm Cas}}_{SO(4)}+\widehat{C}_{\widehat{\Delta}, 0}\right) \prod_{i=1}^2 (-P_i\cdot \mathbb{N}\cdot P_i)^{-\frac{\Delta_i}{2}}\widehat{g}_{\widehat{\Delta}, s}(\xi, \chi)&=0 \;,\\
 \left(\widehat{{\rm Cas}}_{SO(1,1)}+\widehat{C}_{0, s}\right) \prod_{i=1}^2 (-P_i\cdot \mathbb{N}\cdot P_i)^{-\frac{\Delta_i}{2}}\widehat{g}_{\widehat{\Delta}, s}(\xi, \chi)&=0\;.
\end{aligned}
\end{equation}
Here $\widehat{g}_{\widehat{\Delta}, s}$ is the defect channel conformal block for exchanging an operator with dimension $\widehat{\Delta}$ and transverse spin $s$. The Casimir eigenvalue is
\begin{equation}
    \widehat{C}_{a,b}=\widehat{C}_{a,0}+\widehat{C}_{0,b}\;,
\end{equation}
with
\begin{equation}
    \widehat{C}_{a,0}=a^2\;,\quad \widehat{C}_{0,b}=b(b+2)\;.
\end{equation}
The Casimir operators are of the one-particle type (involving either 1 or 2) and are most convenient to write down in embedding space
\begin{equation}\label{Casimirconf}
\begin{split}
   &\widehat{{\rm Cas}}_{SO(4)} = \frac{1}{2} \mathcal{J}^{AB} \mathbb{N}_{BC} \mathcal{J}^{CD} \mathbb{N}_{DA}\;, \\& \widehat{{\rm Cas}}_{SO(1,1)} = \frac{1}{2} \mathcal{J}^{AB} (\delta_{BC}-\mathbb{N}_{BC}) \mathcal{J}^{CD} (\delta_{DA}-\mathbb{N}_{DA})\;,
   \end{split}
\end{equation}
where 
\begin{equation}
    \mathcal{J}_{AB}=P_A \frac{\partial}{\partial P^B}-P_B \frac{\partial}{\partial P^A}\;,
\end{equation}
are the $SO(5,1)$ generators and are projected to the subgroups by  the projectors. The Casimir equations are the easiest to write down in an explicit form and to solve by using the following new variables
\begin{equation}
\label{defrhophi}
    \rho=\frac{\xi+\chi}{2}=\frac{1+V}{2\sqrt{V}}\;,\quad \phi=\frac{\chi}{2}=\frac{1-U+V}{2\sqrt{V}}\;.
\end{equation}
Then we can use separation of variables to turn the two PDEs in (\ref{defectCasimireqs}) into two ODEs 
\begin{eqnarray}
    \label{defectCasimireqs1}
    \left[(1-\rho ^2)\partial_{\rho}^2 - \rho \partial_{\rho}+\widehat{C}_{\widehat{\Delta}, 0} \right]\widehat{g}_{\widehat{\Delta},s}(\rho,\phi)&&=0\;,\\\label{defectCasimireqs2}
\left[(1-\phi^2)
  \partial_{\phi}^2-3 \phi \partial_{\phi} +\widehat{C}_{0, s}\right] \widehat{g}_{\widehat{\Delta},s}(\rho,\phi)&& =0\;.
\end{eqnarray}
Note that these equations are the special case of the Casimir equations derived in \cite{Billo:2016cpy} with defect dimension $p=0$ and co-dimension $q=d-p=4$. For general defect dimensions $p$ and co-dimensions $q$, the solution is \cite{Billo:2016cpy}\footnote{Here we have slightly rewritten the result in a more symmetric form and used a different normalization. The two ${}_2F_1$ factors are related by $\widehat{\Delta}\leftrightarrow -s$, $p\leftrightarrow q-2$. }
\begin{equation}
\begin{split}
\label{pqdefectblock}
  \widehat{g}_{\widehat{\Delta},s}^{(p,q)}(\rho,\phi)={}&\left(2\rho\right)^{-\widehat{\Delta}}{}_2F_1\left(\frac{\widehat{\Delta}}{2},\frac{\widehat{\Delta}+1}{2};\widehat{\Delta}+1-\frac{p}{2};\frac{1}{\rho^2}\right) \\{}&\times (2\phi)^s {}_2 F_1\left( -\frac{s}{2},\frac{1-s}{2};\frac{2-q}{2}-s+1;\frac{1}{\phi^2}\right)\;,
 \end{split}
\end{equation}
with boundary conditions supplied by the defect channel OPE. Upon setting $p=0$, we find that the conformal blocks simplify and exhibit the following piecewise behavior depending on the value of the cross ratio $V$\footnote{Here we have restricted ourselves to $U>0$, $V>0$. The definition of conformal blocks can be extended. But this region will be sufficient for the OPE analysis.} (with an added label $+$ to be explained momentarily) 
\begin{equation}\label{gdefectpiecewise}
\begin{split}
    \widehat{g}^+_{\widehat{\Delta},s}={}&C_s^{(1)}\left(\frac{\chi}{2}\right)\left(\rho+\sqrt{\rho^2-1}\right)^{-\widehat{\Delta}}\\
    ={}&C_s^{(1)}\left(\frac{\chi}{2}\right)\times \left\{\begin{array}{l}V^{\frac{\widehat{\Delta}}{2}}\;,\quad \text{when } 0<V<1\;, \\V^{-\frac{\widehat{\Delta}}{2}}\;,\quad \text{when } V\geq 1\;.\end{array}\right.
\end{split}
\end{equation}
This behavior reflects an important property of defect correlators. In the defect two-point function there are only two inequivalent channels, i.e., bulk and defect channels. This should be contrasted with the three distinct channels in four-point functions. While the bulk channel can be naturally identified with the s-channel in the four-point function, the defect channel is more subtle and should be thought of,  in a sense,  as the simultaneous superposition of both t- and u-channels. More precisely, the t- and u-channels in four-point functions are related by exchanging 3 and 4. However, the symmetry in the $3\leftrightarrow 4$ interchange is built-in for defects. The two branches in (\ref{gdefectpiecewise}) are related by the $3\leftrightarrow 4$ interchange which reflects precisely this symmetry of defects. Moreover, in correspondence with the discussions in the previous subsection, we can also define structures which are antisymmetric under $3\leftrightarrow 4$. This is achieved by changing the sign of the $V^{-\frac{\widehat{\Delta}}{2}}$ factor in (\ref{gdefectpiecewise}) when $V\geq 1$
\begin{equation}
    \widehat{g}_{\widehat{\Delta},s}^-=C_s^{(1)}\left(\frac{\chi}{2}\right)\times \left\{\begin{array}{l}V^{\frac{\widehat{\Delta}}{2}}\;,\quad \text{when } 0<V<1\;, \\-V^{-\frac{\widehat{\Delta}}{2}}\;,\quad \text{when } V\geq 1\;.\end{array}\right.
\end{equation}
As we mentioned in the previous subsection, such structures do not show up if the defect correlator has only dependence on the conformal cross ratios. They do, however, play a role in giant graviton correlators because there is also dependence on the R-symmetry cross ratios which can supply the opposite parity. The existence of these parity odd defect channel conformal blocks is a unique feature of zero dimensional defects.\footnote{The piecewise behavior of defect channel conformal blocks also exists for higher defect dimensions and the two branches are related by t, u crossing (e.g., one can explicitly check this for $p=2$). However, parity odd defect channel conformal blocks do not exist because defect two-point functions must be parity even under $U\to U/V$, $V\to 1/V$, see footnote \ref{footnotehigherp}. }

\vspace{0.5cm}

\noindent{\bf R-symmetry blocks.} Let us now switch our discussion to R-symmetry blocks. Similar to conformal blocks, R-symmetry blocks capture the information of exchanging R-symmetry representations. In the bulk channel, the defect R-symmetry block coincides with that of the four-point functions. This can be seen by using an R-symmetry version of ``OPE'' where the product of the two external representations are projected into various irreducible representations. For exchanging an $SO(6)$ representation with the $SU(4)$ Dynkin label $[\frac{d_2}{2},d_1,\frac{d_2}{2}]$, the R-symmetry block is \cite{Nirschl:2004pa}
\begin{equation}\label{defbulkQ}
    \mathcal{Q}_{\{d_1,d_2\}}^{k_1,k_2}= \mathcal{K}_{d_i,k_{12}} (\sigma\tau)^{-\frac{k_1}{2}}Y^{(\frac{k_{21}}{2}, \frac{k_{21}}{2})}_{
\frac{d_1+d_2+k_{12}}{2}, \frac{d_1+k_{12}}{2}}\;,
\end{equation}
where 
 \begin{equation}
     Y_{m,n}^{(a,b)}=\frac{P_{m+1}^{(a, b)}(y) P_n^{(a, b)}(\bar{y})-P_n^{(a, b)}(y) P_{m+1}^{(a, b)}(\bar{y})}{y-\bar{y}}\;.
 \end{equation}
We have introduced $y=2\alpha-1,\bar{y}=2\bar{\alpha}-1$ and $P_n^{(a,b)}(y)$ is the Jacobi polynomial. The constant coefficient is chosen to be
\begin{equation}
     \mathcal{K}_{d_i,k_{12}} = \frac{2(\frac{d_1+k_{12}}{2})! (\frac{d_1-k_{12}}{2})!(\frac{d_1+d_2+k_{12}+2}{2})! (\frac{d_1+d_2-k_{12}+2}{2})!}{d_1 ! (d_1+d_2+2)!}\;.
\end{equation}
With this normalization, the R-symmetry blocks behave as $\mathcal{Q}_{\{d_1,d_2\}}^{k_1,k_2}\sim\alpha^\frac{d_1+d_2-k_1-k_2}{2} \bar{\alpha}^{\frac{d_1-k_1-k_2}{2}} $ in the bulk channel OPE limit where $\alpha,\bar{\alpha} \to \infty$.

In the defect channel, $SO(6)$ R-symmetry is broken into $SO(4)\times SO(2)$. Again the R-symmetry blocks can be conveniently characterized by the Casimir equations. Let us consider exchanging a defect operator in the representation $(\frac{r_1}{2},\frac{r_1}{2})$ of $SO(4)$ and with $SO(2)$ charge $r_2$, and denote the corresponding R-symmetry block by $\widehat{\mathcal{Q}}^\pm_{r_1,r_2}$.
The Casimir equations take the same form as (\ref{Casimirconf})
\begin{equation}
\begin{aligned}
\label{defectRCasimireqs}
 \left(\widehat{{\rm Cas}}'_{SO(4)}+\widehat{C}_{0, r_1}\right) \prod_{i=1}^{2} \left(-\frac{t_i \cdot \mathbb{M} \cdot t_i}{2} \right)^{\frac{k_i}{2}} \widehat{\mathcal{Q}}_{r_1,r_2}(\sigma,\tau)&=0 \;,\\
 \left(\widehat{{\rm Cas}}'_{SO(2)}+\widehat{C}_{r_2, 0}\right) \prod_{i=1}^{2} \left(-\frac{t_i \cdot \mathbb{M} \cdot t_i}{2} \right)^{\frac{k_i}{2}} \widehat{\mathcal{Q}}_{r_1,r_2}(\sigma,\tau)&=0\;,
\end{aligned}
\end{equation}
where $\widehat{C}_{0,r_1}=r_1(r_1+2)$, $\widehat{C}_{r_2,0}=r_2^2$. The quadratic one-particle Casimirs are defined as
\begin{equation}
\begin{split}
   & \widehat{{\rm Cas}}'_{SO(4)}= \frac{1}{2} \mathcal{K}^{IJ} \mathbb{M}_{JK} \mathcal{K}^{KL} \mathbb{M}_{LI}\;,\\& \widehat{{\rm Cas}}'_{SO(2)} = \frac{1}{2} \mathcal{K}^{IJ} (\delta_{JK}-\mathbb{M}_{JK}) \mathcal{K}^{KL} (\delta_{LI}-\mathbb{M}_{LI})\;,
   \end{split}
\end{equation}
where
\begin{equation}
    \mathcal{K}_{IJ}=t_I \frac{\partial}{\partial t^J}-t_J \frac{\partial}{\partial t^I}\;.
\end{equation}
Similar to the case of conformal blocks, these equations can be decoupled by using a different set of variables 
\begin{equation}
    \rho'=\frac{\bar{\Sigma}-\Sigma}{2}\;,\quad \phi'=\frac{\bar{\Sigma}}{2}\;.
\end{equation}
Then the Casimir equations become the same as (\ref{defectCasimireqs1}) and (\ref{defectCasimireqs2})
\begin{eqnarray}
    \label{defectRCasimireqs1}
    \left[(1-{\rho'}^2)\partial_{\rho'}^2 - 3\rho' \partial_{\rho'}+\widehat{C}_{0, r_1} \right]\widehat{\mathcal{Q}}_{r_1,r_2}(\rho',\phi')&&=0\;,\\\label{defectRCasimireqs2}
\left[(1-\phi'^2)
  \partial_{\phi'}^2- \phi' \partial_{\phi'} +\widehat{C}_{r_2, 0}\right] \widehat{\mathcal{Q}}_{r_1,r_2}(\rho',\phi')&& =0\;.
\end{eqnarray}
These equations are further supplemented by the boundary condition
$\widehat{\mathcal{Q}}_{r_1,r_2} \sim (2\rho')^{r_1} \bar{\Sigma}^{-r_2}$ in the defect OPE limit $\rho',\bar{\Sigma} \to \infty$. The defect channel R-symmetry polynomials are solved to be\footnote{Similar to the conformal blocks, we have also assumed $\sigma>0$ and $\tau>0$. This is also sufficient for the decomposition into different R-symmetry structures.} 
\begin{equation}
    \begin{split}
\widehat{\mathcal{Q}}^+_{r_1,r_2} = {}&(-1)^{r_1}C_{r_1}^{(1)}\left(\frac{\bar{\Sigma}-\Sigma}{2}\right)\left(\frac{\bar{\Sigma}}{2}+\sqrt{\frac{\bar{\Sigma}^2}{4}-1}\right)^{-r_2}\\
    ={}&C_{r_1}^{(1)}\left(\frac{1-\sigma-\tau}{2\sqrt{\sigma \tau}}\right)\times \left\{\begin{array}{l}\left(\frac{\sigma}{\tau} \right)^{\frac{r_2}{2}}\;,\quad \text{when } 0<\frac{\sigma}{\tau}<1\;, \\\left(\frac{\sigma}{\tau} \right)^{-\frac{r_2}{2}}\;,\quad \text{when } \frac{\sigma}{\tau}\geq 1\;.\end{array}\right.
\end{split}
\end{equation}
It is easy to see the similarity with the defect channel conformal blocks (\ref{gdefectpiecewise}) and we see that $\mathcal{\widehat{Q}}^+_{r_1,r_2}$ also has a piecewise behavior. This solution is symmetric under the $3\leftrightarrow 4$ exchange, i.e., under $\sigma\leftrightarrow\tau$. We can also get the antisymmetric combination by adding an overall minus sign to one of the segments of the piecewise function, i.e., 
\begin{equation}\label{defQhatpm}
 \widehat{\mathcal{Q}}^\pm_{r_1,r_2} =C_{r_1}^{(1)}\left(\frac{1-\sigma-\tau}{2\sqrt{\sigma \tau}}\right)\times \left\{\begin{array}{l}\left(\frac{\sigma}{\tau} \right)^{\frac{r_2}{2}}\;,\quad \text{when } 0<\frac{\sigma}{\tau}<1\;, \\\pm\left(\frac{\sigma}{\tau} \right)^{-\frac{r_2}{2}}\;,\quad \text{when } \frac{\sigma}{\tau}\geq 1\;.\end{array}\right.
\end{equation}

\vspace{0.5cm}

\noindent{\bf Superconformal blocks.} Finally, we discuss superconformal blocks. They take the following general form as the sum over the superconformal primary and descendants
\begin{equation}
\begin{split}
    \mathfrak{G}_{\mathcal{M}}={}&\sum_{\mathcal{O}\in\mathcal{M}} \mu_{\mathcal{O}}  g_{\Delta_{\mathcal{O}},\ell_{\mathcal{O}}} \mathcal{Q}_{R_\mathcal{O}}\;,\\  \widehat{\mathfrak{G}}_{\widehat{\mathcal{M}}}={}&\sum_{\widehat{\mathcal{O}}\in\widehat{\mathcal{M}}} \sum_{a=\pm} \mu_{\widehat{\mathcal{O}}}^a  \widehat{g}_{\widehat{\Delta}_{\widehat{\mathcal{O}}},s_{\widehat{\mathcal{O}}}}^a \mathcal{\widehat{Q}}_{\widehat{R}_{\widehat{\mathcal{O}}}}^a\;,
\end{split}
\end{equation}
where $\mathcal{M}$ and $\widehat{\mathcal{M}}$ are superconformal multiplets in bulk and defect channels respectively. 
Note that in the defect case, we also need to sum over parities. The sum is diagonal (i.e., $++$ or $--$) because the superconformal block should have overall positive parity under the $3\leftrightarrow 4$ exchange.
The linear combination coefficients $\mu_{\mathcal{O}}$ and $\widehat{\mu}_{\widehat{\mathcal{O}}}^a$ can be fixed by imposing the superconformal Ward identities (\ref{scfWardid}). Because the bulk channel conformal blocks and R-symmetry blocks are proportional to the four-point function ones, the superconformal blocks are also essentially identical. Only the defect channel superconformal blocks are nontrivial new objects. Note that in defining the defect channel conformal blocks and the R-symmetry blocks we have assumed $U>0$, $V>0$, $\sigma>0$, $\tau>0$. This is nevertheless compatible with the twisting $\alpha=1/z$ in (\ref{scfWardid}), namely, we can find $(z,\bar{z},\bar{\alpha})$ so that all the positivity conditions are met (see (\ref{twistrange})).

For the purpose of extracting data of unprotected operators in Section \ref{Sec:extractingCFTdata}, we will need in particular the superconformal blocks for long multiplets. In the defect channel (which will be our primary focus), the result is particularly simple. The contribution to the correlator from a long multiplet, of which the superconformal primary has quantum numbers $(\widehat{\Delta}, s, r_1, r_2)$, is 
\begin{equation}
\label{Glongpm}
\widehat{\mathfrak{G}}_{\widehat{\Delta},s,r_1,r_2}^{\text{long}}=R(V\sigma\tau)^{-1}\widehat{g}_{\widehat{\Delta}+2, s}^{+} \widehat{\mathcal{Q}}_{r_1,r_2+2}^{+}\;.
\end{equation}
In other words, the contribution to the reduced correlator $\mathcal{H}$ is just the product of a conformal block and R-symmetry block with shifted quantum numbers with $++$ parities. We should emphasize that there is only one type of long superconformal block and replacing $++$ with $--$ does not give rise to a new acceptable solution. This is because $\mathcal{H}$ is separately invariant under $\{U,V\}\to\{U/V,1/V\}$ and $\{\sigma,\tau\}\to\{\tau,\sigma\}$.\footnote{A quick check is to look at the example of $k_1=k_2=2$. Due to the shifted weights mentioned after (\ref{defGandH}), the reduced correlator $\mathcal{H}$ is an R-symmetry singlet (independent of $\sigma$, $\tau$). Invariance under $3\leftrightarrow4$ then implies $\mathcal{H}$ is invariant under $\{U,V\}\to\{U/V,1/V\}$.} The easiest way to see this in general is to use the bulk channel superconformal block decomposition. It is a standard result of four-point functions of $\frac{1}{2}$-BPS operators (with identical operators $\mathcal{O}_3$ and $\mathcal{O}_4$) that the decomposition only consists of super primaries of even spins. On the other hand, each superconformal block contributes to the reduced correlator as a bosonic conformal block with the same parity of spins \cite{Beem:2016wfs}. The contribution consequently has separate positive parities with respect to $\{U,V\}\to\{U/V,1/V\}$ and $\{\sigma,\tau\}\to\{\tau,\sigma\}$, and it follows the reduced correlator also has the same transformation property.

More details of the superconformal blocks are presented in Appendix \ref{App:scfblocks} where we also collect results for other types of superconformal multiplets.

\section{Defects in AdS and supergravity description}\label{Sec:sugraanalysis}
In this section, we review the supergravity description of sphere giant gravitons in AdS$_5\times$S$^5$. The purpose of this section is three-fold. First,  to use bootstrap techniques to compute giant graviton correlators at strong coupling, we need some mild input in addition to the kinematic constraints obtained from last section. This input is supplied by the supergravity analysis. In particular, we will need the spectrum of the defect fluctuations and the schematic form of the couplings between the bulk and defect modes. While the former has been worked out (and we reproduce the result here), an analysis of the latter has not appeared in the literature. Second, an important feature of the classical giant graviton solution is the existence of moduli, and an averaging over the moduli space is needed in order to produce results for physical observables. However, how the averaging is performed can be quite subtle and affects the defect description. We will clarify this issue by highlighting the difference between two prescriptions of averaging which we will interpret as classical and quantum. The correct prescription corresponds to the one which gives rise to a quantum state description of the giant graviton. Finally, we also compute the defect one-point functions which we will not input into the later bootstrap analysis but will serve as a nontrivial consistency check. 

\subsection{D3 brane action and classical solution}\label{Sec:sugraL}

\begin{figure}
 \centering  \includegraphics[width=0.9\linewidth]{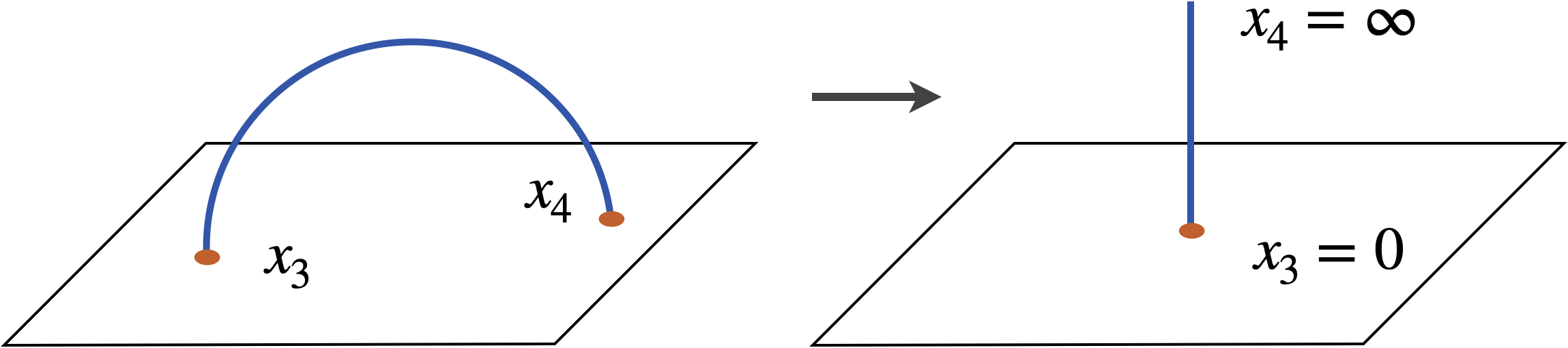}
  \caption{A conformal transformation moves the giant graviton insertions to $0$ and $\infty$, and maps the geodesic in the Poincar\'e patch into a straight line.}
  \label{Fig:Confromaltransform}
\end{figure}

The giant graviton is dual to a D3 brane wrapping an S$^3$ inside the S$^5$ factor of AdS$_5\times$S$^5$. The remaining direction is a geodesic line in AdS$_5$ which intersects the conformal boundary at two points. To simplify the analysis, we can use conformal symmetry to insert the two giant graviton operators at $x_3=0$ and $x_4=\infty$ (see Figure \ref{Fig:Confromaltransform}). Then the geodesic line is the following straight line $\mathbb{R}$ in Poincar\'e coordinates parameterized by the proper length $\mathsf{t}\in(-\infty,\infty)$ 
\begin{equation}\label{geodesicsR}
    z_{i=1,\ldots,4}(\mathsf{t})=x^{i=1,\ldots,4}(\mathsf{t})=0\;,\quad z_0(\mathsf{t})=x^0(\mathsf{t})\equiv x^\mathsf{t}(\mathsf{t})=e^\mathsf{t}\;.
\end{equation}
To discuss its dynamics, we need the action of the D3 brane which is the sum of the Dirac-Born-Infeld action and the Wess-Zumino action
\begin{equation}
    S_{D3}=-\frac{N}{2\pi^2}\int d ^4 \sigma \left( \sqrt{\det h_{\alpha \beta}}-iP[C_4] \right)\;.
\end{equation}
Here $\sigma^{\alpha=0,1,2,3}$ denote the worldvolume coordinates of the D3 brane, and $h_{\alpha\beta}$ is the pull-back of the background metric
\begin{equation}
   h_{\alpha\beta}= \partial_{\alpha} X^{M} \partial_{\beta} X^{N} G_{MN} \;,
\end{equation}
where $M,N=0,1,\ldots,9$ and the  coordinates $X^M$ describe the embedding of the D3 brane in AdS$_5\times $S$^5$ and $P[C_4]$ is the pullback of the background Ramond-Ramond four form. In accordance with the choice (\ref{geodesicsR}), it is convenient to write the AdS$_5\times $S$^5$ coordinates as
\begin{equation}
\label{defX}
    X^{M}:\quad \left(X^{\mu}=(\mathsf{t},\mathsf{x}^i)\;,\;X^{A}=(y^a,\mathsf{q},\Phi)\right)\;,\quad\quad  \mathsf{x}^i\in \mathbb{R}^4\;,\quad y^a\in \mathbb{R}^3\;,
\end{equation}
where $\mu,\nu\ldots=\{0,1,2,3,4\}$ denote components of the AdS$_5$ and $A,B\ldots=\{5,6,7,8,9\}$ represent the components of S$^5$. In this parametrization, the AdS$_5\times $S$^5$ metric becomes
\begin{equation}\label{dsAdS5S5}
    ds^2= \frac{(1+\frac{(\mathsf{x}^i)^2}{4})^2}{(1-\frac{(\mathsf{x}^i)^2}{4})^2} d\mathsf{t}^2+\frac{d\mathsf{x}^i d \mathsf{x}^i}{(1-\frac{(\mathsf{x}^i)^2}{4})^2} + \mathsf{q}^2 \frac{dy^a dy^a}{(1+\frac{(y^a)^2}{4})^2}+\frac{d \mathsf{q}^2}{1-\mathsf{q}^2}+(1-\mathsf{q}^2) d \Phi^2\;,
\end{equation}
where we have set the common radius of AdS$_5$ and S$^5$ to 1. In other words, the AdS$_5$ factor consists of the geodesics $\mathbb{R}$ parameterized by $\mathsf{t}$ and the transverse space  parameterized by $\mathsf{x}^i\in \mathbb{R}^4$. In the S$^5$ factor the D3 brane occupies an S$^3$ with radius $\mathsf{q}\in[0,1]$ which is parameterized by the stereographic coordinates $y^a\in \mathbb{R}^3$. In addition to $\mathsf{q}$, the other transverse direction is $\Phi$ which parameterizes a circle with radius $\sqrt{1-\mathsf{q}^2}$.

The classical giant graviton solution is 
\begin{equation}\label{classicalsol}
    \mathsf{x}^i=0\;,\quad \mathsf{q}=q_0\;,\quad \Phi=i\mathsf{t}\;.
\end{equation}
Here $q_0=1$ corresponds to a {\it maximal} giant graviton while $0<q_0<1$ amounts to a {\it non-maximal} giant graviton. This solution also allows us to explicitly identify the worldvolume coordinates of the D3 brane as 
\begin{equation}
    \sigma^{\alpha}=(\mathsf{t},y^a)\;.
\end{equation}
It is clear from (\ref{dsAdS5S5}) that the D3 brane occupies the subspace
\begin{equation}
\label{worldvolumemetric}
    ds^2_{\rm worldvolume}=q_0^2\left(d\mathsf{t}^2+\frac{dy^a dy^a}{(1+\frac{(y^a)^2}{4})^2}\right)\;,
\end{equation}
which is precisely $\mathbb{R}\times$S$^3$. However, it is important to notice that the giant graviton solution is not unique because new solutions can be obtained from (\ref{classicalsol}) by applying the shifts
\begin{equation}
    \mathsf{t} \to \mathsf{t}+\mathsf{t}_0\;,\quad \Phi\to \Phi+\Phi_0\;,
\end{equation}
which are generated by the actions of $SO(1,1)\subset SO(5,1)$ and $SO(2)\subset SO(6)$. These are the moduli of the giant graviton solutions and correctly averaging over this moduli space is key to reproducing field theory observables from the dual supergravity description. We will clarify this issue in the next subsection.

\subsection{Quantum versus classical averaging}
On top of the classical solution, we can consider small fluctuations. These fluctuations can be from the defect degrees of freedom (of the D3 brane) and from the fluctuation of bulk fields restricted to the defect. As we mentioned in the previous section, the classical solution of giant gravitons has moduli and on general grounds we should expect an averaging over the moduli space to produce physical observables. However, how this averaging is performed is quite subtle and we can sharpen the problem by considering multi-point (more than two) correlations of bulk fluctuations. 

Starting from the D-brane action, once we have expanded it into fluctuations around the classical solution, it appears that we have two ways to proceed. 
\begin{itemize}
    \item {\bf Prescription 1 (moduli averaging last):} We contract the bulk fluctuations on the defect with the sources on the boundary of AdS using propagators. We compute its contribution in terms of Witten diagrams for fixed moduli parameters. Finally, we average the result over the moduli space to obtain the correlator.
    \item {\bf Prescription 2 (moduli averaging first):} We first average the expanded action over moduli space to obtain a new effective action for the fluctuations. Using this new action we compute correlators by contracting with propagators and performing the corresponding Witten diagram integrals. 
\end{itemize}
These two prescriptions differ by at which point in the calculation averaging is performed, and lead to physically inequivalent answers. The reason for the difference is that they correspond to {\it classical} and {\it quantum} averaging respectively. To see the difference more clearly, let us consider the following schematic calculation for two-point correlations. The first prescription with averaging performed at the last step gives a statistical mixture
\begin{equation}
    \langle \mathcal{O}\mathcal{O}\rangle_{\text{Presc 1}}=\int d\Omega \langle \Omega | \mathcal{O}\mathcal{O}|\Omega \rangle ={\rm Tr}\left(\rho_{\rm mixed}\mathcal{O}\mathcal{O}\right)\;, 
\end{equation}
where $\Omega$ labels the moduli space and the averaging gives rise to a density matrix 
\begin{equation}
    \rho_{\rm mixed}=\int d\Omega |\Omega \rangle \langle \Omega |\;.
\end{equation}
The second prescription instead defines a superposition of quantum states by first averaging over the moduli space
\begin{equation}
    |\Psi\rangle =\int d\Omega |\Omega \rangle\;,
\end{equation}
and gives the correlator as the expectation value with respect to this state
\begin{equation}
    \langle \mathcal{O}\mathcal{O}\rangle_{\text{Presc 2}}=\langle \Psi | \mathcal{O}\mathcal{O}|\Psi \rangle \;.
\end{equation}
Very importantly, the second prescription includes the interference terms $\langle \Omega|\mathcal{O}\mathcal{O}|\Omega'\rangle$ that were thrown out by the first prescription. The giant graviton correlators are quantum correlators in the CFT and therefore should be obtained using the second prescription. To phrase it differently, the classical solution at a fixed moduli space point corresponds to a {\it coherent state} in the boundary CFT and does not have definite quantum numbers. The giant graviton state, however, is an eigenstate with specific quantum numbers and is obtained by averaging over the moduli space. 

Another way to convince oneself that Prescription 2 is the correct one is to consider the disconnected defect two-point function
\begin{equation}
    \llangle \mathcal{O}_1\rrangle \llangle \mathcal{O}_2\rrangle=\frac{\langle \mathcal{O}_1\mathcal{D}\mathcal{D}\rangle\langle \mathcal{O}_2\mathcal{D}\mathcal{D}\rangle}{\langle\mathcal{D}\mathcal{D}\rangle^2}\;,
\end{equation}
which is well defined in the boundary CFT and corresponds to the leading large $N$ contribution to the defect two-point function. To reproduce it from bulk supergravity, clearly only the terms linear in the bulk fluctuations are needed. If we follow Prescription 1, the contributions to the correlator from each operator are factorized for each fixed moduli point. However, averaging over the moduli space in the last step of Prescription 1 would couple these contributions and destroy the factorized structure. On the other hand, the Prescription 2 clearly preserves the factorization. 

We should comment that the necessity of averaging over the moduli space has already been emphasized in \cite{Yang:2021kot,Holguin:2022zii}. However, in these works only defect one-point functions were considered where both prescriptions lead to the same answer. It is only when we consider correlation functions which are two-point  or higher, we can start to appreciate the subtlety contained in these two different averaging prescriptions. 

Having established that Prescription 2 is the correct one, we can proceed with outlining the supergravity approach to computing the giant graviton correlators. We start from the D3 brane effective Lagrangian involving both bulk and defect fluctuations around the classical solution and integrate over the moduli space
\begin{equation}
    \int d\Omega\; L_{\rm D3}[\delta \Phi_{\rm bulk},\delta \Phi_{\rm defect}]\;.
\end{equation}
After further integrating over S$^3$, we are left with an effective 1d Lagrangian in AdS$_5$. We can expand it with respect to the fluctuation modes and enumerate the expansion in terms of numbers of fields 
\begin{equation}\label{S1dasLmn}
    S_{\rm 1d}=\int d\mathsf{t} \left({\rm const} + L^{(0,1)}+L^{(2,0)}+L^{(1,1)}+L^{(0,2)}+\ldots\right)\;,
\end{equation}
where $L^{(m,n)}$ involves $m$ defect modes and $n$ bulk modes. These give rise to the vertices ($L^{(2,0)}$ gives the defect propagators) which one could input into the standard AdS perturbation theory to compute correlators using Witten diagrams.

Note while this is in principle possible, implementing the algorithm in practice is very cumbersome. Moreover, it is known in similar but simpler setups that there are unresolved subtleties where the supergravity vertices seemingly lead to defect correlators incompatible with superconformal symmetry \cite{Gimenez-Grau:2023fcy,Zhou:2024ekb}. Later in this paper, we will use a bootstrap approach which does not rely on the explicit details of the vertices. Nevertheless, in the following subsections we will discuss these terms in the effective action in more details. Some of the discussions (one-point functions) will serve as independent checks for the bootstrap results.

\subsection{Defect spectrum}\label{Subsect:defectspec}
Let us first focus on the  fluctuation spectrum of the defect which is encoded in $L^{(2,0)}$ of the expansion (\ref{S1dasLmn}). This has already been analyzed in \cite{Das:2000st} and we review it here for completeness. For the purpose of this subsection, we can temporarily turn off the bulk fluctuations and consider only the fluctuations of the transverse degrees of freedom of the defect\footnote{We have also turned off the gauge fields on the D3 brane. These modes will not be relevant for the computation of the defect two-point correlators.}
\begin{equation}
       \mathsf{x}^i=\epsilon\, x^i(\mathsf{t},y^a)\;,\;\quad \mathsf{q}=q_0(1+\epsilon\, q(\mathsf{t},y^a))\;,\;\quad  \Phi=i(\mathsf{t}+\epsilon\, \phi(\mathsf{t},y^a)) \;.
\end{equation}
Here we have introduced a bookkeeping parameter $\epsilon$ to keep track of the order of the expansion. Substituting this into \eqref{defX}, we find that the induced metric becomes
\begin{equation}
\begin{split}
h_{\alpha \beta}= {}& \frac{(1+\frac{(\epsilon x^i)^2}{4})^2}{(1-\frac{(\epsilon x^i)^2}{4})^2}\partial_{\alpha}\mathsf{t} \partial_{\beta}\mathsf{t}+ \frac{\epsilon^2\partial_{\alpha} x^i \partial_{\beta} x^i}{(1-\frac{(\epsilon x^i)^2}{4})^2}+\frac{q_0^2(1+\epsilon q)^2 \partial_{\alpha}y^a \partial_{\beta} y^a}{(1+\frac{(y^a)^2}{4})^2}\\&+\frac{\epsilon^2\partial_{\alpha} q \partial_{\beta} q}{1-q_0^2(1+\epsilon q)^2}-(1-q_0^2(1+\epsilon q)^2)\partial_{\alpha}(\mathsf{t}+\epsilon \phi)\partial_{\beta}(\mathsf{t}+\epsilon \phi)\;.
\end{split}
\end{equation} 
Further using $\partial_{\alpha} \mathsf{t}=\delta_{\alpha0}\;, \partial_{\alpha} y^a=\delta_{\alpha a}$ in the frame (\ref{geodesicsR}), we can evaluate the determinant term as
\begin{equation}
    \sqrt{\det h_{\alpha \beta}}=q_0^2\sqrt{g} \left(f_0+\epsilon f_1+\epsilon^2 f_2+\mathcal{O}(\epsilon^3)\right)\;,
\end{equation}
where $\sqrt{g}=(1+\frac{(y^a)^2}{4})^{-3}$ and
\begin{equation}
\begin{split}
\nonumber
      &  f_0=q_0^2\;,\\& f_1=4q_0^2q-(1-q_0^2)\partial_{\mathsf{t}} \phi\;,\\& f_2=\frac{1}{2}\partial_{\alpha} x^i \partial^{\alpha} x^i+\frac{1}{2}x^i x^i +\frac{q_0^2 ( \partial_{\alpha}q\partial^{\alpha} q)}{2(1-q_0^2)}+6q_0^2 q^2  -\frac{1-q_0^2}{2q_0^2}(\partial_{\alpha}\phi \partial^{\alpha} \phi)-2(1-2q_0^2)q \partial_{\mathsf{t}} \phi\;.
\end{split}
\end{equation}
In the WZ term, the relevant component of the 4-form potential is
\begin{equation}
C_{ y^1y^2y^3 \Phi}=\mathsf{q}^4 \sqrt{g}\;,
\end{equation}
which leads to 
\begin{equation}
\begin{split}
iP[C_4]={}&\frac{i}{4!} C_{ABCD} \frac{\partial X^{A}}{\partial \sigma^{\alpha_1}} \frac{\partial X^{B}}{\partial \sigma^{\alpha_2}} \frac{\partial X^{C}}{\partial \sigma^{\alpha_3}} \frac{\partial X^{D}}{\partial \sigma^{\alpha_3}} \varepsilon^{\alpha_1 \alpha_2 \alpha_3 \alpha_4}\\={}& q_0^4 \sqrt{g}(1+(4q+\partial_{\mathsf{t}}\phi)\epsilon+(6q^2+4q \partial_{\mathsf{t}} \phi)\epsilon^2+\mathcal{O}(\epsilon^3))\;.
\end{split}
\end{equation}
Therefore, to the order $\mathcal{O}(\epsilon^2)$, the D3 brane effective action gives 
\begin{equation}
    S_{D3}=-\frac{N}{2\pi^2}\int d^4\sigma q_0^2 \sqrt{g} \left( (-\partial_{\mathsf{t}}\phi) \epsilon+(f_2-q_0^2(6q^2+4q\partial_{\mathsf{t}} \phi))\epsilon^2\right)\;.
\end{equation}
Note that the linear order term is a total derivative and vanishes after integrating over $\mathsf{t}$. The nontrivial contribution comes from the quadratic term, which after integrating over S$^3$ gives $L^{(2,0)}$ in the notation of (\ref{S1dasLmn}). Averaging over moduli is trivial at this order. Explicitly, we have
\begin{equation}
    \begin{split}
    \label{L20}
    L^{(2,0)}=-\frac{N}{2\pi^2}\int \sqrt{g}\, dy^a q_0^2 {}&\left( \frac{1}{2}\partial_{\alpha} x^i \partial^{\alpha} x^i+\frac{1}{2}x^i x^i \right.\\{}&\quad \left.+ \frac{1}{2}\partial_{\alpha}q\partial^{\alpha} q -\frac{1}{2}\partial_{\alpha}\phi \partial^{\alpha} \phi-2 q \partial_{\mathsf{t}} \phi\right)\;,
    \end{split}
\end{equation}
where we have already rescaled the fluctuations as 
\begin{equation}
    \frac{q_0}{\sqrt{1-q_0^2}}q\to q\;, \quad \frac{\sqrt{1-q_0^2}}{q_0}\phi\to \phi\;.
\end{equation}
To read off the spectrum, we decompose the fluctuations into  spherical harmonics
\begin{equation}
    \begin{split}
    \label{S3sphharmexpan}
        x^i(\mathsf{t},y^a)&=\sum_{l,m}x^i_{l,m}(\mathsf{t}) \widehat{Y}_{l-1}^m(y^a)\;,\\  q(\mathsf{t},y^a)&=\sum_{l,m}q_{l,m}(\mathsf{t}) \widehat{Y}_{l-1}^m(y^a)\;,\\  \phi(\mathsf{t},y^a)&=\sum_{l,m}\phi_{l,m}(\mathsf{t}) \widehat{Y}_{l-1}^m(y^a)\;,
    \end{split}
\end{equation}
where $\widehat{Y}_{l}^m(y^a)$ is the S$^3$ scalar sphere harmonics of rank $l$ and $m$ labels different states inside the representation. For notational simplicity, we will leave the $m$ index implicit from now on. The integration over S$^3$ can be straightforwardly performed by using the standard properties of the spherical harmonics, which makes the effective action one dimensional.

For the fluctuations $x^i_{l}$, we find that the equation of motion is already diagonalized
\begin{equation}
  \left( \partial_{\mathsf{t}}^2 -l^2 \right) x^i_{l}=0\;.
\end{equation} 
From the standard mass-dimension relation $m^2=\Delta(\Delta-d)$ with $d=0$, we find $x^i_{l}$ has conformal dimension $l$. Note that because of the $i$ index, these modes also have transverse spin $s=1$ under the $SO(4)$ subgroup of the conformal group. For the residual R-symmetry group, these modes have charges $(\frac{r_1}{2},\frac{r_1}{2})$with respect to $SO(4)=SU(2)\times SU(2)$ with $r_1=l-1$ and are singlet ($r_2=0$) under $SO(2)$.

 For the $q_l$ and $\phi_l$ fluctuations, we need the following field redefinitions
\begin{equation}
    q_{l}(\mathsf{t})=\zeta_{l}(\mathsf{t})+\bar{\zeta}_{l}(\mathsf{t})\;,\quad  \phi_{l}(\mathsf{t})=\zeta_{l}(\mathsf{t})-\bar{\zeta}_{l}(\mathsf{t}) \;,
\end{equation}
to diagonalize the equations of motion. In terms of these combinations, the kinetic terms  become that of a complex scalar in a background $U(1)$ gauge field
\begin{equation}
\begin{split}
    &\int_{\text{S}^3} \frac{1}{2}\partial_{\alpha}q\partial^{\alpha} q  - \frac{1}{2}\partial_{\alpha}\phi \partial^{\alpha} \phi -2 q\partial_{\mathsf{t}} \phi= 2\left( (\partial_{\mathsf{t}}+1)\zeta_{l}(\partial_{\mathsf{t}}-1) \bar{\zeta}_{l}+ l^2 \zeta_{l} \bar{\zeta}_{l} \right)\;.
    \end{split}
\end{equation}
The equations of motion are
\begin{equation}
         \left(\partial_{\mathsf{t}}^2 +2\partial_{\mathsf{t}}-(l-1)(l+1)\right)\zeta_{l}=0\;,\quad  \left(\partial_{\mathsf{t}}^2 -2\partial_{\mathsf{t}}-(l-1)(l+1)\right)\bar{\zeta}_{l}=0\;.
\end{equation}
Note that the 1d propagator between $\zeta_l$ and $\bar{\zeta}_l$ has the general form $G(\mathsf{t}_1,\mathsf{t}_2)\sim e^{-\Delta|\mathsf{t}_1-\mathsf{t}_2|}$. From the equations of motion, we can see that if $\zeta_l$ is inserted at an earlier time than $\bar{\zeta}_l$ then $\Delta=l+1$. On the other hand, if the time ordering is reversed then we have $\Delta=l-1$. Said differently, the background $U(1)$ gauge field breaks the symmetry of $\mathsf{t}\to-\mathsf{t}$. The effective dimension is different depending on the alignment of the $U(1)$ charge flow direction with the $\mathsf{t}$ flow direction. There is yet another complication. The equation of motions we have derived so far are implicitly in the center-of-mass frame (CMF) of the D3 brane. However, the D3 brane itself is moving along the $\Phi$ direction. This will give rise to a ``Doppler effect'' which further shifts these dimensions by $\pm 1$ when we switch to the static frame of the AdS target space (where the defect picture lies). It follows from the chain rule ($\tau$ being the static time)
\begin{equation}
    \frac{\partial}{\partial \tau}= \frac{\partial \mathsf{t}}{\partial \tau}\frac{\partial}{\partial \mathsf{t}}+\frac{\partial\Phi}{\partial \tau} \frac{\partial}{\partial \Phi}\;,
\end{equation}
we have the following relation between conformal dimensions
\begin{equation}
    \Delta_{\rm static}=\Delta_{\rm CMF}+r_2\;.
\end{equation}
Here we have replaced the differential operators by their corresponding quantum numbers and $\partial/\partial \Phi$ in particular corresponds to the $U(1)$ charge $r_2$. We then find that for $\zeta_l$ and $\bar{\zeta}_l$ the effective defect dimensions are $l-2$ and $l+2$ in the static frame depending on the $U(1)$ flow direction. Note that this Doppler shift does not happen for $x^i_l$ modes and our earlier result remains valid because their $U(1)$ charge is zero.

Reorganizing the excitations into supermultiplets, we obtain the following Table \ref{table:defectmodes}. The table should be understood in the CFT sense. As we  explained, the propagator of the fluctuation modes $\zeta$ and $\bar{\zeta}$ has different effective conformal dimensions depending on the direction of propagation. It does not make sense to assign two conformal dimensions for a complex scalar. However, we can define defect operators $\chi$ and $\bar{\chi}$ which are obtained from $\zeta$ and $\bar{\zeta}$ and correspond to the two effective dimensions.\footnote{Some labels have also been shifted so that components of a multiplet all have the same label $l$. For the $\zeta_l$ and $\bar{\zeta}_l$ modes, we have found that the quantum numbers $(\Delta,r_1,r_2)$ in the static frame are $\{(l-2,l-1,-1),(l+2,l-1,1)\}$. We shift the label $l$ by $l+1$ and $l-1$ and respectively take the first and last elements of the two sets. They are $\chi_l$ and $\bar{\chi}_l$ in the table.} The results agree with spectrum in \cite{Imamura:2021ytr} obtained from representation theory.

\begin{table}[h]
\centering
\begin{tabular}{||cccc||}
\hline  Defect modes & $\chi_l$ & $x^i_l$ & $\bar{\chi}_l$  \\
\hline \hline 
 $\widehat{\Delta}$ & $l-1$ & $l$ & $l+1$ \rule{0pt}{2.5ex} \\  $s$ & $0$ & $1$ & $0$  \\
 $(r_1,r_2)$ & $(l,-1)$ & $(l-1,0)$ & $(l-2,1)$\\[0.12em]
\hline
\end{tabular}
\caption{Relevant components of a defect supermultiplet corresponding to the exchanged defect fluctuations. The modes are labeled by their conformal dimension $\widehat{\Delta}$, transverse spin $s$ and $SO(4)\times SO(2)$ R-symmetry representation $(r_1,r_2)$.}
\label{table:defectmodes}
\end{table}

\subsection{One-point functions}\label{Sec:sugra1pt}

\begin{table}[h]
\centering
\begin{tabular}{||ccccccc||}
\hline  Bulk modes & $s_k$ & $A_{k, \mu}$ & $\varphi_{k, \mu \nu}$ & $C_{k, \mu}$ & $t_k$ & $r_k$ \\
\hline \hline
 $\Delta$ & $k$ & $k+1$ & $k+2$ & $k+3$ & $k+4$ & $k+2$ \rule{0pt}{2.5ex}\\  $\ell$ & $0$ & $1$ & $2$ & $1$ & $0$ & $0$ \\
 $(d_1,d_2)$ & $(k,0)$ & $(k-2,2)$ & $(k-2,0)$ & $(k-4,2)$ & $(k-4,0)$ & $(k-4,4)$ \\[0.2em]
\hline
\end{tabular}
\caption{Spectrum of bulk fluctuations relevant for computing the defect two-point functions. Fields are labeled by  their conformal dimensions $\Delta$, Lorentz spins $\ell$ and R-symmetry representations $[\frac{d_2}{2},d_1,\frac{d_2}{2}]$ ($SU(4)$ Dynkin labels). }
\label{table:bulkmodes}
\end{table}

In this section, we examine the $L^{(0,1)}$ terms in the effective Lagrangian. This will give us the one-point functions which we will later use as an independent check of the bootstrap results. The insertion of the giant graviton operators does not alter the bulk spectrum because they are not heavy enough to back-react on the geometry.  We have recorded the relevant results from \cite{Kim:1985ez} in Table \ref{table:bulkmodes}. For the computation of one-point functions, we will focus on the superconformal primaries $s_k$ and the superconformal descendants $t_k$.

The relevant fluctuations of the bulk supergravity fields are\footnote{All the other modes in Table \ref{table:bulkmodes} are turned off since they are not needed in the computation of one-point functions.}
\begin{equation}
    \begin{aligned}
    \label{bulkexcitation}
\delta G_{\mu \nu}= {}&\epsilon  \Big(\frac{4}{k+1}\big(\nabla_\mu \nabla_\nu-\frac{1}{2} k(k-1) \bar{G}_{\mu \nu}\big) s_k\\&+\frac{4}{k+3}\big(\nabla_\mu \nabla_\nu-\frac{1}{2}(k+4)(k+5) \bar{G}_{\mu \nu}\big) t_{k+4}\Big) Y_k\;,\\ \delta G_{AB} = {}& 2 \bar{G}_{AB}\epsilon (  k  s_k+(k+4)t_{k+4})Y_k\;,\\
        \delta C_{ABCD} ={}& 4 \epsilon_{ABCDE} \epsilon ( s_k- t_{k+4})\nabla^{E} Y_k\;,
    \end{aligned}
\end{equation}
where $\bar{G}_{MN}$ denotes the background metric \eqref{dsAdS5S5}. The fluctuations are proportional to the scalar modes $s_k^I$ and $t_{k+4}^I$, and $Y_k^I$ are the S$^5$ scalar sphere harmonics of rank $k$. The index $I$ labeling different states in the representation is suppressed in these expressions for notational simplicity. Plugging these fluctuations into the D3 brane action gives us the $L^{(0,1)}$ vertices for the $s_k$ and $t_k$ fields. 

Let us first focus on the $s_k$ fields. After simplification, we find that the one-point vertices are\footnote{We have discarded terms of the form $\partial_{\tau}^2 s_k$ which are total derivatives of the $\tau$-integral.}
\begin{equation}
\label{L01sk}
    L^{(0,1)}(s_k) = -\frac{N}{2\pi^2} \int_{\rm S^3}  q_0^2\left(2k\Big(2q_0^2-\frac{k}{k+1}\Big)  +4 q_0(1-q_0^2) \partial_{q_0}  \right)Y_ks_k\;.
\end{equation}
Recall that $s_k$ are dual to the to the superconformal primaries of the $\frac{1}{2}$-BPS multiplet. If we use the same R-symmetry index labeling as $s_k^I$, the dual operator can be written as
\begin{equation}
    \mathcal{O}_k(x, t)=\frac{1}{k!}t^{i_1} \ldots t^{i_k} \mathcal{C}_{i_1 \ldots i_k}^I \mathcal{O}_k^I(x)\;.
\end{equation}
Here $\mathcal{C}_{i_1 \ldots i_k}^I$ is the $SO(6)$  invariant tensor converting between the $I$ index and the $SO(6)$ indices, and satisfies \cite{Lee:1998bxa}
\begin{equation}
    \begin{aligned}
& C_{i_1 \ldots i_k}^I C_{i_1 \ldots i_k}^J=\delta^{I J}\;, \\
& C_{i_1 \ldots i_k}^I C_{j_1 \ldots j_k}^I=\delta^{i_1 \ldots i_k,\left(j_1 \ldots j_k\right)}+(\text{mixing } i, j)\;.
\end{aligned}
\end{equation}
The S$^5$ scalar spherical harmonics $Y_k$ can also be expressed in terms of $C_{i_1 \ldots i_k}^I$ as
\begin{equation}\label{S5sphericalharmonics}
    Y_k^I = \frac{1}{\sqrt{Z(k)}}C_{i_1 \ldots i_k}^I n^{i_1}\ldots n^{i_k}\;,\quad Z(k) = \frac{\pi^3}{2^{k-1} (k+1)(k+2)}\;,
\end{equation}
where $n^i$ is the 6d embedding coordinates of $S^5$ satisfying $n^in^i=1$. To compute the one-point function, we replace $s_k$ by the bulk-to-boundary propagator (more details of propagators and Witten diagrams will be discussed in Section \ref{Sec:treelevelWittendiagrams})
\begin{equation}
    G^k_{B\partial}(x,z(\tau)) = \left(\frac{e^{\tau}}{e^{2 \tau}+x^2}\right)^k\;,
\end{equation}
which connects the operator insertion at boundary point $x$ and the bulk position along the 1d defect at value $\tau$. Note that in (\ref{L01sk}) the bulk fluctuations $s_k$ are restricted to the world volume of the D3 brane. Therefore, we should also restrict the S$^5$ coordinates in (\ref{S5sphericalharmonics}) to S$^3$
\begin{equation}
    n^i = \left(q_0 \frac{y^a}{1+\frac{1}{4} y^2}, q_0 \frac{1-\frac{1}{4} y^2}{1+\frac{1}{4} y^2}, \sqrt{1-q_0^2} \cos \Phi_0, \sqrt{1-q_0^2} \sin \Phi_0\right)\;.
\end{equation}
This gives us the following expression for the one-point functions 
\begin{equation}
\label{1ptsupergravity}
    \llangle \mathcal{O}_k(x,t)\rrangle = \mathcal{N}_k^s  \int_{-\infty}^{\infty} d \tau \int_0^{2\pi} \frac{d\Phi_0}{2\pi} \, \mathcal{L}^{(0,1)}\;,
\end{equation}
with 
\begin{equation}
    \mathcal{L}^{(0,1)} = -\frac{N}{2\pi^2} \int_{\rm S^3} \,\frac{ q_0^2 }{\sqrt{Z(k)}}\left(2k\Big( 2q_0^2-\frac{k}{k+1} \Big)+4q_0 (1-q_0^2) \partial_{q_0} \right) (t\cdot n)^k  G^k_{B\partial}(x,z(\tau))\;,
\end{equation}
where the integral over $\Phi_0$ is the moduli average. The overal factor $\mathcal{N}_k^s$ receives contributions from several parts
\begin{equation}
    \mathcal{N}_k^s = (\xi_{s,k})^{-\frac{1}{2}} N_k^{\rm  prop} = \frac{\pi ^{\frac{3}{2}} (k+1)}{2 N \sqrt{2 k (k+1) (k+2)}}\;.
\end{equation}
Here $\xi_{s, k}$ is the overall coefficients in the quadratic action of $s_k$ \cite{Arutyunov:1998hf} in AdS$_5$
\begin{equation}
S(s_k)=\int_{\rm AdS_5} \xi_{s, k}\left(-\frac{1}{2} \nabla_{\mu} s_k \nabla^{\mu} s_k-\frac{1}{2} k(k-4) s_k^2\right)\;, \quad \xi_{s, k}=\frac{128 N^2 k(k-1)(k+2)}{(2 \pi)^5(k+1)}\;,
\end{equation}
and $N_k^{\rm  prop}$  is the normalization for bulk-to-boundary propagator 
\begin{equation}
    N_k^{\rm  prop} = \pi^{-2}(k-2)(k-1) (N_k^{2 \mathrm{pt}})^{-\frac{1}{2}}\;,\quad   N_k^{2 \mathrm{pt}}=\frac{\Gamma(k)(2 k-4)}{\pi^{2} \Gamma(k-2)}\;,
\end{equation}
which ensures the two-point function of $\mathcal{O}_k$ are unit normalized. Then the computation ultimately boils down to integrals. The first integral is
\begin{equation}
    \int_{-\infty}^\infty d \tau \left(\frac{e^{\tau}}{e^{2\tau}+x^2}\right)^k =  \frac{\sqrt{\pi }\,  \Gamma (\frac{k}{2})}{2^k\Gamma (\frac{k+1}{2})} \frac{1}{\left(-P \cdot\mathbb{N}\cdot P\right)^{\frac{k}{2}}}\;,
\end{equation}
which gives the conformal part of the defect one-point function. The  second integral is
\begin{equation}
\nonumber
      \int_0^{2\pi} \frac{d \Phi_0}{2\pi} \int_{\rm S^3} (t\cdot n)^k = \frac{(1+(-1)^k) \pi^{\frac{3}{2}}\Gamma(\frac{k+1}{2})}{q_0^2 \Gamma(\frac{k}{2}+2)} {}_2 F_1 \left( -\frac{k+2}{2},\frac{k+2}{2};1;1-q_0^2 \right)(t\cdot \mathbb{M}\cdot t)^{\frac{k}{2}}\;,
\end{equation}
which generates the correct R-symmetry structure. Combining these results, we arrive at the following form of the one-point function
\begin{equation}
\label{1ptforsk}
    \llangle \mathcal{O}_k(x,t)\rrangle = a_{s_k} \frac{(\frac{1}{2}t\cdot \mathbb{M} \cdot t)^{\frac{k}{2}}}{\left(P \cdot\mathbb{N}\cdot P\right)^{\frac{k}{2}}}\;,
\end{equation}
where
\begin{equation}\label{sk1ptcoef}
    a_{s_k} = -\frac{(1+(-1)^k)(-1)^{\frac{k}{2}}}{\sqrt{k}}  {}_2F_1\left(-\frac{k}{2},\frac{k}{2};1;1-q_0^2\right)\;.
\end{equation}
This reproduces the result in \cite{Yang:2021kot}.

Following the same procedure, we can compute the defect one-point function for the operators dual to $t_{k}$, which we denote by $\mathcal{T}_k$. The supergravity analysis gives the following one-point vertices
\begin{equation}
\label{L01tk}
    L^{(0,1)}(t_k) = -\frac{N}{2\pi^2} \int_{\rm S^3}  q_0^2\left(2k\Big(2q_0^2-\frac{k}{k-1}\Big)   - 4 q_0(1-q_0^2) \partial_{q_0}  \right) Y_{k-4} t_{k}\;.
\end{equation}
We find that the one-point function is
\begin{equation}
\label{1ptfortk}
    \llangle \mathcal{T}_{k}(x,t)\rrangle = a_{t_{k}} \frac{(\frac{1}{2}t\cdot \mathbb{M} \cdot t)^{\frac{k-4}{2}}}{\left(P \cdot\mathbb{N}\cdot P\right)^{\frac{k+4}{2}}}\;,
\end{equation}
with 
\begin{equation}\label{ratioasat}
    a_{t_k}=\frac{ k^{2} (k^2-4)}{256 (k^2-9)^{\frac{1}{2}} (k^2-1)^{\frac{3}{2}}} a_{s_k}\;.
\end{equation}
The ratio $a_{t_k}/a_{s_k}$ can be independently verified from supersymmetry because these two fields belong to the same superconformal multiplet. More precisely, we can consider the defect two-point function $\llangle \mathcal{O}_p\mathcal{O}_p\rrangle$ in which both $s_k$ and $t_k$ are exchanged in the bulk channel. The bulk channel superconformal block, which is determined by superconformal symmetry, fixes the relative ratio of their contributions $(C_{\mathcal{O}_p \mathcal{O}_p \mathcal{T}_k}a_{t_k})/(C_{\mathcal{O}_p  \mathcal{O}_p  \mathcal{O}_k}a_{s_k})$. On the other hand, the ratio of three-point functions can be read off from the superconformal block of the same multiplet appearing in the defect-free four-point function $\langle \mathcal{O}_p \mathcal{O}_p\mathcal{O}_p \mathcal{O}_p\rangle$. Performing this check, we not only confirm (\ref{ratioasat}) is the correct one-point function ratio, but also find the following relation with the three-point function ratio
\begin{equation}
    \frac{C_{\mathcal{O}_p \mathcal{O}_p \mathcal{T}_k} }{C_{\mathcal{O}_p  \mathcal{O}_p  \mathcal{O}_k} }=\frac{a_{t_k}}{a_{s_k}}\;.
\end{equation}

\subsection{Other couplings}
To conclude this section, let us give some qualitative comments on other couplings in the effective 1d action.

In the previous subsection, we considered in detail the one-point vertices of $s_k$ and $t_k$. Other component fields in the super multiplet can be similarly analyzed and the full $L^{(0,1)}$ vertices take the following schematic form
\begin{equation}
    L^{(0,1)}= 
\sum_k\left(c_1 s_k+c_2 t_{k+4}+c_3 A_{k+1, \tau}+c_4 C_{k+3, \tau}+c_5 r_{k+2}+c_6 \varphi_{k+2, \tau \tau}\right)\;,
\end{equation}
where $c_i$ denote R-symmetry structures (and the coefficient of each vertex). The important feature to notice is that the spinning bulk fields all appear with indices restricted to the direction of the geodesic. This shows that the effective action at this order has the form of a probe particle minimally coupled to the background fields, which is precisely what we expect from the semiclassical picture. 

The bulk-defect coupling $L^{(1,1)}$ takes the form 
\begin{equation}
     L^{(1,1)} = \sum_{k, l}\left(d_1 x_l^i \partial_i s_k+d_2 \chi_l s_k+d_3 \bar{\chi}_l s_k+d_4 \partial_\tau \chi_l s_k+d_5 \partial_\tau \bar{\chi}_l s_k\right)\;.
\end{equation}
Here the important feature is that the zero transverse spin defect fields $\chi$ and $\bar{\chi}$ can couple to the bulk fields with zero or one derivative. Note that this is a special feature of 1d defects in AdS. Rotation invariance in the bulk no longer requires the derivatives to be contracted into pairs. 

Finally, we will not write down the contact vertices $L^{(2,0)}$. As we mentioned, they seem to suffer from a generic subtlety that a naive supergravity analysis leads to vertices which violate superconformal symmetry in the defect two-point function. This problem has been discussed in detail, for example, in the case of $\frac{1}{2}$-BPS Wilson loop in $\mathcal{N}=4$ SYM \cite{Gimenez-Grau:2023fcy}, although a solution is currently not available. Note that this issue does not signal any underlying inconsistency in these models but merely indicates a more careful treatment is needed if one were to follow the traditional diagrammatic algorithm. Fortunately, this subtlety does not present an obstacle for us because these contact vertex contributions will all be fixed from the bootstrap.

\section{Tree-level Witten diagrams}\label{Sec:treelevelWittendiagrams}

\begin{figure}
 \centering  \includegraphics[width=0.9\linewidth]{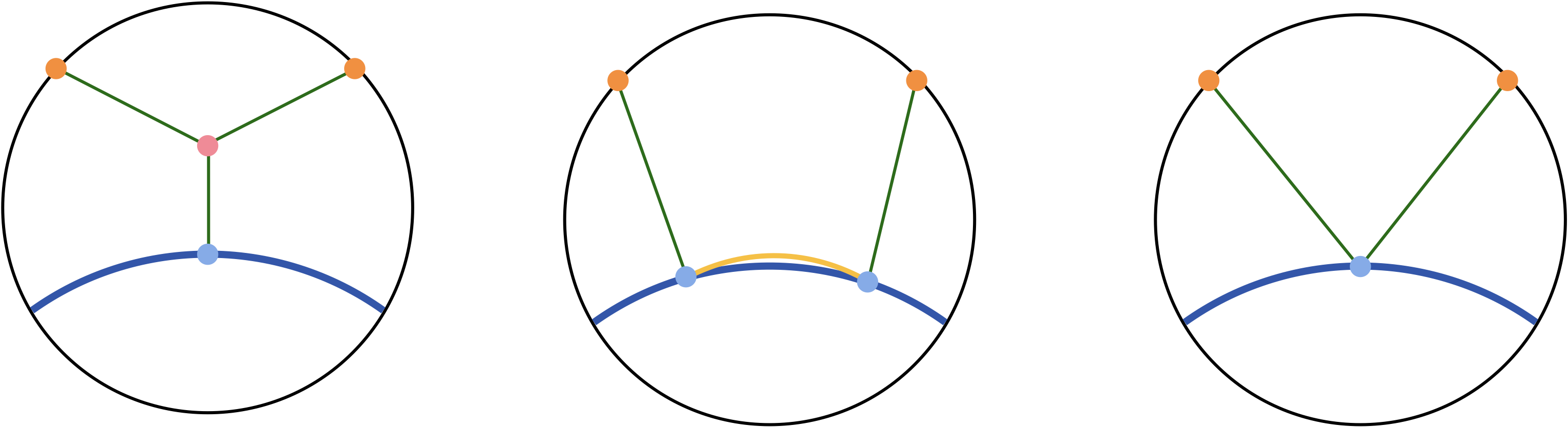}
  \caption{Three types of tree-level Witten diagrams.}
  \label{Fig:Wittendiagrams}
\end{figure}

Before applying the bootstrap strategy to compute the giant graviton defect two-point functions, let us first perform a detailed analysis of the underlying Witten diagrams. They are divided into three groups, namely bulk channel exchange Witten diagrams, defect channel exchange Witten diagrams and contact Witten diagrams (see Figure \ref{Fig:Wittendiagrams}). The results from this analysis form the necessary ingredients for implementing the bootstrap algorithm which we will describe in Section \ref{Sec:2ptfunbootstrap}. They will also be useful for establishing an analytical functional method for bootstrap studies in general, as we will briefly discuss in Section \ref{Sec:defectforLLHH}. While largely similar to the ones appearing in probe-brane setups with generic defect dimensions \cite{Gimenez-Grau:2023fcy}, the Witten diagrams considered in this section have some special features which are only present when the AdS defect is one dimensional. These features also allow us to more efficiently evaluate the Witten diagrams. In particular, in the defect channel we are able to directly evaluate the integrals and express the defect channel exchange Witten diagrams with general internal and external dimensions in terms of the Meijer $G$-functions. To our knowledge, no such closed form results are known for higher dimensional defects. 

\subsection{Contact Witten diagrams}
Let us first discuss the simplest contact Witten diagrams. A special feature of the $p=0$ dimensional defect is that the contact Witten diagram can have any number (both odd and even) of derivatives. We define these contact Witten diagrams as\footnote{We have also introduced the notation that calligraphic capital letters denote functions of cross ratios where a product of one-point functions is factored out.}
\begin{equation}
\label{defcontact}
 C_{\Delta_1 \Delta_2}^{(n)}=\frac{\mathcal{C}_{\Delta_1 \Delta_2}^{(n)}(U,V)}{\prod_{i=1}^2 (-P_i\cdot \mathbb{N}\cdot P_i)^\frac{\Delta_i}{2}}=\int_\Gamma d \tau\, G^{\Delta_1}_{B\partial}(x_1,z(\tau))(\partial_{\tau})^n G^{\Delta_2}_{B\partial}(x_2,z(\tau))\;,
\end{equation}
where $\Gamma(\tau)$ is the geodesic line connecting the two giant graviton insertions on the boundary and $\tau$ is the (Euclidean) proper time of the trajectory. In general, one can also consider contact Witten diagrams with derivatives transverse to the geodesic defect. But such diagrams will only be relevant for stringy corrections beyond supergravity so we do not include them here. It is convenient to define \begin{equation}
    K_1(x_1,\tau)=\frac{z_0(\tau)}{z_0^2(\tau)+(\vec{z}(\tau)-\vec{x}_1)^2}\;,\quad  K_2(x_2,\tau)=\frac{z_0(\tau)}{z_0^2(\tau)+(\vec{z}(\tau)-\vec{x}_2)^2}\;,
\end{equation}
so that the bulk-to-boundary propagators can be written as
\begin{equation}
    G^{\Delta_1}_{B\partial}(x_1,z(\tau))=K_1^{\Delta_1}(x_1,\tau)\;,\quad G^{\Delta_1}_{B\partial}(x_2,z(\tau))=K_2^{\Delta_2}(x_2,\tau)\;.
\end{equation}
When we use conformal invariance to choose $\Gamma$ to be the straight line $\Gamma_0$ perpendicular to the conformal boundary in the Poincar\'e coordinates, we have $z_0(\tau)=e^\tau$ and
\begin{equation}
    K_1\big|_{\Gamma=\Gamma_0}=\frac{z_0}{z_0^2+x_1^2}\;,\quad K_2\big|_{\Gamma=\Gamma_0}=\frac{z_0}{z_0^2+x_2^2}\;.
\end{equation}

For the case without derivatives, i.e., $n=0$, it is not difficult to directly evaluate the integral (\ref{defcontact}). We find that the result coincides with the general defect dimension result of \cite{Gimenez-Grau:2023fcy} after setting $p=0$\footnote{The $p$-dependence is in fact quite trivial and only enters via the overall factor.}
\begin{equation}
    \mathcal{C}_{\Delta_1 \Delta_2}^{(0)}=\frac{\pi^{\frac{1}{2}}\Gamma( \tfrac{\Delta_1+\Delta_2}{2})}{2^{\Delta_1+\Delta_2}\Gamma( \tfrac{\Delta_1+\Delta_2+1}{2})} {}_2 F_1 \left(\Delta_1,\Delta_2,\frac{\Delta_1+\Delta_2+1}{2},-\frac{(1-\sqrt{V})^2}{4\sqrt{V}} \right)\;.
\end{equation}
For the $n>0$ cases involving derivatives, we can use the following identities of bulk-to-boundary propagators when the bulk-point is restricted to the geodesic defect
\begin{equation}
\label{deridentity}
\begin{split}
    {}&\partial_{\tau} G^{\Delta_1}_{B\partial}=\frac{\Delta_1}{1-V}\left(-2G^{\Delta_1+1}_{B\partial}\left(K_2\right)^{-1}+(1+V) G^{\Delta_1}_{B\partial}\right)\;,\\
    {}& \partial_{\tau} G^{\Delta_1}_{B\partial}\partial_{\tau} G^{\Delta_2}_{B\partial}=\Delta_1 \Delta_2 \left(-2(x_1^2+x_2^2) G^{\Delta_1+1}_{B\partial} G^{\Delta_2+1}_{B\partial} + G^{\Delta_1}_{B\partial} G^{\Delta_2}_{B\partial}\right)\;.
\end{split}
\end{equation}
Using these identities, it is not difficult to show that the results can be expressed in terms of zero-derivative contact Witten diagrams with shifted weights
\begin{equation}
\begin{split}
   \mathcal{C}_{\Delta_1 \Delta_2}^{(1)}={}&\Delta_1 \left(\frac{2\sqrt{V}}{1-V}  \mathcal{C}_{\Delta_1+1,\Delta_2-1}^{(0)}-\frac{1+V}{1-V} \mathcal{C}_{\Delta_1 \Delta_2}^{(0)}\right)\;,\\
     \mathcal{C}_{\Delta_1 \Delta_2}^{(2)}={}&\Delta_1 \Delta_2 \left(\frac{2(1+V)}{\sqrt{V}}  \mathcal{C}_{\Delta_1+1,\Delta_2+1}^{(0)}-\mathcal{C}_{\Delta_1 \Delta_2}^{(0)}\right)\;.
\end{split}
\end{equation}
More generally, it is more convenient to use the following recursion relation
\begin{equation}
    \mathcal{C}^{(n)}_{\Delta_1\Delta_2}=\left(-2V\frac{d}{dV}\right)^n\mathcal{C}^{(0)}_{\Delta_1\Delta_2}\;.
\end{equation}
This can be proven by noting
\begin{equation}
    \partial_\tau^n G^{\Delta_2}_{B\partial}=\left(z_0\frac{\partial}{\partial z_0}\right)^nG^{\Delta_2}_{B\partial}=\left(-\Delta_2-x_2^i\frac{\partial}{\partial x_2^i}\right)^nG^{\Delta_2}_{B\partial}\;.
\end{equation}

\subsection{Exchange Witten diagrams in the bulk channel}
Next let us consider the bulk channel exchange Witten diagrams. These diagrams can be collectively defined as 
\begin{equation}\label{EDeltaell}
   E_{\Delta_1\Delta_2}^{\Delta,\ell}=\frac{ \mathcal{E}_{\Delta_1\Delta_2}^{\Delta,\ell}(U,V)}{\prod_{i=1}^2 (-P_i\cdot \mathbb{N}\cdot P_i)^\frac{\Delta_i}{2}}=\int_\Gamma d \tau\, I^{\mu_1\ldots\mu_{\ell}}_{\Delta}(x_1,x_2,\tau) \delta_{\mu_1,\tau} \cdots \delta_{\mu_\ell,\tau}\;,
\end{equation}
where $I^{\mu_1\ldots\mu_{\ell}}_{\Delta}$ is a three-point integral defined as the product of two bulk-to-boundary propagators and a spin-$\ell$ bulk-to-bulk propagator integrated at the cubic vertex in AdS$_5$. The tensor indices of this integral are then restricted to the component along $\Gamma$ and is integrated over the geodesic line. Explicitly, we have for $\ell\leq2$
\begin{equation}
\begin{aligned}
\label{defIintegral}
    I_{\Delta}(x_1,x_2,\tau)&=\int \frac{d^5 w}{w_0^5} G_{B\partial}^{\Delta_1}(x_1,w) G_{B\partial}^{\Delta_2}(x_2,w)G_{BB}^{\Delta,0}\left(w,z(\tau)\right)\;,\\  I_{\Delta}^{\mu}(x_1,x_2,\tau)&=\int \frac{d^5 w}{w_0^5}  G_{B\partial}^{\Delta_1}(x_1,w) \overleftrightarrow{\nabla}_{\nu} G_{B\partial}^{\Delta_2}(x_2,w)   G_{BB}^{\Delta,1;\mu \nu}\!\left(w, z(\tau)\right)\;, \\  I_{\Delta}^{\mu\nu}(x_1,x_2,\tau)&=\int \frac{d^5 w}{w_0^5}  \left(\frac{1}{2} \nabla_{(\rho} G^{\Delta_1}_{B\partial} (x_1,w) \nabla_{\sigma)} G^{\Delta_2}_{B\partial} (x_2,w)-\frac{1}{2}g_{\rho\sigma} \Big(\nabla_\rho G^{\Delta_1}_{B\partial} (x_1,w)  \right.\\&\left. \quad \; \times \nabla^\rho G^{\Delta_2}_{B\partial} (x_2,w) +\eta \,G^{\Delta_1}_{B\partial} (x_1,w) G^{\Delta_2}_{B\partial} (x_2,w)\Big) \right)  G_{BB}^{\Delta,2;\mu \nu\rho\sigma}\!\left(w, z(\tau)\right)\;,
    \end{aligned}
\end{equation}
where $G^{\Delta,\ell}_{BB}$ is the bulk-to-bulk propagator for a spin-$\ell$ field and $\eta=\frac{1}{2}(\sum_{i=1}^2\Delta_i(\Delta_i-4)-\Delta(\Delta-4))$. The three-point integrals truncate to finite sums of products of bulk-to-boundary propagators with shifted dimensions, provided the following condition is met \cite{DHoker:1999mqo,Berdichevsky:2007xd,Rastelli:2017udc,Gimenez-Grau:2023fcy}
\begin{equation}
    \Delta_1+\Delta_2-\Delta+\ell\in 2\mathbb{Z}_+\;.
\end{equation}
This is always the case in our theory thanks to the structure of its spectrum.\footnote{This condition can also be satisfied in other theories, such as 11d supergravity in AdS$_7\times$S$^4$.} Consequently, the bulk channel exchange Witten diagrams can be expressed as a finite sum of defect contact Witten diagrams. 

For $\ell=0,1$, the reduction to contact Witten diagrams can be explicitly written down in a compact form
\begin{equation}
    \mathcal{E}_{\Delta_1 \Delta_2}^{\Delta_1+\Delta_2+\ell-2m,\ell}=\sum_{i=0}^{m-1} \frac{U^{i-m}}{V^{\frac{i-m}{2}}}  \mathfrak{a}_i^{(\ell)} \mathcal{C}^{(\ell)}_{\Delta_1+i-m,\Delta_2+i-m}\,,
\end{equation}
where $m\in\mathbb{Z}_+$ and
\begin{equation}
    \begin{aligned}
 \mathfrak{a}_{i}^{(\ell)} =  \frac{\Gamma (m)\Gamma \left(i-m+\Delta _1\right) \Gamma \left(i-m+\Delta _2\right)  \Gamma \left(\Delta _1+\Delta _2-m+\ell-2\right)}{2^{2-\ell} \Gamma (\Delta _1) \Gamma (\Delta
   _2) \Gamma (i+1) \Gamma \left(\Delta _1+\Delta _2+i-2 m+\ell-1\right)}\;.
\end{aligned}
\end{equation}
The case of $\ell=2$ is technically more involved and it is difficult to write down a similar closed form formula. Nevertheless, it is straightforward to follow the steps given in \cite{Rastelli:2017udc,Gimenez-Grau:2023fcy} to evaluate $I_\Delta^{\mu\nu}$ and we repeat the prescription here for reader's convenience. It was shown that $I_\Delta^{\mu\nu}$ can always be expressed as a linear combination of the following tensors  
\begin{equation}
    \{g_{\mu\nu}\;,\;\; Q_{\mu}Q_{\nu}\;,\;\; Q_{\mu}R_{\nu}\;,\;\; R_{\mu}Q_{\nu}\;,\;\; R_{\mu}R_{\nu}\}\;,
\end{equation}
with explicit combination coefficients given in \cite{Rastelli:2017udc,Gimenez-Grau:2023fcy}. For our case, the vector variables $Q_\mu$ and $R_\mu$ are further restricted to the geodesic and become explicitly
\begin{equation}
    Q_\mu(\tau) =-\frac{(z(\tau)-x_1)_\mu}{(z(\tau)-x_1)^2}+\frac{(z(\tau)-x_2)_\mu}{(z(\tau)-x_2)^2}\;,\quad R_\mu(\tau) =\frac{\delta_{0\mu}}{z_0(\tau)}-2 \frac{(z(\tau)-x_1)_\mu}{(z(\tau)-x_1)^2}\;.
\end{equation}
 Contracting the indices according to \eqref{EDeltaell}, we find the following simple replacement rules for the basic structures in terms of the bulk-to-boundary propagators
\begin{equation}
    \begin{aligned}
{}&    g_{\tau\tau}=1\;,\quad \quad 
R_{\tau} R_{\tau}=1-4 x_1^2 K^2_1\;,\\
{}&Q_{\tau} Q_{\tau}=x_{12}^4 \left( \frac{1-V}{U}\right)^2 K^2_1 K^2_2\;,\\
{}& Q_{\tau} R_{\tau}=-x_1^2 K^2_1 +x_2^2  K^2_2+x_{12}^4 \left( \frac{1-V}{U}\right)^2 K^2_1K^2_2\;.
    \end{aligned}
\end{equation}
Then it is straightforward to assemble the result into a sum of contact Witten diagrams.

\subsection{Exchange Witten diagrams in the defect channel}\label{Subsec:defectexW}
\subsubsection{Without derivatives}
\begin{figure}
  \centering \includegraphics[width=0.8\linewidth]{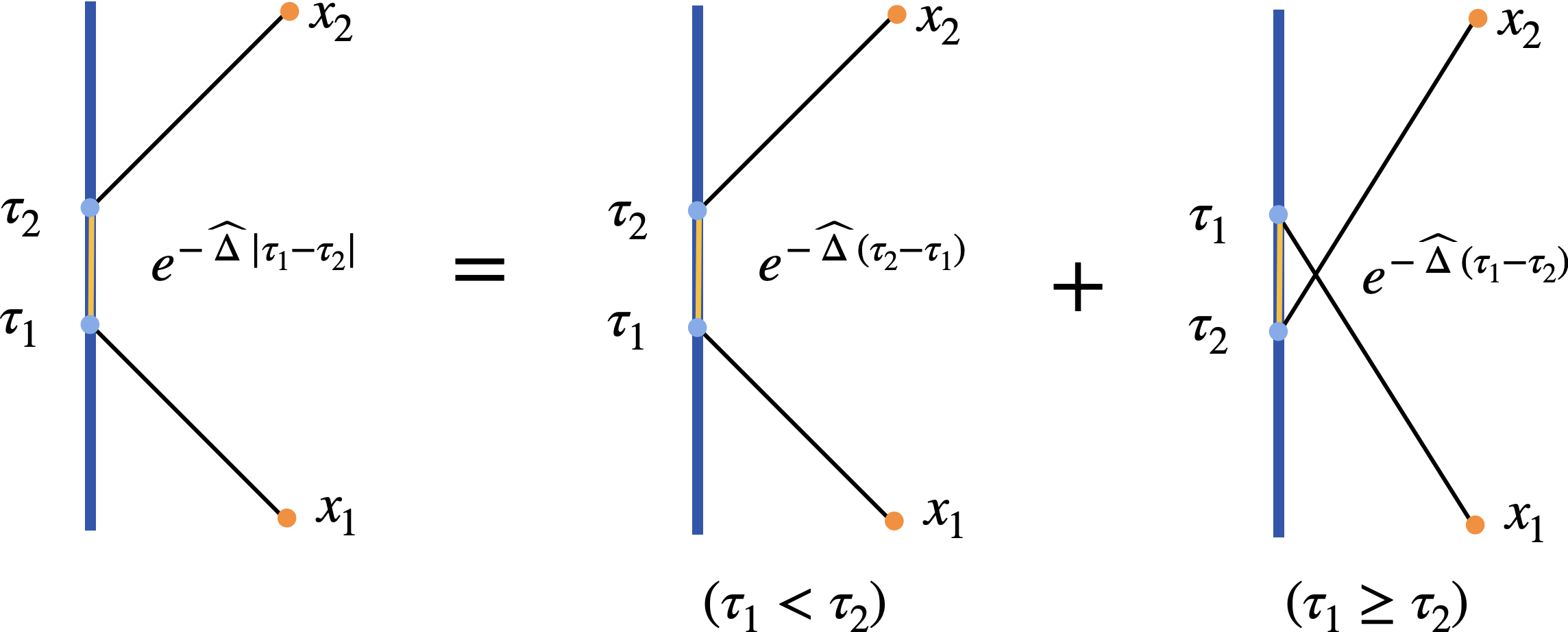}
  \caption{The integral of the defect channel exchange Witten diagram splits into two parts according to different time orderings of the two inserted vertices.}
  \label{Fig:Defectsplit}
\end{figure}

We now consider the defect channel exchange Witten diagrams. Let us first focus on the case where there are no derivatives in the vertices. The Witten diagram is defined by 
\begin{equation}
    \widehat{E}^{\widehat{\Delta},0}_{\Delta_1\Delta_2}= \frac{\widehat{\mathcal{E}}^{\widehat{\Delta},0}_{\Delta_1\Delta_2}(U,V)}{\prod_{i=1}^2 (-P_i\cdot \mathbb{N}\cdot P_i)^\frac{\Delta_i}{2}}=\int d\tau_1 d\tau_2 G^{\Delta_1}_{B\partial}(x_1,z(\tau_1)) \widehat{G}_{BB}^{\widehat{\Delta}}(z(\tau_1),z(\tau_2)) G^{\Delta_2}_{B\partial}(x_2,z(\tau_2))\;, 
\end{equation}
where $\widehat{G}_{BB}^{\widehat{\Delta}}(z(\tau_1),z(\tau_2))$ is the 1d propagator of a scalar field living on the geodesic line. We can also use conformal symmetry to map the geodesic line to the straight line defined by $\vec{z}=0$ in the Poincar\'e coordinates and $\tau=\log(z_0)$. Then explicitly, the geodesic propagator can be written as 
\begin{equation}\label{geoprop}
\widehat{G}_{BB}^{\widehat{\Delta}}(z,w)=e^{-\widehat{\Delta} |\log(\frac{z_0}{w_0})|}=\left\{\begin{array}{c}\left(\frac{z_0}{w_0}\right)^{\widehat{\Delta}}\;,\quad z_0<w_0 \\\left(\frac{w_0}{z_0}\right)^{\widehat{\Delta}}\;,\quad z_0\geq w_0\end{array}\right.\;.
\end{equation}
The defect channel exchange Witten diagram can therefore be expressed as the sum of two basic integrals different in the time ordering of the two vertices inserted on the geodesic (Figure \ref{Fig:Defectsplit})
\begin{equation}\label{EItu}  \widehat{E}^{\widehat{\Delta},0}_{\Delta_1\Delta_2}=\widehat{I}^{(u)}_{\widehat{\Delta}}+\widehat{I}^{(t)}_{\widehat{\Delta}}\;,
\end{equation}
where
\begin{equation}\label{intIu}
\widehat{I}_{\widehat{\Delta}}^{(u)}=\int_0^\infty \frac{dw_0}{w_0} \int_0^{w_0} \frac{dz_0}{z_0} \left(\frac{z_0}{z_0^2+x_1^2}\right)^{\Delta_1}\left(\frac{z_0}{w_0}\right)^{\widehat{\Delta}} \left(\frac{w_0}{w_0^2+x_2^2}\right)^{\Delta_2}\;,
\end{equation}
and 
\begin{equation}\label{intIt}
\widehat{I}_{\widehat{\Delta}}^{(t)}=\int_0^\infty \frac{dw_0}{w_0} \int_{w_0}^\infty \frac{dz_0}{z_0} \left(\frac{z_0}{z_0^2+x_1^2}\right)^{\Delta_1}\left(\frac{w_0}{z_0}\right)^{\widehat{\Delta}} \left(\frac{w_0}{w_0^2+x_2^2}\right)^{\Delta_2}\;.
\end{equation}
Is is easy to see that these two integrals are simply related by crossing symmetry
\begin{equation}\label{ItandIu}
\widehat{I}_{\widehat{\Delta}}^{(t)}=\widehat{I}_{\widehat{\Delta}}^{(u)}\big|_{1\leftrightarrow 2}\;,
\end{equation}
where $1\leftrightarrow 2$ means both $x_1$, $x_2$ and $\Delta_1$, $\Delta_2$ are interchanged.

To compute this defect integral, we can make the change of variables $t_1=z_0^2$, $t_2=w_0^2$ followed by $t_1=\kappa t_2$. The integral becomes
\begin{equation}
\widehat{I}_{\widehat{\Delta}}^{(u)}=\frac{1}{4}\int_0^\infty dt_2 \int_0^{1} d\kappa\, (\kappa t_2)^{\frac{\Delta_1+\widehat{\Delta}}{2}-1}t_2^{\frac{\Delta_2-\widehat{\Delta}}{2}}(\kappa t_2+x_1^2)^{-\Delta_1}(t_2+x_2^2)^{-\Delta_2}\;.
\end{equation}
We further rescale $t_2$ by $t_2\to \frac{1}{\kappa}x_1^2t_2$ and rewrite the integral as 
\begin{equation}
\widehat{I}_{\widehat{\Delta}}^{(u)}=\frac{1}{4}(x_1^2)^{-\frac{\Delta_1+\Delta_2}{2}}\int_0^\infty dt_2 \int_0^{1} d\kappa\, \kappa^{\frac{\widehat{\Delta}+\Delta_2}{2}-1} t_2^{\frac{\Delta_1+\Delta_2}{2}-1}(t_2+1)^{-\Delta_1}\left(t_2+\kappa\frac{x_2^2}{x_1^2}\right)^{-\Delta_2}\,.
\end{equation}
By using the integral representation of ${}_2F_1$
\begin{equation}
{}_2F_1(a,b;c;z)=\frac{\Gamma(c)}{\Gamma(b)\Gamma(c-b)}\int_0^\infty dt\, t^{-b+c-1}(t+1)^{a-c}(t-z+1)^{-a}\;,
\end{equation}
we can integrate out $t_2$ and obtain 
\begin{equation}\label{Iu2F1}
\widehat{I}_{\widehat{\Delta}}^{(u)}=\frac{\Gamma^2(\frac{\Delta_1+\Delta_2}{2})}{4\Gamma(\Delta_1+\Delta_2)}(x_1^2)^{-\frac{\Delta_1+\Delta_2}{2}}\int_0^{1} d\kappa\, \kappa^{\frac{\widehat{\Delta}+\Delta_2}{2}-1}{}_2F_1\left(\Delta_2,\frac{\Delta_1+\Delta_2}{2};\Delta_1+\Delta_2;1-\kappa\frac{x_2^2}{x_1^2}\right)\;.
\end{equation}
It is convenient to extract a factor of $x_1^2$, $x_2^2$ so that the integral can be written as a function of a single cross ratio  
\begin{equation}
\widehat{\mathcal{I}}^{(u)}_\Delta(r)=(x_1^2)^{\frac{\Delta_1}{2}}(x_2^2)^{\frac{\Delta_2}{2}}\widehat{I}^{(u)}_{\widehat{\Delta}}\;,
\end{equation}
where we have defined $r^2=x_2^2/x_1^2$ and we note $V=r^2$ in terms of the cross ratio $V$ defined in \eqref{crossratios4pt}. Using the contour integral representation of ${}_2F_1$, we can rewrite the integral as 
\begin{equation}
\begin{split}
\widehat{\mathcal{I}}_{\widehat{\Delta}}^{(u)}={}&\frac{r^{\Delta_2}}{4\Gamma(\Delta_1)\Gamma(\Delta_2)}\int_0^1d\kappa\kappa^{\frac{\widehat{\Delta}+\Delta_2}{2}-1-s}\int_{\gamma-i\infty}^{\gamma+i\infty}\frac{ds}{2\pi i}r^{-2s}\\
{}&\times \Gamma(s)\Gamma\left(s+\frac{\Delta_1-\Delta_2}{2}\right)\Gamma(\Delta_2-s)\Gamma\left(\frac{\Delta_1+\Delta_2}{2}-s\right)\;,
\end{split}
\end{equation}
where the contour is placed at $\max\{0,\frac{\Delta_2-\Delta_1}{2}\}<\gamma<\min\{\Delta_2,\frac{\Delta_1+\Delta_2}{2}\}$. The $\kappa$ integral can be easily performed and gives
\begin{equation}
\widehat{\mathcal{I}}_{\widehat{\Delta}}^{(u)}=\frac{1}{2\Gamma(\Delta_1)\Gamma(\Delta_2)}\int_{\gamma-i\infty}^{\gamma+i\infty}\frac{ds}{2\pi i}r^{-2s+\Delta_2}\frac{\Gamma(s)\Gamma(s+\frac{\Delta_1-\Delta_2}{2})\Gamma(\Delta_2-s)\Gamma(\frac{\Delta_1+\Delta_2}{2}-s)}{\widehat{\Delta}+\Delta_2-2s}\;.
\end{equation}
This allows us to write the defect integral as a Meijer $G$-function
\begin{equation}
\widehat{\mathcal{I}}_{\widehat{\Delta}}^{(u)}=\frac{r^{\Delta_2}}{4\Gamma(\Delta_1)\Gamma(\Delta_2)}G^{3,2}_{3,3}\left[\left.\begin{array}{c}\frac{2+\Delta_2-\Delta_1}{2}, 1, \frac{2+\widehat{\Delta}+\Delta_2}{2} \\\Delta_2,\frac{\Delta_1+\Delta_2}{2},\frac{\widehat{\Delta}+\Delta_2}{2}\end{array}\right.\bigg|\frac{1}{r^2}\right]\;,
\end{equation}
where we recall the Meijer $G$-functions are defined as
\begin{equation}
G^{m,n}_{p,q}\left[\left.\begin{array}{c}a_1,\ldots,a_p \\b_1,\ldots b_q\end{array}\right.\bigg|z\right]= \int \frac{ds}{2\pi i} z^s \frac{\prod_{j=1}^m \Gamma(b_j-s)\prod_{j=1}^n\Gamma(1-a_j+s)}{\prod_{j=m+1}^q\Gamma(1-b_j+s)\prod_{j=n+1}^p\Gamma(a_j-s)}\;.
\end{equation}
Therefore, for general $\Delta_1$, $\Delta_2$ and $\widehat{\Delta}$, we can write the defect channel exchange Witten diagram in a closed form as 
\begin{equation}
\begin{split}
\widehat{\mathcal{E}}^{\widehat{\Delta},0}_{\Delta_1\Delta_2}={}&\frac{1}{4\Gamma(\Delta_1)\Gamma(\Delta_2)}\left\{r^{\Delta_2}G^{3,2}_{3,3}\left[\left.\begin{array}{c}\frac{2+\Delta_2-\Delta_1}{2}, 1, \frac{2+\widehat{\Delta}+\Delta_2}{2} \\\Delta_2,\frac{\Delta_1+\Delta_2}{2},\frac{\widehat{\Delta}+\Delta_2}{2}\end{array}\right.\bigg|\frac{1}{r^2}\right]\right.\\
{}&\quad\quad\quad\quad\quad\quad\quad\left.+\frac{1}{r^{\Delta_1}}G^{3,2}_{3,3}\left[\left.\begin{array}{c}\frac{2+\Delta_1-\Delta_2}{2}, 1, \frac{2+\widehat{\Delta}+\Delta_1}{2} \\\Delta_1,\frac{\Delta_1+\Delta_2}{2},\frac{\widehat{\Delta}+\Delta_1}{2}\end{array}\right.\bigg|r^2\right]\right\}\;.
\end{split}
\end{equation}
For the defect spectrum of the giant graviton, these Witten diagrams simplify greatly and evaluate to functions spanned by $1$ and $\log V$ with rational function coefficients of the cross ratios. As a consistency check, we can also verify that the defect channel exchange Witten diagram satisfy the equation of motion identity. It is straightforward to show 
\begin{equation}
    \left(-r^2 \partial_{r}^2-r \partial_{r} + \widehat{\Delta}^2\right) \widehat{\mathcal{E}}^{\widehat{\Delta},0}_{\Delta_1\Delta_2}= 2 \widehat{\Delta}\,\mathcal{C}_{\Delta_1\Delta_2}^{(0)}\;,
\end{equation}
where the LHS is essentially the defect channel Casimir equation and we recall $\mathcal{C}_{\Delta_1\Delta_2}^{(0)}$ is the contact Witten diagram with zero derivatives. 

Before proceeding to the discussion of the defect channel exchange Witten diagrams involving derivatives, let us point out a few interesting properties satisfied by these defect integral. These properties should be useful for establishing an analytic functional approach \cite{Mazac:2016qev,Mazac:2018mdx,Mazac:2018ycv,Zhou:2018sfz,Kaviraj:2018tfd,Mazac:2018biw,Mazac:2019shk,Giombi:2020xah,Ghosh:2021ruh,Antunes:2025vvl}, but we will not pursue this further in the current paper. The proofs of these properties are given in Appendix \ref{App:moreWittendiag}.
\begin{itemize}
    \item {\bf Weight exchanging symmetry.} The integrals $\widehat{\mathcal{I}}^{(u)}_{\widehat{\Delta}}$ and $\widehat{\mathcal{I}}^{(t)}_{\widehat{\Delta}}$ are separately symmetric in exchanging the external dimensions $\Delta_1\leftrightarrow \Delta_2$
\begin{equation}\label{Ire1}
    \widehat{\mathcal{I}}^{(u)}_{\widehat{\Delta}}=\widehat{\mathcal{I}}^{(u)}_{\widehat{\Delta}}\big|_{\Delta_1\leftrightarrow \Delta_2}\;,\quad \widehat{\mathcal{I}}^{(t)}_{\widehat{\Delta}}=\widehat{\mathcal{I}}^{(t)}_{\widehat{\Delta}}\big|_{\Delta_1\leftrightarrow \Delta_2}\;.
\end{equation}
Here we emphasize that unlike the crossing symmetry (\ref{ItandIu}), $x_1$ and $x_2$ are not exchanged.
\item {\bf Differential equations.} The defect integrals satisfy the following third order ODEs
\begin{equation}\label{Ire2}
\begin{split}
{}&\left((\varpi-\Delta_2)(\varpi-\Delta_1)(\varpi+\widehat{\Delta})-r^2(\varpi+\Delta_2)(\varpi+\Delta_1)(\varpi+\widehat{\Delta})\right)\widehat{\mathcal{I}}^{(u)}_{\widehat{\Delta}}=0\;,\\
{}&\left((\varpi-\Delta_2)(\varpi-\Delta_1)(\varpi-\widehat{\Delta})-r^2(\varpi+\Delta_2)(\varpi+\Delta_1)(\varpi-\widehat{\Delta})\right)\widehat{\mathcal{I}}^{(t)}_{\widehat{\Delta}}=0\;,
\end{split}
\end{equation}
where we have defined $\varpi=r d/dr$.
\item {\bf Shadow integral relation.} The defect integral of the shadow operator, i.e., a defect operator with dimension $-\widehat{\Delta}$, is simply related the original integral by 
\begin{equation}\label{Ire3}
\widehat{\mathcal{I}}^{(u)}_{\widehat{\Delta}}(r)+\widehat{\mathcal{I}}^{(u)}_{-\widehat{\Delta}}(1/r)=r^{-\widehat{\Delta}}\frac{\Gamma(\frac{\Delta_1-\widehat{\Delta}}{2})\Gamma(\frac{\Delta_2-\widehat{\Delta}}{2})\Gamma(\frac{\Delta_1+\widehat{\Delta}}{2})\Gamma(\frac{\Delta_2+\widehat{\Delta}}{2})}{4\Gamma(\Delta_1)\Gamma(\Delta_2)}\;.
\end{equation}
Recalling $V=r^2$, we note that the RHS of the identity is proportional to the defect channel conformal block (\ref{gdefectpiecewise}). 
\end{itemize}

\subsubsection{With one or two derivatives}
We now consider the exchange Witten diagrams with derivatives. For the supergravity calculation, we will only need these diagrams with up to two derivatives. When the integral contains a single derivative, we have two possibilities to extend the definition (\ref{intIu})
\begin{equation}
\begin{split}
{}&\widehat{I}_{\widehat{\Delta}}^{(u)|1,0}=\int_0^\infty \frac{dw_0}{w_0} \int_0^{w_0} \frac{dz_0}{z_0} z_0\frac{\partial}{\partial z_0}\left[\left(\frac{z_0}{z_0^2+x_1^2}\right)^{\Delta_1}\right]\left(\frac{z_0}{w_0}\right)^{\widehat{\Delta}} \left(\frac{w_0}{w_0^2+x_2^2}\right)^{\Delta_2}\;,\\
{}&\widehat{I}_{\widehat{\Delta}}^{(u)|0,1}=\int_0^\infty \frac{dw_0}{w_0} \int_0^{w_0} \frac{dz_0}{z_0} \left(\frac{z_0}{z_0^2+x_1^2}\right)^{\Delta_1}\left(\frac{z_0}{w_0}\right)^{\widehat{\Delta}} w_0\frac{\partial}{\partial w_0}\left[\left(\frac{w_0}{w_0^2+x_2^2}\right)^{\Delta_2}\right]\;.
\end{split}
\end{equation}
Using integration by parts, it is easy to see that they reduce to the ones without derivatives
\begin{equation}
\widehat{I}_{\widehat{\Delta}}^{(u)|1,0}=C_{\Delta_1\Delta_2}^{(0)}-\widehat{\Delta} \widehat{I}_{\widehat{\Delta}}^{(u)}\;,\quad \widehat{I}_{\widehat{\Delta}}^{(u)|0,1}=-C_{\Delta_1\Delta_2}^{(0)}+\widehat{\Delta} \widehat{I}_{\widehat{\Delta}}^{(u)}\;.
\end{equation}
 Similarly, for the t-channel we can define
\begin{equation}
\begin{split}
{}&\widehat{I}_{\widehat{\Delta}}^{(t)|0,1}=\int_0^\infty \frac{dw_0}{w_0} \int_{w_0}^\infty \frac{dz_0}{z_0} \left(\frac{z_0}{z_0^2+x_1^2}\right)^{\Delta_1}\left(\frac{w_0}{z_0}\right)^{\widehat{\Delta}} w_0\frac{\partial}{\partial w_0}\left[\left(\frac{w_0}{w_0^2+x_2^2}\right)^{\Delta_2}\right]\;,\\
{}&\widehat{I}_{\widehat{\Delta}}^{(t)|1,0}=\int_0^\infty \frac{dw_0}{w_0} \int_{w_0}^\infty \frac{dz_0}{z_0} z_0\frac{\partial}{\partial z_0}\left[\left(\frac{z_0}{z_0^2+x_1^2}\right)^{\Delta_1}\right]\left(\frac{w_0}{z_0}\right)^{\widehat{\Delta}} \left(\frac{w_0}{w_0^2+x_2^2}\right)^{\Delta_2}\;,
\end{split}
\end{equation}
which satisfy analogous relations
\begin{equation}\label{I1derre}
\widehat{I}_{\widehat{\Delta}}^{(t)|1,0}=-C_{\Delta_1\Delta_2}^{(0)}+\widehat{\Delta} \widehat{I}_{\widehat{\Delta}}^{(t)}\;,\quad \widehat{I}_{\widehat{\Delta}}^{(t)|0,1}=C_{\Delta_1\Delta_2}^{(0)}-\widehat{\Delta} \widehat{I}_{\widehat{\Delta}}^{(t)}\;.
\end{equation}
From these relations, it follows that the defect channel exchange Witten diagrams with a single derivative are given by
\begin{equation}
\begin{split}
\widehat{E}^{1,0}_{\widehat{\Delta}}={}&\widehat{I}_{\widehat{\Delta}}^{(u)|1,0}+\widehat{I}_{\widehat{\Delta}}^{(t)|1,0}=-\widehat{\Delta} (\widehat{I}_{\widehat{\Delta}}^{(u)}-\widehat{I}_{\widehat{\Delta}}^{(t)})\;,\\
\widehat{E}^{0,1}_{\widehat{\Delta}}={}&\widehat{I}_{\widehat{\Delta}}^{(u)|0,1}+\widehat{I}_{\widehat{\Delta}}^{(t)|0,1}=\widehat{\Delta} (\widehat{I}_{\widehat{\Delta}}^{(u)}-\widehat{I}_{\widehat{\Delta}}^{(t)})\;,
\end{split}
\end{equation}
where the distributions of derivatives in the vertices are indicated by the superscripts. From this result, it is manifest that the single-derivative exchange Witten diagrams are antisymmetric under exchanging 1 and 2. 

For the exchange Witten diagrams with two derivatives, we can define the following defect integral
\begin{equation}
\widehat{I}_{\widehat{\Delta}}^{(u)|2,0}=\int_0^\infty \frac{dw_0}{w_0} \int_0^{w_0} \frac{dz_0}{z_0} z_0\frac{\partial}{\partial z_0}\left[z_0\frac{\partial}{\partial z_0}\left[\left(\frac{z_0}{z_0^2+x_1^2}\right)^{\Delta_1}\right]\right]\left(\frac{z_0}{w_0}\right)^{\widehat{\Delta}} \left(\frac{w_0}{w_0^2+x_2^2}\right)^{\Delta_2}\;,
\end{equation}
and similarly for the other integrals. Using integration by parts and (\ref{I1derre}), it is straightforward to find
\begin{equation}
\begin{split}
\widehat{I}_{\widehat{\Delta}}^{(u)|2,0}={}&-C_{\Delta_1\Delta_2}^{(1)}-\widehat{\Delta} \widehat{I}_{\widehat{\Delta}}^{(u)|1,0}=-C_{\Delta_1\Delta_2}^{(1)}-\widehat{\Delta} C_{\Delta_1\Delta_2}^{(0)}+\widehat{\Delta}^2 \widehat{I}^{(u)}_{\widehat{\Delta}}\;,\\
\widehat{I}_{\widehat{\Delta}}^{(t)|2,0}={}&C_{\Delta_1\Delta_2}^{(1)}+\widehat{\Delta} \widehat{I}_{\widehat{\Delta}}^{(t)|1,0}=C_{\Delta_1\Delta_2}^{(1)}-\widehat{\Delta} C_{\Delta_1\Delta_2}^{(0)}+\widehat{\Delta}^2 \widehat{I}^{(t)}_{\widehat{\Delta}}\;,\\
\widehat{I}_{\widehat{\Delta}}^{(u)|1,1}={}&C_{\Delta_1\Delta_2}^{(1)}-\widehat{\Delta} \widehat{I}_{\widehat{\Delta}}^{(u)|0,1}=C_{\Delta_1\Delta_2}^{(1)}+\widehat{\Delta} C_{\Delta_1\Delta_2}^{(0)}-\widehat{\Delta}^2 \widehat{I}^{(u)}_{\widehat{\Delta}}\;,\\
\widehat{I}_{\widehat{\Delta}}^{(t)|1,1}={}&-C_{\Delta_1\Delta_2}^{(1)}+\widehat{\Delta} \widehat{I}_{\widehat{\Delta}}^{(t)|0,1}=-C_{\Delta_1\Delta_2}^{(1)}+\widehat{\Delta} C_{\Delta_1\Delta_2}^{(0)}-\widehat{\Delta}^2 \widehat{I}^{(t)}_{\widehat{\Delta}}\;,\\
\widehat{I}_{\widehat{\Delta}}^{(u)|0,2}={}&-C_{\Delta_1\Delta_2}^{(1)}+\widehat{\Delta} \widehat{I}_{\widehat{\Delta}}^{(u)|0,1}=-C_{\Delta_1\Delta_2}^{(1)}-\widehat{\Delta} C_{\Delta_1\Delta_2}^{(0)}+\widehat{\Delta}^2 \widehat{I}^{(u)}_{\widehat{\Delta}}\;,\\
\widehat{I}_{\widehat{\Delta}}^{(t)|0,2}={}&C_{\Delta_1\Delta_2}^{(1)}-\widehat{\Delta} \widehat{I}_{\widehat{\Delta}}^{(t)|0,1}=C_{\Delta_1\Delta_2}^{(1)}-\widehat{\Delta} C_{\Delta_1\Delta_2}^{(0)}+\widehat{\Delta}^2 \widehat{I}^{(t)}_{\widehat{\Delta}}\;.
\end{split}
\end{equation}
Combining these results, we get
\begin{equation}\label{E2derto0der}
    \begin{split}
        {}&\widehat{E}^{2,0}_{\widehat{\Delta}}=\widehat{I}_{\widehat{\Delta}}^{(u)|2,0}+\widehat{I}_{\widehat{\Delta}}^{(t)|2,0}=\widehat{\Delta}^2 \widehat{E}_{\widehat{\Delta}}-2\widehat{\Delta} C_{\Delta_1\Delta_2}^{(0)}\;,\\
        {}&\widehat{E}^{1,1}_{\widehat{\Delta}}=\widehat{I}_{\widehat{\Delta}}^{(u)|1,1}+\widehat{I}_{\widehat{\Delta}}^{(t)|1,1}=-\widehat{\Delta}^2 \widehat{E}_{\widehat{\Delta}}+2\widehat{\Delta} C_{\Delta_1\Delta_2}^{(0)}\;,\\
         {}&\widehat{E}^{0,2}_{\widehat{\Delta}}=\widehat{I}_{\widehat{\Delta}}^{(u)|0,2}+\widehat{I}_{\widehat{\Delta}}^{(t)|0,2}=\widehat{\Delta}^2 \widehat{E}_{\widehat{\Delta}}-2\widehat{\Delta} C_{\Delta_1\Delta_2}^{(0)}\;,\\
    \end{split}
\end{equation}
which tells us that the two-derivative defect channel exchange Witten diagrams are proportional to the zero-derivative ones up to contact Witten diagrams.

\section{Defect two-point functions for all supergravitons}\label{Sec:2ptfunbootstrap}

\subsection{Bootstrap algorithm}
The most efficient way to compute these giant graviton correlators is bootstrap. The bootstrap approach avoids using the explicit details of the effective Lagrangian and fixes directly the correlators using symmetries and consistency conditions. This strategy was first developed in defect-free CFTs for four-point functions in \cite{Rastelli:2016nze,Rastelli:2017udc} (see \cite{Bissi:2022mrs} for a review) and was extended to defect two-point functions in \cite{Gimenez-Grau:2023fcy} (see also \cite{Chen:2023yvw,Zhou:2024ekb}). Our algorithm here will be similar to \cite{Gimenez-Grau:2023fcy,Chen:2023yvw,Zhou:2024ekb}. However, it also has a few distinct features which arise from the special 1d nature of the giant graviton geodesic. Schematically, the algorithm can be summarized into three steps
\begin{enumerate}
    \item For a particular two-point function $G_{k_1k_2}$, we write down an ansatz which is a linear combination of all the possible Witten diagrams with unfixed coefficients. 
    \item We evaluate the ansatz so that it becomes a function with explicit dependence on the cross ratios. 
    \item We impose the superconformal Ward identities and solve for the unknown coefficients.
\end{enumerate}
The evaluation of the Witten diagrams has already been discussed in detail in Section \ref{Sec:treelevelWittendiagrams} and further imposing the superconformal Ward identities is also straightforward. Therefore, in what follows we will focus on explaining the first step of this algorithm. 

The ansatz can be naturally divided into three parts according to the topology of the underlying Witten diagrams
\begin{equation}
    \mathcal{G}_{k_1 k_2}^{\rm ansatz}=\mathcal{G}^{\rm bulk}_{k_1 k_2}+\mathcal{G}^{\rm defect}_{k_1 k_2}+\mathcal{G}^{ \rm contact}_{k_1 k_2}\;,
\end{equation}
The three terms respectively correspond to the exchange Witten diagrams in the bulk and defect channels and contact Witten diagrams. The bulk channel exchange contribution is quite straightforward and can be written more explicitly as 
\begin{equation}\label{defGbulk}
\mathcal{G}^{\rm bulk}_{k_1 k_2} =  \sum_{\mathcal{X}\in \mathcal{S}} \mu_{\mathcal{X}} \mathcal{E}_{k_1 k_2}^{\Delta_{\mathcal{X}},\ell_{\mathcal{X}}} \mathcal{Q}^{k_1,k_2}_{\mathcal{X}}\;.
\end{equation}
Here the exchange Witten diagrams are multiplied by the bulk channel R-symmetry blocks defined in (\ref{defbulkQ}) which capture the exchange of R-symmetry information, and  $\mu_{\mathcal{X}}$ are unknown parameters. The set of  bulk fields $\mathcal{S}$ allowed to be exchanged is selected from Table \ref{table:bulkmodes} based on two criteria. The first is the R-symmetry selection rule, i.e., the representation of the field $\mathcal{X}$ must be inside the tensor product of $[0,k_1,0]\times [0,k_2,0]$. Because the number of irreducible representations appearing in the tensor product is finite, this already ensures the finiteness of the bulk exchange part of the ansatz. The second requirement is that the cubic couplings must be non-extremal \cite{Rastelli:2017udc}. This is essentially a consistency requirement and is needed for the supergravity effective action to be finite. This requirement translates into the inequality $\Delta_{\mathcal{X}}-\ell_{\mathcal{X}}<k_1+k_2$. 

\begin{figure}
  \centering \includegraphics[width=0.95\linewidth]{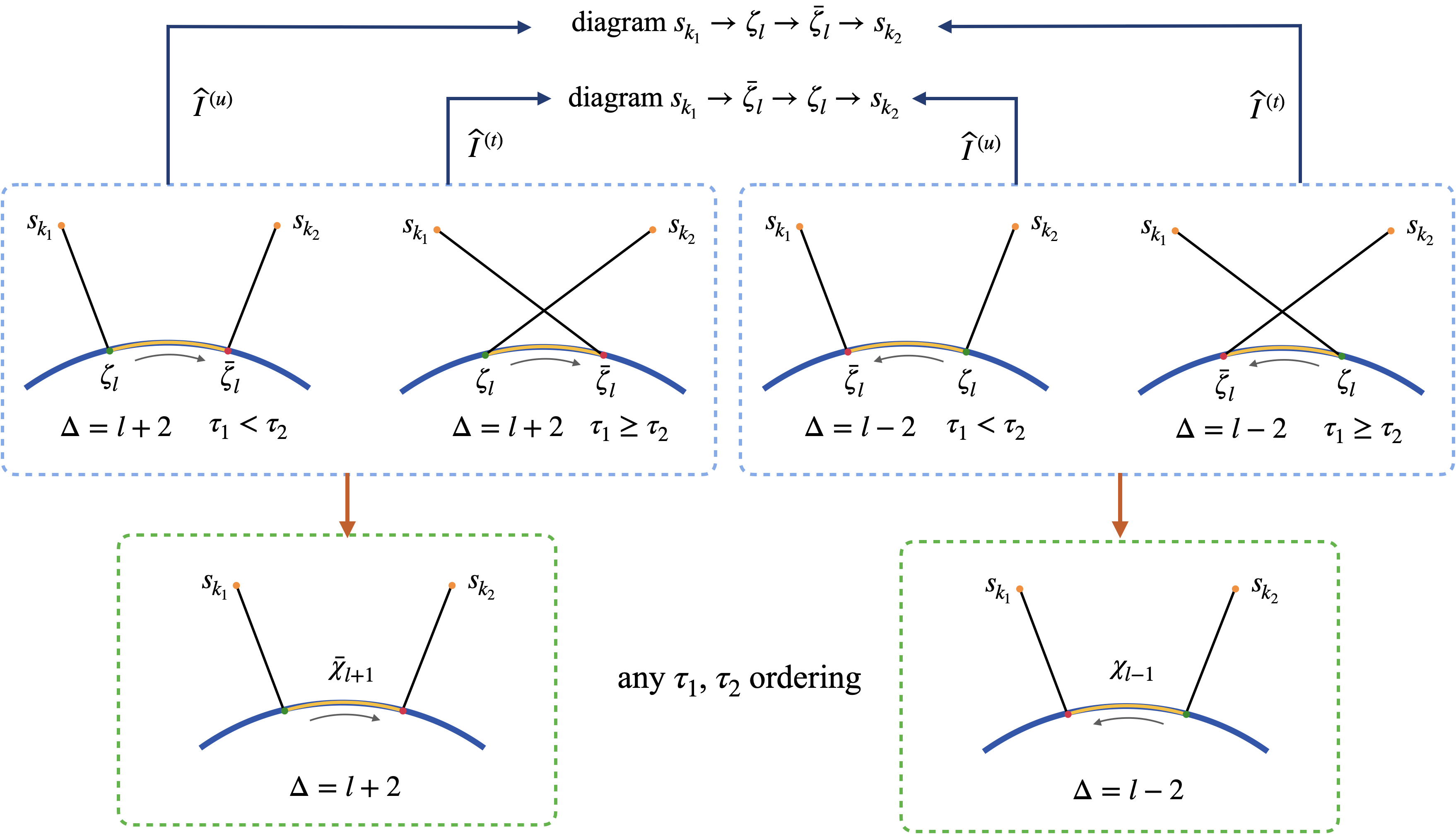}
  \caption{The two diagrams with different $\zeta$, $\bar{\zeta}$ insertion orders reorganize into two defect channel exchange Witten diagrams. The first and last diagrams of the first row correspond to the AdS diagram with the fields connected by propagators in the order $s_{k_1}$-$\zeta_l$-$\bar{\zeta}_l$-$s_{k_2}$. The middle two diagrams of the first row corresponds to the other order $s_{k_1}$-$\bar{\zeta}_l$-$\zeta_l$-$s_{k_2}$. The arrow labels the flow of the $U(1)$ charge. They reorganize into two defect channel exchange Witten diagrams in the second row. The two diagrams have internal propagators of dimensions $l
  +2$ and $l-2$ respectively, corresponding to exchanging the modes $\bar{\chi}_{l+1}$ and $\chi_{l-1}$.}
  \label{Fig:Exchangerecombine}
\end{figure}

The defect channel exchange contribution can also be cast into a similar form 
\begin{equation}\label{Gdefect}
  \mathcal{G}^{\rm defect}_{k_1 k_2} = \!\!\!\!  \sum_{\mathcal{Y}\in \widehat{\mathcal{S}},i=0,1,2}\!\!\!\!\!\! \widehat{\mu}_{\mathcal{Y},i}  \widehat{\mathcal{E}}_{k_1 k_2,(i)}^{\Delta_{\mathcal{Y}},0} \widehat{\mathbb{Q}}^{(-1)^i}_{\mathcal{Y}}+\! \sum_{\mathcal{Z}\in \widehat{\mathcal{S}}} \widehat{\mu}_{\mathcal{Z}} \widehat{\mathcal{E}}_{k_1 k_2}^{\Delta_{\mathcal{Z}},1} \widehat{\mathbb{Q}}^+_{\mathcal{Z}}\;,
\end{equation}
with the defect channel exchange Witten diagrams defined in Section \ref{Subsec:defectexW}. 
Here the set $\widehat{\mathcal{S}}$ is taken from the defect spectrum Table \ref{table:defectmodes}, and is subject to similar R-symmetry selection rules and non-extremal conditions $\widehat{\Delta}-s\leq \min\{k_1,k_2\}$. The index $i$ labels the number of derivatives in the defect channel exchange Witten diagrams because the transverse spin zero fields can have zero or one derivative at the bulk-defect couplings. The parameters $\widehat{\mu}_{\mathcal{Y}}$, $\widehat{\mu}_{\mathcal{Z}}$ are unknowns. However, there are two technical subtleties which need to be addressed in order to write the defect exchange ansatz in this nice form
\begin{itemize}
    \item Recall in Table \ref{table:defectmodes}, the modes $\chi$, $\bar{\chi}$ are extracted from the effective conformal dimensions of the propagators and are not exactly the field fluctuations $\zeta$, $\bar{\zeta}$ which directly enter the diagrammatic perturbation calculation. How do the diagrams formed with the latter  become equivalent to the defect channel exchange Witten diagrams associated with $\chi$, $\bar{\chi}$?
    \item The other issue is the R-symmetry dependence. Generally, we expect the correlator not to be piecewise defined (i.e., depending on $\sigma\leq \tau$ or $\sigma>\tau$), but rather a single polynomial in $\sigma$ and $\tau$ defined over the whole range (after factoring out some overall power of $\sigma$ and $\tau$). But the defect channel R-symmetry blocks for a given set of R-symmetry charges are only piecewise defined. How does this piecewise behavior disappear in the correlator?
\end{itemize}
Let us start with the second point by defining the following functions used in the ansatz
\begin{equation}\label{Qpolypm}
    \begin{split}
\widehat{\mathbb{Q}}^+_{r_1,r_2}(\sigma,\tau)={}& \frac{1}{2}(\widehat{\mathcal{Q}}^+_{r_1,r_2}(\sigma,\tau)+\widehat{\mathcal{Q}}^+_{r_1,-r_2}(\sigma,\tau))\;,\\
\widehat{\mathbb{Q}}^-_{r_1,r_2}(\sigma,\tau)={}& \frac{1}{2}(\widehat{\mathcal{Q}}^-_{r_1,r_2}(\sigma,\tau)-\widehat{\mathcal{Q}}^-_{r_1,-r_2}(\sigma,\tau))\;.
    \end{split}
\end{equation}
Using the definition of the R-symmetry blocks (\ref{defQhatpm}), one can easily check that these functions are defined on the union of both intervals as a single polynomial. Moreover, these functions have definite parities under $3\leftrightarrow 4$ exchange, as are labeled by the $\pm$ indices. Note when the $U(1)$ charge $r_2$ is zero, $\widehat{\mathbb{Q}}^-_{r_1,0}$ vanishes and  $\widehat{\mathbb{Q}}^+_{r_1,0}$ is identical to $\widehat{\mathcal{Q}}^+_{r_1,0}$ (which does not have a piecewise behavior). This justifies the second term in (\ref{Gdefect}) which corresponds to exchanging the fields $x_l^i$. One can also check that this term has the correct overall positive parity because the transverse spin one exchange Witten diagrams are also parity even. To justify the first term, which is where the complexities lie, we first recall from Section \ref{Subsect:defectspec} that the conformal dimension of the internal propagator changes according to the  direction of the $U(1)$ charge flow. A single diagram (say propagators connect fields as $s_{k_1}$-$\zeta$-$\bar{\zeta}$-$s_{k_2}$) will have different internal dimensions for the two pieces $\widehat{I}^{(u)}$, $\widehat{I}^{(t)}$ of the $t$ integral (see (\ref{EItu})). Nevertheless, if we combine it with the other diagram $s_{k_1}$-$\bar{\zeta}$-$\zeta$-$s_{k_2}$ where the $\zeta$, $\bar{\zeta}$ insertion order is reversed, they reorganize into two defect channel exchange Witten diagrams $\widehat{\mathcal{E}}$ defined in Section \ref{Subsec:defectexW} where inside each Witten diagram $\widehat{I}^{(u)}$, $\widehat{I}^{(t)}$ have same internal dimension.\footnote{It is easiest to see this by fixing the flow direction of the $U(1)$ charge (which is captured by an R-symmetry block). One sees that one Witten diagram contributes $\widehat{I}^{(u)}$, and the other one contributes $\widehat{I}^{(t)}$, making a full defect channel exchange Witten diagram.} The recombination is illustrated by Figure \ref{Fig:Exchangerecombine}. This gives rise to the factorized structure as in (\ref{Gdefect}) where the spacetime part is already $\widehat{\mathcal{E}}$ (also with derivatives) but the R-symmetry part so far is only the R-symmetry block. However, the defect channel exchange Witten diagrams have definite parities under $3\leftrightarrow 4$. The exchange Witten diagrams with even (odd) number of derivatives are parity even (odd), i.e., the parity is $(-1)^i$. To achieve overall positive parity, each parity even (odd) diagram must be multiplied with parity even (odd) R-symmetry blocks, and the latter will be automatically symmetrized (anti-symmetrized) into the R-symmetry polynomials according to (\ref{Qpolypm}).

Finally, the contact part of the ansatz takes the form 
\begin{equation}
   G^{ \rm contact}_{k_1 k_2}=  C_{k_1 k_2}^{(0)} \sum_{i,j \geq 0}^{i+j\leq \min\{k_1,k_2\}}\bar\mu_{ij}\sigma^{i-\frac{k_1}{2}} \tau^{j-\frac{k_1}{2}} \;,
\end{equation}
where we sum over all possible R-symmetry structures with unknown coefficients $\bar{\mu}_{ij}$. We have also made the assumption that only contact Witten diagrams with zero derivatives in the vertices can appear in the ansatz. This is because adding derivatives enhances the growth at the high energy limit. We do not expect the dominant contribution at high energy to come from contact interactions in the supergravity limit, which would be the case if we include higher-derivative contact Witten diagrams. 

A quick comment is that the ansatz can be evaluated as elementary functions of the cross ratios. This should be obvious for the R-symmetry dependence, as well as for the bulk channel exchange and contact contributions. The bulk channel exchange Witten diagrams can all be written as finite sums of contact Witten diagrams which upon evaluation reduce to functions with at most logarithms. The defect channel exchange Witten diagrams are evaluated as Meijer $G$ functions in Section \ref{Subsec:defectexW}. However, for our case they also reduce to simple logarithms. This trivializes the implementation of the superconformal Ward identities (\ref{scfWardid}) which gives rise to linear equations for the unknown coefficients in the ansatz.

It is instructive to illustrate this algorithm with an explicit example. Let us consider the simplest case with $k_1=k_2=2$ which corresponds to the stress tensor multiplet. 
The bulk channel exchange part contains only three terms from the same multiplet
\begin{equation}
    \mathcal{G}_{22}^{\rm bulk}= \mu_{s_2} \mathcal{E}_{2,2}^{2,0} \mathcal{Q}_{\{2,0\}}^{2,2} +  \mu_{A_2} \mathcal{E}_{2,2}^{3,1} \mathcal{Q}_{\{0,2\}}^{2,2} + \mu_{\varphi_2}\mathcal{E}_{2,2}^{4,2} \mathcal{Q}_{\{0,0\}}^{2,2} \;.
\end{equation}
The bulk channel exchange Witten diagrams read\footnote{Compared with the Witten diagrams defined in Section \ref{Sec:treelevelWittendiagrams}, we changed the normalization in this example such that $\mathcal{E}_{2,2}^{\Delta,\ell} \to z^{\frac{\Delta-\ell}{2}-2}  \bar{z}^{\frac{\Delta+\ell}{2}-2}$ in the bulk OPE limit.}
\begin{equation}
\begin{split}
    & \mathcal{E}_{2,2}^{2,0}=\frac{V \log V}{U(V-1)}\;, \quad \mathcal{E}_{2,2}^{3,1} =\frac{6 V (-2 V+(V+1) \log V+2)}{U (V-1)^2}\;, \\& \mathcal{E}_{2,2}^{4,2} = \frac{30 V (-3 V^2+(V^2+4 V+1) \log V+3)}{U
   (V-1)^3}\;,
\end{split}
\end{equation}
and R-symmetry blocks are explicitly given by
\begin{equation}
   \mathcal{Q}_{\{2,0\}}^{2,2} =  \frac{3\sigma+ 3\tau-1}{6\sigma\tau}\;,\quad  \mathcal{Q}_{\{0,2\}}^{2,2} =  \frac{\sigma-\tau}{\sigma \tau}\;,\quad \mathcal{Q}_{\{0,0\}}^{2,2} = \frac{1}{\sigma\tau}\;.
\end{equation}
The defect exchange part of the ansatz is
\begin{equation}
\label{defectansatz1}
    \mathcal{G}_{22}^{\rm defect} =   \widehat{\mu}_{\chi_1,0} \mathcal{\widehat{E}}_{2,2,(0)}^{0,0} \widehat{\mathbb{Q}}_{1,-1}^{+}+ \widehat{\mu}_{\chi_1,1} \mathcal{\widehat{E}}_{2,2,(1)}^{0,0} \widehat{\mathbb{Q}}_{1,-1}^{-}  +  \widehat{\mu}_{x_1}\mathcal{\widehat{E}}_{2,2}^{1,1}  \widehat{\mathbb{Q}}_{0,0}^{+}\;.
\end{equation}
Here we recall from (\ref{E2derto0der}) that the two-derivative exchange Witten diagram differs from the zero-derivative one only by a contact Witten diagram. Therefore, we can also eliminate it from the ansatz at the expense of changing the contact part of the ansatz. These defect exchange Witten diagrams evaluate to be 
\begin{equation}
\begin{split}
   & \mathcal{\widehat{E}}_{2,2,(0)}^{0,0} = 1\;,\quad \mathcal{\widehat{E}}_{2,2,(1)}^{0,0} = \frac{-V^2+2 V \log V+1}{(V-1)^2}\;,\\&  \mathcal{\widehat{E}}_{2,2}^{1,1} = \frac{(-U+V+1) \left(V^2-2 V \log V-1\right)}{(V-1)^3}\;,
    \end{split}
\end{equation}
where we also normalized each diagram such as $\mathcal{\widehat{E}}_{2,2}^{\widehat{\Delta},s} \to r^{\widehat{\Delta}} C_s^{(1)}(\chi/2)$ in the defect OPE limit. The associated defect channel R-symmetry polynomials are
\begin{equation}
\begin{split}
  &\widehat{\mathbb{Q}}_{0,0}^{+} = 1\;,\quad 
     \widehat{\mathbb{Q}}_{1,-1}^{\pm}=  \mp\frac{(\sigma \pm\tau )(\sigma +\tau -1) }{2\sigma  \tau }\;.
\end{split}
\end{equation}
Finally, the contact part of the ansatz is
\begin{equation}
     \mathcal{G}_{22}^{\rm contact}= \mathcal{C}_{2,2}^{(0)}\sum_{i,j \geq 0}^{i+j\leq 2}\bar\mu_{i,j}\sigma^{i-1} \tau^{j-1}\;,
\end{equation}
where
\begin{equation}
    \mathcal{C}_{2,2}^{(0)} = \frac{V (-2 V+(V+1) \log V+2)}{2 (V-1)^3}\;.
\end{equation}
Imposing the superconformal Ward identities (\ref{scfWardid}), we find that all the coefficients are uniquely fixed up to an overall factor\footnote{This is expected because the superconformal Ward identities are homogeneous.} 
\begin{equation}
\begin{aligned}
   &  \mu_{A_2} = \frac{1}{12} \mu_{s_2} \;,\quad \mu_{\varphi_2} = \frac{1}{180} \mu_{s_2} \;,\quad \widehat{\mu}_{\chi_1,0}=\widehat{\mu}_{\chi_1,1}=\widehat{\mu}_{x_1}=\frac{1}{2}\mu_{s_2} \;,\\& \bar{\mu}_{0,1}=\bar{\mu}_{1,0}=-\bar{\mu}_{0,2}=-\bar{\mu}_{2,0}=-\mu_{s_2}\;,\quad \bar{\mu}_{0,0}=\bar{\mu}_{1,1}=0\;.
\end{aligned}
\end{equation}
The result of the $k_1=k_2=2$ two-point function can be compactly written as
\begin{equation}
    G_{22,{\rm free}}= \mu_{s_2}\frac{ (\frac{1}{2}t_1 \cdot \mathbb{M}\cdot t_1) (\frac{1}{2}t_2 \cdot \mathbb{M}\cdot t_2) (\tau -\sigma  \tau  U+\sigma  V)}{2  (P_1\cdot \mathbb{N}\cdot P_1)(P_2\cdot \mathbb{N}\cdot P_2) U \sigma \tau }\;,
\end{equation}
\begin{equation}
    H_{22}= \mu_{s_2}\frac{V \left(V^2-2 V \log V-1\right)}{2(P_1\cdot \mathbb{N}\cdot P_1)^2(P_2\cdot \mathbb{N}\cdot P_2)^2U (1-V)^3}\;.
\end{equation}

Similarly, correlators with higher $k_1$ and $k_2$ can also be fixed by the bootstrap algorithm up to an overall coefficient, with structurally similar results but increasingly more terms. However, these overall coefficients cannot be independent because the coefficients $\mu_{\mathcal{X}}$ in (\ref{defGbulk}) have interpretations in terms of the CFT data of single-particle operators (similarly in the defect channel (\ref{Gdefect})). For example, focusing on the supergravitons, we have
\begin{equation}
    \mu_{s_k}=a_{s_k}\lambda_{k_1k_2k}\;,
\end{equation}
where 
\begin{equation}
    \lambda_{k_1k_2k}=\frac{\sqrt{k_1k_2k}}{N}\;,
\end{equation}
are the three-point function coefficients of $\langle\mathcal{O}_{k_1}\mathcal{O}_{k_2}\mathcal{O}_k \rangle$ in the defect-free theory. The same field $s_k$ can be exchanged in different two-point functions $\llangle \mathcal{O}_{k_1}\mathcal{O}_{k_2} \rrangle$ (e.g., $s_2$ appears in all $\llangle \mathcal{O}_k\mathcal{O}_k \rrangle$). This allows us to relate all the overall coefficients and reduce them a single one which appears in the simplest correlator with $k_1=k_2=2$. This remaining degree of freedom can then be fixed in multiple ways. For example, we can determine it by computing the one-point function of $\llangle \mathcal{O}_2\rrangle$, either from supergravity or from the free theory using its protected nature.

\subsection{All two-point functions and hidden symmetry}\label{Subsec:hiddensymm}
The general two-point function can also be written down in a closed form for arbitrary $k_1$ and $k_2$. We find the free part of (\ref{defGandH}) is
\begin{equation}
\begin{aligned}
    G_{k_1k_2,{\rm free}}= {}&\frac{\prod_{i=1}^2 (\frac{1}{2}t_i\cdot \mathbb{M}\cdot t_i)^\frac{k_i}{2}}{N\prod_{i=1}^2 (P_i\cdot \mathbb{N}\cdot P_i)^\frac{k_i}{2}}(-1)^{\frac{k_{21}}{2}}\sqrt{k_1 k_2}\\&\times U^{-k_1}V^{\frac{k_1}{2}}(\sigma \tau)^{-\frac{k_1}{2}}\bigg((V\sigma+\tau)\sum_{s=1}^{\lfloor \frac{k_1+1}{2}\rfloor}U^{2s-1}V^{-s}(\sigma\tau)^{s-1}\\&-2\sum_{s=1}^{\lfloor\frac{k_1}{2}\rfloor}U^{2s}V^{-s}(\sigma \tau)^{s}+ \frac{1+(-1)^{k_1}}{2}U^{k_1}V^{-\frac{k_1}{2}}(\sigma \tau)^{\frac{k_1}{2}}\bigg)\;.
    \end{aligned}
\end{equation}
On the other hand, the general reduced correlators $H_{k_1k_2}$ can be most conveniently written down in terms of a generating function. This is similar to the hidden conformal symmetry observed in \cite{Caron-Huot:2018kta} but now extended to include defects. To achieve this, we first group together $P$ and $t$ to form 
\begin{equation}
    Z_i=(P_i,t_i)\;, \label{defZ}
\end{equation}
which is a vector in the twelve dimensional embedding space for $\mathbb{R}^{10}$. Note that the reduced defect two-point function $H_{22}$ is R-symmetry neutral and depends only on $x$. Therefore, it can be uniquely expressed in terms of the variables
\begin{equation}
    P_1\cdot P_2\;,\quad P_1\cdot \mathbb{N}\cdot P_1\;,\quad P_1\cdot \mathbb{N}\cdot P_2\;,\quad P_2\cdot \mathbb{N}\cdot P_2\;.
\end{equation}
We will take $H_{22}$ as a seed function and promote it into a generating function by uplifting it into ten dimensions. With the correct overall factor, the seed function is 
\begin{equation}
    H_{22}= \frac{2V \left(V^2-2 V \log V-1\right)}{N(P_1\cdot \mathbb{N}\cdot P_1)^2(P_2\cdot \mathbb{N}\cdot P_2)^2U (1-V)^3}\;.
\end{equation}
To do the uplifting, we define the following replacement rules $\mathbf{T}$
\begin{eqnarray}
       &&P_1\cdot P_2\to Z_1\cdot Z_2\;, \label{repla1}\\
       && P_a\cdot \mathbb{N}\cdot P_b\to Z_a\cdot (\mathbb{N}+\mathbb{M})\cdot Z_b\;,\;\; a,b=1,2\;. \label{repla2}
\end{eqnarray}
This gives us the generating function \begin{equation}
 {\bf H}(x_i;t_i;\lambda_1,\lambda_2)={\bf T}[H_{22}]\big|_{t_1\to\frac{\lambda_1}{\sqrt{2}}t_1,t_2\to\frac{\lambda_2}{\sqrt{2}}t_2}\;,
\end{equation}
where we have also rescaled $t_1$ and $t_2$ by $\lambda_1$ and $\lambda_2$ to more conveniently keep track of the dependence. The general reduced correlator is obtained by the Taylor expansion
\begin{equation}\label{Hk1k2fromHbf}
    H_{k_1k_2}=\frac{\sqrt{k_1k_2}}{2}{\bf H}(x_i;t_i;\lambda_1,\lambda_2)\big|_{\lambda_1^{k_1-2}\lambda_2^{k_2-2}}\;.
\end{equation}
Here $(\ldots)\big|_{\lambda_1^a\lambda_2^b}$ stands for taking the coefficient of $\lambda_1^a\lambda_2^b$ in the expansion which selects all the terms with $a$ copies of $t_1$ and $b$ copies of $t_2$. 

Compared with \cite{Caron-Huot:2018kta}, the new feature of this generalized version of the hidden conformal symmetry is the  replacement rule (\ref{repla2}). Instead of promoting the individual embedding vectors $P_3$ and $P_4$, we promote only the projector $\mathbb{N}$ formed by these two vectors. Recall that $\mathbb{N}$ is defined in (\ref{defN}) as a projector which projects to the line defect $\mathbb{R}$ in the AdS component. The R-symmetry projector $\mathbb{M}$ defined in (\ref{defM}) projects to the part orthogonal to $t_1$ and $t_2$ in $S^5$. This is the $S^3$ which together with $\mathbb{R}$ form the world volume of the D3 brane. Therefore, it is natural to combine them in (\ref{repla2}) to obtain a projector in ten dimensions which projects to the defect. We should view (\ref{Hk1k2fromHbf}) as the manifestation of the 10d hidden conformal symmetry when it is partially broken by the 4d defect.     

As was pointed out in \cite{Chen:2025yxg}, this hidden symmetry also appears to hold at weak coupling for the Lagrangian insertion integrands, generalizing a similar observation for supergraviton four-point functions \cite{Caron-Huot:2021usw}. Denoting the $\ell$-th loop correction at weak coupling as $H_{k_1k_2}^{(\ell)}$, the Lagrangian insertion method allows us to write it as the integral
\begin{equation}
    H_{k_1k_2}^{(\ell)}=\int \prod_{p=5}^{\ell+4} d^4x_p I_{k_1k_2}^{(\ell)}\;.
\end{equation}
The integrand $I_{k_1k_2}^{(\ell)}$ depends on $x_1$, $x_2$ and the insertion points $x_{p=5,\ldots,\ell+4}$. The hidden conformal symmetry can be made manifest by introducing
\begin{equation}\label{defZp}
    Z_p=(x_p,0)\;,
\end{equation}
for the insertion points where the R-symmetry components are set to zero. At each loop level, the seed function is the integrand $I_{22}^{(\ell)}$ for $k_1=k_2=2$ which is a R-symmetry neutral $(\ell+2)$-point defect correlator and has dimension 4 at each point. The replacement rules $\mathbf{T}$ need to be extended into a larger set of rules $\mathbf{T}'$ to account for the additional points. The extension is obtained by simply requiring (\ref{repla2}) to apply to the insertion points as well
\begin{equation}
    P_m\cdot \mathbb{N}\cdot P_n\to Z_m\cdot (\mathbb{N}+\mathbb{M})\cdot Z_n\;,\;\; m,n=1,2,5,\ldots,\ell+4\;.
\end{equation}
Because of (\ref{defZp}), the replacement is only nontrivial when both $m$, $n$ are 1 or 2. Implementing the extended replacement rules on the seed function, we get the generating function for the integrands
\begin{equation}
    \mathbf{I}^{(\ell)}=\mathbf{T}'[I_{22}^{(\ell)}]\big|_{t_1\to \frac{\lambda_1}{\sqrt{2}}t_1,t_2\to \frac{\lambda_2}{\sqrt{2}}t_2}\;.
\end{equation}
We then Taylor expand and get the general integrand
\begin{equation}
    I_{k_1k_2}^{(\ell)}=\frac{\sqrt{k_1k_2}}{2}\mathbf{I}^{(\ell)}\big|_{\lambda_1^{k_1-2}\lambda_2^{k_2-2}}\;.
\end{equation}
We checked this proposal in \cite{Chen:2025yxg} up to two loops. More details about the weak coupling computation of giant graviton correlators can be found in \cite{Wu:2025ott}. 

\subsection{Giant graviton correlators in Mellin space}
A useful formalism for studying holographic defect correlators is the Mellin space representation \cite{Rastelli:2017ecj,Goncalves:2018fwx}. For a generic defect two-point function of scalar operators, the Mellin representation takes the form
\begin{equation}\label{defMellingen}
    \frac{\llangle \mathcal{O}_{\Delta_1} (x_1) \mathcal{O}_{\Delta_2} (x_2) \rrangle}{\llangle \mathcal{O}_{\Delta_1}(x_1) \rrangle \llangle \mathcal{O}_{\Delta_2}(x_2) \rrangle}=\int \frac{d\delta d\gamma}{(2\pi i)^2}\xi^{-\delta}\chi^{-\gamma+\delta}\mathcal{M}_{\Delta_1\Delta_2}(\delta,\gamma)\Gamma(\delta)\Gamma(\gamma-\delta)\prod_{i=1}^2\Gamma\left(\frac{\Delta_i-\gamma}{2}\right)\;,
\end{equation}
where the defect conformal cross ratios $\xi$ and $\chi$ are defined in (\ref{defxichi}). When the theory has a local holographic dual,  $\mathcal{M}_{\Delta_1\Delta_2}(\delta,\gamma)$ has the interpretation of a form factor of two particles scattering with an extended object in AdS. The dual variables $\delta$ and $\gamma$ can be thought of as the Mandelstam variables in the bulk and defect channels respectively. In particular, Witten diagrams admit simple representations in Mellin space. The zero-derivative contact Witten diagrams become just constants
\begin{equation}
    \mathcal{M}^{\rm con}_{\Delta_1\Delta_2}=\frac{\pi^{p/2}\Gamma(\frac{\Delta_1+\Delta_2-p}{2})}{4\Gamma(\Delta_1)\Gamma(\Delta_2)}\;,
\end{equation} 
where $p=0$ for the zero dimensional defect of giant gravitons. The exchange Witten diagrams in the defect channel with transverse spin zero have Mellin amplitudes which are just sums over simple poles
\begin{equation}
    \mathcal{M}_{\Delta_1\Delta_2}^{{\rm defect},\widehat{\Delta}}=\sum_{n=0}^\infty \frac{f_n}{\gamma-\widehat{\Delta}-2n}\;,
\end{equation}
and the residues $f_n$ are constants. In the bulk channel, we can consider exchange Witten diagrams of fields with even spin $\ell$. Their Mellin amplitudes have the form 
\begin{equation}
    \mathcal{M}_{\Delta_1\Delta_2}^{{\rm bulk},\Delta,\ell}=\sum_{n=0}^\infty\frac{h_n(\gamma,\delta)}{\delta-\frac{\Delta_1+\Delta_2-\Delta+\ell-2n}{2}}\;,
\end{equation}
where $h_n(\gamma,\delta)$ are polynomials of degree $\frac{\ell}{2}$. Apparently, the analytic structures of these Witten diagrams are very similar to the corresponding flat-space Feynman diagrams in momentum space.

Unfortunately, this formalism does not directly apply to the giant graviton correlators which we are considering, at the level of full correlators. The reason is that the special one dimensional nature of the AdS defect allows for diagrams which are odd under $3\leftrightarrow 4$ exchange (or $\{U,V\}\to\{U/V,1/V\}$ in terms conformal cross ratios).\footnote{Examples are bulk channel exchange Witten diagrams where the exchanged fields are vectors (more generally any odd spin). Contact Witten diagrams and defect channel exchange Witten diagrams with odd numbers of derivatives are also parity odd.} By contrast, the Mellin representation (\ref{defMellingen}) is always manifestly even with respect to this transformation and therefore cannot be used to represent these Witten diagrams. On the other hand, as we explained below (\ref{Glongpm}) the reduced correlator $\mathcal{H}_{k_1k_2}$ is always invariant under $\{U,V\}\to\{U/V,1/V\}$. This implies that the reduced correlators admit such a representation in Mellin space.\footnote{Physically, this is because the consequence of supersymmetry has already been taken into account and the ``reduced'' fields which are exchanged become essentially just scalars.} It further turns out that the most natural definition of the Mellin amplitudes for the reduced correlators is
\begin{equation}
\label{defMellinH}
H_{k_1k_2}= \prod_{i=1}^{2}\frac{(\frac{1}{2} t_i \cdot \mathbb{M} \cdot t_i)^{\frac{k_i-2}{2}}}{(P_i \cdot \mathbb{N} \cdot P_i)^{\frac{k_i+2}{2}}} \int \frac{d\delta d\gamma}{(2\pi i)^2} \xi^{-\delta}\chi^{-\gamma+\delta-1}\mathcal{M}_{k_1k_2}(\delta,\gamma;\sigma,\bar{\sigma}) \widetilde{\Gamma}_{k_1 k_2}(\delta,\gamma)\;,
\end{equation} 
with 
\begin{equation}
   \widetilde{\Gamma}_{k_1 k_2}(\delta,\gamma)= \Gamma(\delta)\Gamma(\gamma-\delta+1)\prod_{i=1}^2\Gamma\left(\frac{k_i+1-\gamma}{2}\right)\;,
\end{equation}
where we have included some shifts for later convenience.

To show how it works, let us start with the simplest case where $k_1=k_2=2$. We note that the reduced correlator is essentially a contact Witten diagram (in this subsection we have removed the overall $1/N$ factor in the reduced correlators for convenience)
\begin{equation}
H_{22}=-\frac{1}{\prod_{i=1}^2(P_i\cdot \mathbb{N} \cdot P_i)^2 } \frac{8}{\xi} \mathcal{C}^{(0)}_{1,3}\;.
\end{equation}
Using the Mellin representation of the contact Witten diagram, it follows from (\ref{defMellinH}) that the Mellin amplitude of the reduced correlator is
\begin{equation}
\mathcal{M}_{22}=\frac{2}{(\gamma-1)(\delta-1)}\;.
\end{equation}
To obtain the Mellin amplitude of a general reduced correlator, we use the hidden conformal symmetry (\ref{Hk1k2fromHbf}). We simply apply the replacement rules (\ref{repla1}) and (\ref{repla2}) to the Mellin representation of $H_{22}$ to get the generating function. Let us define $\bm{\xi}$ and $\bm{\chi}$ as the cross ratios $\xi$ and $\chi$ after the replacements
\begin{equation}
\begin{split}
    \bm{\xi}  ={}&\frac{-2 Z_1 \cdot Z_2}{(\prod_{i=1}^{2}Z_i\cdot(\mathbb{N}+\mathbb{M})\cdot Z_i)^{\frac{1}{2}}}=\frac{x_{12}^2-\lambda_1 \lambda_2 t_{12}}{\prod_{i=1}^{2} \big(\frac{x_{i3}^2x_{i4}^2}{x_{34}^2}+\lambda_i^2 \frac{t_{i3}t_{i4}}{t_{34}}\big)^{\frac{1}{2}}}\;, \\ \bm{\chi}={}&\frac{2Z_1\cdot Z_2- 2Z_1\cdot (\mathbb{N}+\mathbb{M}) \cdot Z_2}{(\prod_{i=1}^{2}Z_i\cdot(\mathbb{N}+\mathbb{M})\cdot Z_i)^{\frac{1}{2}}}= \frac{\frac{x_{13}^2 x_{24}^2+x_{14}^2 x_{23}^2-x_{12}^2x_{34}^2}{x_{34}^2}+\lambda_1 \lambda_2 \frac{t_{13}t_{24}+t_{14} t_{23}}{t_{34}}}{\prod_{i=1}^{2} \big(\frac{x_{i3}^2x_{i4}^2}{x_{34}^2}+\lambda_i^2 \frac{t_{i3}t_{i4}}{t_{34}}\big)^{\frac{1}{2}}}\;.
    \end{split}
\end{equation}
Then the generating function reads
\begin{equation}
\label{Mellingen}
    \mathbf{H}=\frac{1}{\prod_{i=1}^2\left(Z_i\cdot(\mathbb{N}+\mathbb{M})\cdot Z_i\right)^2}\int\frac{d\delta d\gamma}{(2\pi i)^2} \bm{\xi}^{-\delta}\bm{\chi}^{-\gamma+\delta-1}\mathcal{M}_{22}(\delta,\gamma) \widetilde{\Gamma}_{22}(\delta,\gamma)\;.
\end{equation}
Expanding in $\lambda_1$ and $\lambda_2$, it is not difficult to find
\begin{equation}
\begin{split}
   & \frac{\bm{\xi}^{-\delta}\bm{\chi}^{-\gamma+\delta-1}}{\prod_{i=1}^2\left(Z_i\cdot(\mathbb{N}+\mathbb{M})\cdot Z_i\right)^2}\bigg|_{\lambda_1^{k_1-2}\lambda_2^{k_2-2}}= \prod_{i=1}^{2}\frac{(\frac{1}{2} t_i \cdot \mathbb{M} \cdot t_i)^{\frac{k_i-2}{2}}}{(P_i \cdot \mathbb{N} \cdot P_i)^{\frac{k_i+2}{2}}}\\&\times \sum_{m,n} \binom{-\delta }{m} \binom{-\gamma+\delta-1 }{n} \prod_{i=1}^2\binom{\frac{\gamma-3}{2}}{\frac{k_i-m-n-2}{2}} (-\Sigma)^{m} \bar{\Sigma}^{n} \xi^{-\delta-m}\chi^{-\gamma+\delta-1-n} \;.
    \end{split}
\end{equation}
Comparing with the definition \eqref{defMellinH}, we still need to shift $\delta$, $\gamma$ by $\delta\to \delta-m$, $\gamma\to \gamma-m-n$ and extract a new Gamma function factor $\widetilde{\Gamma}_{k_1 k_2}$. This gives
\begin{equation}\label{Mk1k2gen}
\mathcal{M}_{k_1k_2}=\sum_{m,n}\frac{ \sqrt{k_1k_2} }{m! n! \big(\frac{k_1-m-n-2}{2}\big)!
  \big(\frac{k_2-m-n-2}{2}\big)! }\frac{(-\Sigma)^{m} \bar{\Sigma}^{n}}{ (\gamma -m-n-1)(\delta -m-1)}\;,
\end{equation}
where the sums over $m$, $n$ need to satisfy
\begin{equation}
0\leq m,n\leq k_1-2\;,\quad m+n\leq k_1-2\;,\quad k_1-m-n\in 2\mathbb{Z}_{+}\;.
\end{equation}

It is also useful to see the effect of the hidden conformal symmetry from a slightly different perspective where we use the generalized Mellin representation introduced in \cite{Aprile:2020luw}. In this formalism, not only the conformal cross ratios are Mellin transformed but the R-symmetry cross ratios are transformed as well. This in a sense defines a Mellin amplitude in the full 10d spacetime. To extend this formalism to the giant graviton correlators, we rewrite the reduced correlator (\ref{defMellinH}) as
\begin{equation}
\begin{split}
H_{k_1k_2}={}& \prod_{i=1}^{2}\frac{(\frac{1}{2} t_i \cdot \mathbb{M} \cdot t_i)^{\frac{k_i-2}{2}}}{(P_i \cdot \mathbb{N} \cdot P_i)^{\frac{k_i+2}{2}}} \sum_{\tilde{\delta},\tilde{\gamma}}\int \frac{d\delta d\gamma}{(2\pi i)^2} \xi^{-\delta}\chi^{-\gamma+\delta-1}(-\Sigma)^{-\tilde{\delta}}\bar{\Sigma}^{-\tilde{\gamma}+\tilde{\delta}-1}\\
{}&\times\sqrt{k_1k_2}\,\boldsymbol{\mathcal{M}}_{k_1k_2}(\delta,\gamma;\sigma,\bar{\sigma}) \widetilde{\Gamma}_\otimes(\delta,\gamma;\tilde{\delta},\tilde{\gamma})\;,
\end{split}
\end{equation}
where
\begin{equation}
\widetilde{\Gamma}_\otimes(\delta,\gamma;\tilde{\delta},\tilde{\gamma})=\frac{\Gamma(\delta)\Gamma(\gamma-\delta+1)\Gamma\left(\frac{k_1+1-\gamma}{2}\right)\Gamma\left(\frac{k_2+1-\gamma}{2}\right)}{\Gamma(1-\tilde{\delta})\Gamma(\tilde{\delta}-\tilde{\gamma})\Gamma\left(\frac{k_1+1+\tilde{\gamma}}{2}\right)\Gamma\left(\frac{k_2+1+\tilde{\gamma}}{2}\right)}\;.
\end{equation}
Here we have essentially only performed a change of variables from $m$, $n$ to $\tilde{\delta}$, $\tilde{\gamma}$ and we also need to sum over all the R-symmetry structures. This sum can further be rewritten as contour integrals \`a la Sommerfeld but we will not explicitly do it here. In this formalism, hidden conformal symmetry is manifested partially by the statement that the Mellin amplitude $\boldsymbol{\mathcal{M}}_{k_1k_2}$ is independent of $k_1$ and $k_2$. It is not difficult to check using (\ref{Mk1k2gen}) that the new Mellin amplitude is given by
\begin{equation}
\boldsymbol{\mathcal{M}}_{k_1k_2}(\delta,\gamma;\tilde{\delta},\tilde{\gamma})=\frac{1}{({\boldsymbol \delta}-1)({\boldsymbol \gamma}-1)}\;,
\end{equation}
where 
\begin{equation}
{\boldsymbol \delta}=\delta+\tilde{\delta}\;,\quad {\boldsymbol \gamma}=\gamma+\tilde{\gamma}+1\;.
\end{equation}
Note that instead of depending separately on $\delta$, $\gamma$ and $\tilde{\delta}$, $\tilde{\gamma}$, the 10d Mellin amplitude depends only on the boldfaced variables which are their diagonal combinations. This is also a key feature shared by the case of four-point functions of light supergravitons \cite{Aprile:2020luw}. Note that to achieve the independence on $k_1$ and $k_2$, we have extracted a factor of $\sqrt{k_1k_2}$ in the definition. We can get rid of this factor by redefining the normalization of the operators $\mathcal{O}_{k_1}$ and $\mathcal{O}_{k_2}$. We note that such a normalization redefinition was also needed in the four-point function case.

\subsection{Consistency checks}
There are several nontrivial consistency checks which we can perform on our results. These checks use results obtained from different independent methods in different regimes and limits. Therefore, their agreement provides nontrivial support. 

First, superconformal symmetry allows us to split the giant graviton correlators into a free part and a coupling dependent part as in (\ref{defGandH}). The free theory corrector can also be computed by using Wick contractions and this has been done for $k_1=k_2=p$ in \cite{Jiang:2019xdz}. We find that our result agrees with \cite{Jiang:2019xdz} for odd $p$ but differs for even $p$ 
\begin{equation}
   G_{\rm free}-G_{{\rm free},\; \text{\cite{Jiang:2019xdz}}}=\frac{1+(-1)^p}{2}\frac{p}{N}\frac{\prod_{i=1}^2 (\frac{1}{2}t_i\cdot \mathbb{M}\cdot t_i)^\frac{p}{2}}{\prod_{i=1}^2 (P_i\cdot \mathbb{N}\cdot P_i)^\frac{p}{2}}\;,
\end{equation}
which has the form of the product of two one-point functions. This difference arises from two sources. The free theory calculation \cite{Jiang:2019xdz} considered $U(N)$ gauge group while the bootstrap calculation in the supergravity limit corresponds to $SU(N)$. Moreover, the supergravity calculation are naturally in the single-particle basis but the field theory calculation was performed in the single-trace basis.

Second, from our giant graviton correlators we can extract one-point functions of various fields. Note that we have not used these one-point functions as input in our bootstrap algorithm. We only used the fact that the same one-point functions can appear in different two-point functions in order to relate the overall coefficients for correlators of different $k_1$ and $k_2$. On the other hand, these one-point functions can be independently computed using the supergravity effective action as we have shown in Section \ref{Sec:sugra1pt}. We find that the results from the two calculations are in complete agreement, upon setting $q_0=1$ in (\ref{sk1ptcoef}). 

Third, we can also examine the flat-space limit of our results. Because Mellin amplitudes cannot be defined for the full correlator, we will focus on those of the reduced correlator.\footnote{It is also possible to analyze the flat-space limit of the full correlator in position space, see \cite{Chen:2025cod}.} It was shown in \cite{Alday:2024srr} that the flat-space amplitude is related to the high-energy limit of the Mellin amplitude which according to (\ref{Mk1k2gen}) is 
\begin{equation}
    \mathcal{M}_{k_1,k_2}\sim \sum_{m,n}\frac{(-\Sigma)^m\bar{\Sigma}^n}{m!n!(\frac{k_1-2-m-n}{2})!(\frac{k_2-2-m-n}{2})!}\frac{1}{\gamma\delta}\;.
\end{equation}
The $1/(\gamma\delta)$ behavior agrees with the leading small $\ell_s$ expansion of the amplitude of strings scattering with a D-brane \cite{Klebanov:1995ni,Hashimoto:1996bf,Garousi:1996ad}.\footnote{The fact that we are looking at the Mellin amplitude of the reduced correlator removes a quadratic factor in $\delta$ and $\gamma$ compared to the case of the full correlator considered in \cite{Alday:2024srr}.} Moreover, the factorized dependence on the R-symmetry cross ratios for correlators of higher KK levels is also expected as in the defect-free cases \cite{Alday:2021odx}. 

Finally, we can compare our result with supersymmetric localization \cite{Brown:2024tru} which computed the integrated correlator for $k_1=k_2=2$. Using the same normalization, it boils down to computing the integral
\begin{equation}
    -\frac{2N}{\pi} \int_0^{\infty} d r \int_0^\pi d \theta \frac{r^3 \sin ^2 \theta}{V^2} \mathcal{H}_{22}(U,V) =1\;,
\end{equation}
where $U=1-2r\cos \theta + r^2$ and $V=r^2$. This agrees with the leading term of the prediction of \cite{Brown:2024tru} expanded at strong coupling.

\section{Extracting CFT data from giant graviton correlators}\label{Sec:extractingCFTdata}

The giant graviton correlators contain a wealth of data which can be extracted from an OPE analysis. At tree level, one finds not only the single-particle states which already appeared as the exchanged fields in the Witten diagrams, but also double-particle operators. Before proceeding, let us characterize such double-particle operators in more detail. 

In the bulk channel, these operators are double-particle in the literal sense and are normal ordered products of two single-particle operators dressed with derivatives. In the defect channel, the notion is more subtle. For extended defects (with defect dimensions $p\geq 1$), we can create such defect double-particle operators by restricting the bulk local single-particle operators to the defect and dress them with transverse derivatives. We usually further ``subtract'' the defect and think of them as excitations living on the defect. For our case, the zero dimensional nature of the defect makes this procedure  harder to visualize. However, we can instead imagine such manipulations in the dual AdS picture where the defect has an extended dimension. Alternatively, we can exploit the fact that giant graviton operators are also local operators use the four-point function OPE. These defect double-particle operators are related to the operators in the $\mathcal{O}_{k_1}\times \mathcal{D}$ OPE (and similarly in the $\mathcal{O}_{k_2}\times \mathcal{D}$ OPE) of the schematic form $:\mathcal{O}_{k_1}\square^n\partial^s\mathcal{D}:$. They are defined by further ``removing'' the heavy background $\mathcal{D}$.  We denote the defect channel double-trace operators from $\mathcal{O}_k$ as $\widehat{[\mathcal{O}_k]}_{n,s}$. They have engineering dimensions $\widehat{\Delta}=k+2n+s$ where $2n$ comes from the $n$ pairs of contracted derivatives and $s$ is from the $s$ uncontracted derivatives. They further carry quantum numbers $r_1$, $r_2$ with respect to the unbroken $SO(4)$ and $U(1)$ R-symmetry subgroups, which we suppress in this notation. Note that ``subtracting'' or ``removing'' the defect background is difficult to formalize in both pictures. However, this becomes automatic in terms of the defect conformal blocks and R-symmetry blocks.

Our primary focus will be the defect channel double-particle operators. Unlike the bulk channel double-particle operators, which already show up in four-point functions of light operators, the defect ones can only be accessed by studying giant graviton correlators. We are interested in extracting the CFT data associated with these operators. A difficulty is that these operators are degenerate in their engineering dimensions. The conformal block decomposition of a certain defect two-point function only gives the average of the data over the degenerate operators. We will show that it is however possible to unmix the data by considering all different correlators. The strategy is similar to \cite{Aprile:2017xsp} in the defect-free case and to \cite{Chen:2024orp} in the defect case. 

To help a casual reader navigate through this somewhat technical section, we highlight the main results of the analysis as follows.
\begin{itemize}
    \item We obtain the complete tree-level supergravity anomalous dimensions of the defect channel double-particle operators. The result is given by (\ref{defectanomalousdim}).
    \item The mixing structure of double-particle operators in the defect channel also has imprints from the partially broken 10d hidden conformal symmetry. We show in Section \ref{Subsec:higherddefectdecom} how this symmetry can be used to automatically diagonalize the unmixing problem.
    \item We also find a useful fourth order differential operator (\ref{Delta4}). Using this operator, we obtain all-loop results for leading logarithmic singularities in the defect channel.
\end{itemize}

\subsection{Defect channel superconformal block decomposition}
Let us start by reviewing the structure of decomposition in the defect channel and also set up some notations for the rest of the section. Nonperturbatively, the unprotected part of the two-point correlators, which consists of long superconformal multiplets, can be expanded into defect channel superconformal blocks as
\begin{equation}
\label{longexpan}
   \mathcal{G}_{k_1 k_2}\big|_{\rm long} = \sum_{\mathcal{R}} B_{\mathcal{R}}  \,\widehat{\mathfrak{G}}^{\rm long}_{\mathcal{R}}\;.
\end{equation}
Here $\mathcal{R}=\{\widehat{\Delta},s ,r_1,r_2\}$ is a collection of quantum numbers of the superconformal primaries of the long multiplets. The decomposition coefficients have the form of products of bulk-defect two-point function coefficients
\begin{equation}
   B_{\mathcal{R}}= \sum_{\widehat{\mathcal{O}} \in \mathcal{R}} b_{k_1 \widehat{\mathcal{O}}} b_{k_2 \widehat{\mathcal{O}}}\;,
\end{equation}
and we have included the possibility of degenerate operators $\widehat{\mathcal{O}}$ with the same quantum numbers. In the infinite $N$ limit, where the theory is essentially generalized free fields, the decomposition also takes this form. The long operators are the defect channel double-particle operators $\widehat{[\mathcal{O}_k]}_{n,s}$ we introduced earlier. Operator degeneracy is common in this limit. 

We then consider the $1/N$ corrections at infinite 't Hooft coupling. This corresponds to turning on the supergravity interactions in AdS. The conformal dimensions and OPE coefficients of the double-particle long operators $\widehat{\mathcal{O}}$ have expansions of the form 
\begin{equation}
\widehat{\Delta}_{\widehat{\mathcal{O}}} =  \widehat{\Delta}_{\widehat{\mathcal{O}}}^{(0)}+\frac{2}{N} \widehat{\gamma}_{\widehat{\mathcal{O}}}^{(1)}+\mathcal{O}\left(\frac{1}{N^2}\right)\;,\quad  b_{k_i \widehat{\mathcal{O}}} = b_{k_i \widehat{\mathcal{O}}}^{(0)}+\frac{1}{N} b_{k_i \widehat{\mathcal{O}}}^{(1)}+\mathcal{O}\left(\frac{1}{N^2}\right)\;.
\end{equation}
In particular, the correction to the conformal dimension can be viewed as the binding energy between the supergraviton and the giant graviton. We will truncate at tree-level supergravity and the correction to the two-point defect correlator is of order $\mathcal{O}(1/N)$. From the $\widehat{\Delta}$ dependence of the conformal blocks (\ref{gdefectpiecewise}), it is clear that the $1/N$ expansion gives rise to $\log V$ singularities. The decomposition (\ref{longexpan}) then up to this order is corrected as
\begin{equation}\label{Gk1k2longsubleading}
    \mathcal{G}_{k_1 k_2}\big|_{\rm long} = \sum_{\mathcal{R}_0} B_{\mathcal{R}_0}^{(0)}  \widehat{\mathfrak{G}}^{\rm long}_{\mathcal{R}_0} + \frac{1}{N}\left((\log V) \sum_{\mathcal{R}_0} \Omega_{\mathcal{R}_0}^{(1)}  \widehat{\mathfrak{G}}^{\rm long}_{\mathcal{R}_0}+\sum_{\mathcal{R}_0} B_{\mathcal{R}_0}^{(1)}  \widehat{\mathfrak{G}}^{\rm long}_{\mathcal{R}_0}\right)+\mathcal{O}\left(\frac{1}{N^2}\right)\;,
\end{equation}
where $\mathcal{R}_0 = \{\widehat{\Delta}_{\widehat{\mathcal{O}}}^{(0)},s,r_1,r_2\}$ are the engineering quantum numbers
\begin{equation}
    \begin{aligned}
    \label{quadsystems}
        B_{\mathcal{R}_0}^{(0)}= {}&\sum_{\widehat{\mathcal{O}} \in \mathcal{R}_0} b_{k_1 \widehat{\mathcal{O}}}^{(0)} b_{k_2 \widehat{\mathcal{O}}}^{(0)}\;, \\   \Omega_{\mathcal{R}_0}^{(1)}= {}&\sum_{\widehat{\mathcal{O}} \in \mathcal{R}_0} b_{k_1 \widehat{\mathcal{O}}}^{(0)} b_{k_2 \widehat{\mathcal{O}}}^{(0)} \widehat{\gamma}_{\widehat{\mathcal{O}}}^{(1)}\;,
    \end{aligned}
\end{equation}
and 
\begin{equation}
    B_{\mathcal{R}_0}^{(1)}=\sum_{\widehat{\mathcal{O}} \in \mathcal{R}_0} b_{k_1 \widehat{\mathcal{O}}}^{(0)} b_{k_2 \widehat{\mathcal{O}}}^{(1)}+b_{k_1 \widehat{\mathcal{O}}}^{(1)} b_{k_2 \widehat{\mathcal{O}}}^{(0)}\;.
\end{equation}
Note that these decomposition coefficients can be explicitly computed from GFF correlators and the tree-level defect two-point functions.
The task of this section is to solve these constraints and extract the unmixed CFT data. In particular, we want to extract the anomalous dimensions from the system (\ref{quadsystems}).

Before we proceed to unmixing, which is the task of the next subsection, let us look at the decomposition in more detail. We first consider the zeroth order where the two-point function is just a free-propagator in AdS
\begin{equation}
    G_{kk}^{(0)}=\left(\frac{t_{12}}{x_{12}^2}\right)^k\;,
\end{equation}
and two-point functions with unequal $k_1$, $k_2$ vanish. Rewriting them in terms of cross ratios, we have
\begin{equation}
\mathcal{G}_{kk}^{(0)}=\left(\frac{\Sigma}{\xi}\right)^{k}\;.
\end{equation}
The GFF two-point functions can be decomposed into superconformal blocks as\footnote{Strictly speaking, this decomposition identity is only valid in the regions $\{0<V<1,0<\sigma/\tau<1\}$ and $\{V\geq 1, \sigma/\tau\geq 1\}$ which are in the defect OPE limit. In the other regions $\{0<V<1,\sigma/\tau\geq 1\}$, $\{V\geq 1,0<\sigma/\tau<1\}$, we find additional superconformal decomposition elements (see Appendix \ref{App:scfblocks} for more comments). But these elements vanish in the first two regions and therefore do not participate in the OPE analysis.}
\begin{equation}
\begin{aligned}
\label{superexpanoffreepropagator}
    \mathcal{G}_{kk}^{(0)} = {}&\sum_{i=0}^{k} A^{(0)}_{k,i} \widehat{\mathfrak{G}}_{i,k-i}^{\rm sh} +   \sum_{i=0}^{k-1} \sum_{s=0}^{\infty} C^{(0)}_{k,s,i}\widehat{\mathfrak{G}}_{s,i,k-i-2}^{\rm se} \\{}& +\sum _{r_1=0}^{k-2} \sum_{j=0}^{k-r_1-2} \sum_{m,s=0}^{\infty} B^{(0)}_{k,m,s,r_1,j} \widehat{\mathfrak{G}}^{\rm long}_{k+2m+s,s,r_1,-k+r_1+2j}\;,
    \end{aligned}
\end{equation}
where the coefficients of the short and semi-short superconformal multiplets are
\begin{equation}
    A_{k,i}^{(0)}=\binom{k}{i}\;,\quad    C_{k,s,i}^{(0)}=\frac{\Gamma (k+s+2)}{(s+i+2) \Gamma (i+1) \Gamma (s+2) \Gamma (k-i)}\;.
\end{equation}
The coefficients of the long multiplets, which we are most interested in, read
\begin{align}
\label{longcoef}
 B^{(0)}_{k,m,s,r_1,j} ={}&\frac{\left(r_1+1\right) (s+1) (m+1)_k (m+s+2)_k}{(k-j+m)
   (k-j+m-r_1-1) (k-j+m+s+1) (k-j+m-r_1+s)} \nonumber\\ & \times \frac{1}{\Gamma (j+1) \Gamma (k-j) \Gamma (j+r_1+2) \Gamma(k-j-r_1-1) }\;.
   \end{align}

We then consider the decomposition of the tree-level supergravity correction, focusing on the part proportional to the anomalous dimensions. This amounts to looking at the $\log V$ coefficients of the tree-level defect two-point functions. Thanks to the nice factorized form of the long superconformal blocks (\ref{Glongpm}), the decomposition reduces to expanding the $\log V$ part of the reduced two-point function into bosonic conformal and R-symmetry blocks with shifted quantum numbers
\begin{equation}
\label{treelogdecomphigherk}
\begin{split}
    \mathcal{H}_{k_1 k_2}\big|_{\log V } ={}& \sum_{r_1=0}^{k_1-2}\sum_{j=0}^{k_1-r_1-2}  \sum_{m,s = 0}^{\infty} \Omega^{(1)}_{k_1,k_2,r_1,j,m,s}\frac{\widehat{\mathfrak{G}}^{\rm long}_{k_2+2m+s,s,r_1,-k_1+r_1+2j}}{R(V\sigma\tau)^{-1}}\\
    ={}& \sum_{r_1=0}^{k_1-2}\sum_{j=0}^{k_1-r_1-2}  \sum_{m,s = 0}^{\infty} \Omega^{(1)}_{k_1,k_2,r_1,j,m,s}\widehat{g}_{k_2+2m+s+2,s}^+ \widehat{\mathcal{Q}}_{r_1,-k_1+r_1+2j+2}^+\;.
 \end{split}
\end{equation}
Here the $1/N$ factor in the tree-level reduced correlators has also been extracted outside of the decomposition as in (\ref{Gk1k2longsubleading}). Unlike the previous GFF case, it is difficult to obtain a closed form formula for the $\Omega^{(1)}_{k_1,k_2,r_1,j,m,s}$ coefficients.\footnote{However, this is easy for low-lying correlators. For example, for $\mathcal{H}_{2k}$ we found
\begin{equation}
    \mathcal{H}_{2k}\big|_{\log V } = -\sum_{m,s = 0}^{\infty} \frac{\sqrt{2k}(m+1)_k}{\Gamma \left(\frac{k}{2}+1\right) \Gamma \left(\frac{k}{2}\right)}\widehat{g}_{k+2m+s+2,s}^+\;.
\end{equation}} This does not present a problem for us because we can always obtain sufficiently many coefficients to perform the unmixing which we will detail in the next subsection.

\subsection{Unmixing defect channel double-particle operators}
To understand more precisely what we are unmixing, let us first analyze the degeneracy pattern of the defect double-particle operators. We consider the degenerate operators with a given twist $\widehat{\tau}=\widehat{\Delta}-s$, transverse spin $s$ and R-symmetry charges $(r_1,r_2)$. These are the double-particle operators of the form $\widehat{[\mathcal{O}_p]}_{\frac{\widehat{\tau}-p}{2},s}$. The allowed values of $p$ are 
\begin{equation}\label{pandi}
    p = r_1+|r_2+2|+2i\;, \quad    i = 1,2,\ldots,\frac{\widehat{\tau}-r_1-|r_2+2|}{2}\;.
\end{equation}
Note that $\widehat{\tau}$ has a lower bound $\widehat{\tau}\geq p_{\min}=r_1+|r_2+2|+2$. The correlation between allowed range of $p$ and  $(r_1,r_2)$ can be qualitatively understood as follows. The original single-particle operator $\mathcal{O}_p$ carries charge $p$ with respect to the full $SO(6)$ R-symmetry. When branched into the $SO(4)\times SO(2)$ subgroups, $p$ must be large enough so that $(r_1,r_2)$ is contained in the spectrum. Moreover, $p$ must also be capped from above because the number of contracted derivatives used to construct such operators must be non negative.

Since different $(r_1,r_2)$ sectors of the reduced correlator are independent, we can consider unmixing in each sector separately.

Let us first consider the singlet channel\footnote{Here singlet is from the point of view of the reduced correlator. Notice there is a shift of $r_2$ in (\ref{Glongpm}). The contribution to the reduced correlator does not carry R-symmetry charge, even though the superconformal primary is charged.} where $r_1=0$, $r_2=-2$. The allowed twists in this sector are $\widehat{\tau}=2+2n$ with $n=0,1,2\ldots$. We can analyze these degenerated operators with increasing twists. 

\noindent{\bf Twist 2:} At twist 2, there is a single operator and degeneracy is absent. The equations (\ref{quadsystems}) become\footnote{Here the notation is $b_{pi}^{(0)}$ and $\widehat{\gamma}^{(1)}_i$. The label $p$ denotes the operator $\mathcal{O}_p$ and the index $i$ runs over the diagonalized degenerate long operators that can be exchanged.}
\begin{equation}
    (b_{21}^{(0)})^2 = s+1\;,\quad (b_{21}^{(0)})^2 \widehat{\gamma}^{(1)}_1 = -4\;.
\end{equation}
Solving these equations, we get the anomalous dimension
\begin{equation}
    \widehat{\gamma}^{(1)}_1 = -\frac{4}{s+1}\;.
\end{equation}

\noindent{\bf Twist 4:} The problem becomes nontrivial at twist 4 where double-particle operators from both $\mathcal{O}_2$ and $\mathcal{O}_4$ appear. The two-fold degeneracy leads to six CFT data, which in the diagonalized basis are denoted by $b_{21}^{(0)}$, $b_{22}^{(0)}$, $b_{41}^{(0)}$, $b_{42}^{(0)}$ and $\widehat{\gamma}^{(1)}_1$, $\widehat{\gamma}^{(1)}_2$. It is more convenient to organize them in matrix form
\begin{equation}
    \mathbf{B} = \left(\begin{array}{cc}
b_{21}^{(0)} & b_{22}^{(0)} \\
b_{41}^{(0)} & b_{42}^{(0)}
\end{array}\right)\;, \quad \mathbf{\Gamma} = \left(\begin{array}{cc}
\widehat{\gamma}_1^{(1)} & 0 \\
0 & \widehat{\gamma}_2^{(1)}
\end{array}\right)\;.
\end{equation}
Then (\ref{quadsystems}) can be written as 
\begin{equation}
\begin{aligned}
\label{quadraticeq}
 \mathbf{B} \mathbf{B}^T={} \mathbf{N}\;, \quad   \mathbf{B} \mathbf{\Gamma} \mathbf{B}^T = \mathbf{\Omega}\;,
\end{aligned}
\end{equation}
with
\begin{equation}
\label{NandOmega}
 \mathbf{N} =  \left(\begin{array}{cc}
s+1 & 0 \\
0 & (s+1)(s+2)(s+5)
\end{array}\right)\;,\quad \mathbf{\Omega} = \left(\begin{array}{cc}
-12 & -24\sqrt{2} \\
-24\sqrt{2} & -12 (12 + s (5 + s))
\end{array}\right)\;,
\end{equation}
computed from \eqref{superexpanoffreepropagator}, \eqref{treelogdecomphigherk}. Because the GFF matrix $\mathbf{N}$ is always diagonal, we can define a new matrix $\mathbf{C}$ via
\begin{equation}
    \mathbf{B} = \sqrt{\mathbf{N}} \mathbf{C}\;,
\end{equation} 
where $\sqrt{\mathbf{N}}$ is the diagonal matrix defined with the square roots of the elements of $\mathbf{N}$.
Note that $\mathbf{C}^T = \mathbf{C}^{-1}$ and the second equations in \eqref{quadraticeq} can be rewrite as 
\begin{equation}
\mathbf{C} \mathbf{\Gamma} \mathbf{C}^{-1} = (\sqrt{\mathbf{N}})^{-1} \mathbf{\Omega} (\sqrt{\mathbf{N}})^{-1}\;.    
\end{equation}
This implies the diagonalized anomalous dimension $\widehat{\gamma}^{(1)}_1,\widehat{\gamma}^{(1)}_2$ are the eigenvalues of
\begin{equation}
\label{mixingmatrix}
    (\sqrt{\mathbf{N}})^{-1} \mathbf{\Omega} (\sqrt{\mathbf{N}})^{-1}  \equiv  \left(\begin{array}{cc}
\frac{(b_{21}^{(0)})^2 \widehat{\gamma}_1^{(1)}+(b_{22}^{(0)})^2 \widehat{\gamma}_2^{(1)}}{(b_{21}^{(0)})^2+(b_{22}^{(0)})^2} &  \frac{b_{21}^{(0)} b_{41}^{(0)} \widehat{\gamma}_1^{(1)}+b_{22}^{(0)} b_{42}^{(0)} \widehat{\gamma}_2^{(1)}}{\sqrt{(b_{21}^{(0)})^2+(b_{22}^{(0)})^2} \sqrt{(b_{41}^{(0)})^2+(b_{42}^{(0)})^2}} \\
\frac{b_{21}^{(0)} b_{41}^{(0)} \widehat{\gamma}_1^{(1)}+b_{22}^{(0)} b_{42}^{(0)} \widehat{\gamma}_2^{(1)}}{\sqrt{(b_{21}^{(0)})^2+(b_{22}^{(0)})^2} \sqrt{(b_{41}^{(0)})^2+(b_{42}^{(0)})^2}} &   \frac{(b_{41}^{(0)})^2 \widehat{\gamma}_1^{(1)}+(b_{42}^{(0)})^2 \widehat{\gamma}_2^{(1)}}{(b_{41}^{(0)})^2+(b_{42}^{(0)})^2}
\end{array}\right)\;,
\end{equation}
which we can easily solve and get
\begin{equation}
\label{example}
\widehat{\gamma} _1^{(1)} = -\frac{12 (s+4)}{(s+1) (s+2)}\;,\quad  \widehat{\gamma}_2^{(1)} = -\frac{12}{s+5}\;.
\end{equation}

\noindent{\bf Twist 6:} We can similarly proceed to twist 6. Here the contributing $\mathcal{O}_p$ are $p=2,4,6$ and we have a three-fold degeneracy with $i=1,2,3$. The matrix $ (\sqrt{\mathbf{N}})^{-1} \mathbf{\Omega} (\sqrt{\mathbf{N}})^{-1} $, expressed in terms of the transverse spin $s$, reads
\begin{equation}
\nonumber
    \frac{-24}{s+1}\times\left(
\begin{array}{ccc}
 1 & \frac{2 \sqrt{5}}{\sqrt{(s+3) (s+6)}} & \frac{6 \sqrt{5}}{\sqrt{(s+2) (s+3) (s+6) (s+7)}} \\
 \frac{2 \sqrt{5}}{\sqrt{(s+3) (s+6)}} & \frac{5 s^2+31 s+126}{5 (s+3) (s+6)} & \frac{6 (3 s^2+19 s+66)}{5 (s+3) (s+6) \sqrt{(s+2) (s+7)}} \\
 \frac{6 \sqrt{5}}{\sqrt{(s+2) (s+3) (s+6) (s+7)}} & \frac{6 (3 s^2+19 s+66)}{5 (s+3) (s+6) \sqrt{(s+2) (s+7)}} & \frac{5 s^4+74 s^3+445 s^2+1192 s+1716}{5 (s+2) (s+3) (s+6) (s+7)}\\
\end{array}
\right)\;.
\end{equation}
Diagonalizing this matrix gives the anomalous dimensions
\begin{equation}
    \widehat{\gamma}^{(1)}_1 =-\frac{24 (s+4) (s+5)}{(s+1) (s+2) (s+3)}\;,\quad\widehat{\gamma}^{(1)} _2 = -\frac{24}{s+3}\;,\quad\widehat{\gamma}^{(1)} _3 = -\frac{24
   (s+4)}{(s+6) (s+7)}\;.
\end{equation}
\noindent{\bf Higher twists:} After experimenting with higher values of twists, we arrive at the following formula for general anomalous dimensions in the $(0,-2)$ sector
\begin{equation}
    \widehat{\gamma}^{(1)}_{i} = -\frac{\widehat{\tau } \left(\widehat{\tau }+2\right) \left(\widehat{\tau }+2s+2\right) \left(\widehat{\tau
   }+2s+4\right)}{8(s+2 i-1)_3}\;.
\end{equation}

Similar analyses can be performed in other R-symmetry sectors and lead to closed form formulas for the anomalous dimensions. For example, in the $(0,0)$ sector we find 
\begin{equation}
    \widehat{\gamma}^{(1)}_i = -\frac{(\widehat{\tau}-2)\widehat{\tau}(\widehat{\tau}+2s)(\widehat{\tau}+2s+2)}{8(s+2 i+1)_3}\;,
\end{equation}
and in the $(1,-2)$ sector
\begin{equation}
   \widehat{\gamma}^{(1)}_i =  -\frac{(\widehat{\tau}-1)(\widehat{\tau}+3)(\widehat{\tau}+2s+1)(\widehat{\tau}+2s+5)}{8(s+2i-1)_3 }\;.
\end{equation}
By tracking the dependence of the formulas on the R-symmetry charges, we obtain the following general result. For the degenerate defect channel double-particle operators $\{\widehat{[\mathcal{O}_p]}_{\frac{\widehat{\tau}-p}{2},s}\}$ with R-symmetry charges $(r_1,r_2)$, the diagonalized anomalous dimensions are given by
\begin{equation}
\label{defectanomalousdim}
    \widehat{\gamma}^{(1)}_p =  -  \frac{2 M_t^{(2)} M_{t+s+1}^{(2)} }{\left( s+p-r_1-1\right)_3}\;,
\end{equation}
where 
\begin{equation}
    M_t^{(2)} = (t-1)(t+r_1)\;,\quad  t=\frac{\widehat{\tau}-r_1-r_2}{2}\;,
\end{equation}
and we recall $p$ is related to the degeneracy label $i$ via (\ref{pandi}). We also point out that this result is very similar to anomalous dimensions of the bulk channel double-trace operators obtained from unmixing defect-free four-point functions \cite{Aprile:2018efk}.

\subsection{Implication of hidden symmetry}\label{Subsec:higherddefectdecom}
In Section \ref{Subsec:hiddensymm} we pointed out the existence of a remarkable hidden symmetry which allows us to unify all different KK two-point functions. We emphasized that in contrast to the 10d hidden conformal symmetry observed in  defect-free four-point functions, the inclusion of giant gravitons acts as 4d defects in the 10d total spacetime and partially breaks this symmetry. In this subsection, we keep developing this idea and we show that this symmetry can be used to automatically diagonalize the mixing problem in the defect channel.

We start with the analogue of the GFF two-point function in ten dimensions. Similar to the defect-free case, we should take the operators to be the Lagrangian of $\mathcal{N}=4$ SYM which has dimension 4. The two-point function is
\begin{equation}
    \mathbb{H}^{(0)}_{22} = \frac{V^2}{U^4}\;,
\end{equation}
where the cross ratios defined from 10d distances. We then decompose it into defect channel conformal blocks where the defect is a 4d one living in 10d flat space. These conformal blocks are given by (\ref{pqdefectblock}) with dimension and co-dimension set to $p=4$ and $q=6$. The decomposition reads
\begin{equation}
     \mathbb{H}_{22}^{(0)} = \sum_{s=0}^{\infty} \mathbb{a}_{s}^{(0)}  \widehat{g}_{4+s,s}^{(4,6)}\;,
\end{equation}
where the coefficients are given by
\begin{equation}\label{a0coef}
   \mathbb{a}_{s}^{(0)}  = \frac{1}{6} (s+1) (s+2) (s+3)\;.
\end{equation}
A remarkable feature of this decomposition is that all the defect channel conformal blocks have the same twist.

We then move on to the tree-level correlators. Hidden symmetry implies that it is sufficient to consider only the lowest KK two-point function which gives rise to the generating function ${\bf H}$. We extract the $\log V$ part of the generating function\footnote{Here we are slightly abusing the notation by dropping the one-point function factors in $\mathbb{H}$.}
\begin{equation}
\label{treelog}
    \mathbf{H}\big|_{\log V} = \frac{4 V^2}{U(V-1)^3}\;,
\end{equation}
where the cross ratios are also ten dimensional. Decomposing it into defect channel conformal blocks gives 
\begin{equation}
\label{10ddecomofH22}
   \mathbf{H}\big|_{\log V} = \sum_{s=0}^{\infty} \mathbb{a}_{s}^{(0)} \mathbb{\Gamma}_{s}^{(1)}  \widehat{g}_{4+s,s}^{(4,6)}\;,
\end{equation}
where we also find just a single twist in the expansion and the coefficients are 
\begin{equation}
   \mathbb{a}_{s}^{(0)} \mathbb{\Gamma}_{s}^{(1)} = -4\;.
\end{equation}
Comparing it with (\ref{a0coef}), we obtain the 10d anomalous dimensions
\begin{equation}
     \mathbb{\Gamma}_{s}^{(1)} = -\frac{24}{(s+1)_3}\;.
\end{equation}

We now show this defect decomposition in 10d diagonalizes the mixing problem. We illustrate this with an example where the diagonalization is in the $(0,-2)$ sector and the operators are two-fold degenerate with twist 4. This example has been considered in the previous subsection. Here we will approach it from a different angle to reproduce the results (\ref{example}). Our starting point is the matrix
\begin{equation}
   \left(\begin{array}{cc}
\widehat{\mathcal{D}}_{22} & \widehat{\mathcal{D}}_{24} \\
\widehat{\mathcal{D}}_{42} & \widehat{\mathcal{D}}_{44}
\end{array}\right)  \mathbb{a}_{s}^{(0)} \mathbb{\Gamma}_{s}^{(1)}\widehat{g}^{(4,6)}_{4+s,s}\;,
\end{equation}
with
\begin{equation}
    \begin{aligned}
    \nonumber
     \widehat{\mathcal{D}}_{22} = {}& 1\;,
    \\ \widehat{\mathcal{D}}_{24} = {}& \widehat{\mathcal{D}}_{42} =  \sqrt{2} \left(\frac{1+V}{1-V} V\partial_V + \frac{V}{1-V} U\partial_{U}-2 \right)\;,
    \\ \widehat{\mathcal{D}}_{44} = {}&   V (V-1)^{-2} \left(2 (U^2+U-3) V+U^2-2 U+3
   V^2+3\right)  \partial_{U}^2\\&  + 2 V (V-1)^{-3} \left((V (V (5 V-4)-10)+3)
   \partial_V +(V-1) V (V (V+4)+1) \partial_V^2 \right.\\&\left.+\left(U (5
   (V-2) V+2)+(V-1)^2\right) \partial_U+2 (V-1) V (U
   (V+2)+V-1) \partial_U \partial_V\right)+8\;.
    \end{aligned}
\end{equation}
These differential operators are obtained from Taylor expanding the generating function and then projecting to the $(0,-2)$ sector. We then decompose their actions on the 10d defect channel conformal blocks into 4d ones where the defect is 0d (here we omitted the $+$ parity label)
\begin{equation}
    \begin{aligned}
       \widehat{\mathcal{D}}_{22}\, \widehat{g}^{(4,6)}_{4+s,s} = {}& \widehat{g}_{4+s,s}+ \frac{2(s+4)}{s+3}\widehat{g}_{6+s,s} + \frac{s-1}{s+1} \widehat{g}_{6+s-2,s-2} + \mathcal{O}((1-z)^4)\;,
        \\  \widehat{\mathcal{D}}_{24}\, \widehat{g}^{(4,6)}_{4+s,s} = {}& \frac{\sqrt{2}(s+4)(s+5)}{s+3} \widehat{g}_{6+s,s} - \frac{\sqrt{2} (s-1)s}{s+1} \widehat{g}_{6+s-2,s-2} + \mathcal{O}((1-z)^4)\;, \\ \widehat{\mathcal{D}}_{44}\, \widehat{g}^{(4,6)}_{4+s,s}  ={}&  \frac{(s+4)(s+5)^2}{s+3} \widehat{g}_{6+s,s} + \frac{2(s-1)s^2}{s+1} \widehat{g}_{6+s-2,s-2} + \mathcal{O}((1-z)^4)\;.
    \end{aligned}
\end{equation}
Since we are focusing on the twist 4 double-particle operators, we should extract only the twist 4 contribution for which the 4d conformal blocks have the form $\widehat{g}_{6+s,s}$. The conformal blocks $\widehat{g}_{6+s-2,s-2}$ also contribute, but only after a shift $s\to s+2$. These contributions form a $2\times 2$ matrix which is $\mathbf{\Omega}$ in \eqref{quadraticeq}. We then multiply both sides of this matrix by $(\sqrt{\mathbf{N}})^{-1}$ where we recall $\mathbf{N}$ was defined in \eqref{NandOmega}. This gives
\begin{equation}
\begin{aligned}
 (\sqrt{\mathbf{N}})^{-1} \mathbf{\Omega} (\sqrt{\mathbf{N}})^{-1} = {}& \frac{s+4}{(s+1)(s+3)}\times\left(\begin{array}{cc}
2 &\sqrt{\frac{2(s+5)}{s+2}} \\
\sqrt{\frac{2(s+5)}{s+2}}  & \frac{s+5}{s+2} 
\end{array}\right) \times\mathbb{a}_{s}^{(0)} \mathbb{\Gamma}_{s}^{(1)} \\& +\frac{1}{s+3} \times \left(\begin{array}{cc}
1 & -\sqrt{\frac{2(s+2)}{s+5}}\\
 -\sqrt{\frac{2(s+2)}{s+5}} & \frac{2(s+2)}{s+5} 
\end{array}\right)\times\mathbb{a}_{s+2}^{(0)} \mathbb{\Gamma}_{s+2}^{(1)}\;.
\end{aligned}
\end{equation}
After we plug in the values of $\mathbb{a}_{s}^{(0)}$ and $\mathbb{\Gamma}_{s}^{(1)}$, we get
\begin{equation}
\begin{aligned}
\nonumber
   & \frac{-12(s+4)}{(s+1)(s+2)}\times\left(\begin{array}{cc}
 \frac{2(s+2)}{3(s+3)} &\frac{\sqrt{2(s+2)(s+5)}}{3(s+3)} \\
\frac{\sqrt{2(s+2)(s+5)}}{3(s+3)}  & \frac{s+5}{3(s+3)} 
\end{array}\right)  +\frac{-12}{s+5} \times \left(\begin{array}{cc}
\frac{s+5}{3(s+3)} & -\frac{\sqrt{2(s+2)(s+5)}}{3(s+3)}\\
-\frac{\sqrt{2(s+2)(s+5)}}{3(s+3)} & \frac{2(s+2)}{3(s+3)} 
\end{array}\right)\,.
\end{aligned}
\end{equation}
We recognize that the two matrices  are in fact mutually orthogonal projectors. This provides an automatic diagonalization of the mixing problem with the coefficients of the matrices being the diagonalized anomalous dimension \eqref{example}. This nicely generalizes the observation of \cite{Caron-Huot:2018kta} to the case with partially broken hidden conformal symmetry.

\subsection{Leading logarithms at any loop}\label{Subsec:Delta4}
Much similar to the defect-free case, we also find the existence of certain differential operators with remarkable properties to simplify loop calculations in AdS. For four-point functions of all light supergravitons, an eighth order differential operator was discovered in \cite{Aprile:2018efk,Caron-Huot:2018kta}. This operator is characterized by the property that the long superconformal blocks are its eigenfunctions and the eigenvalues correspond to the numerator of the anomalous dimension formula. This was shown to facilitate the computation of leading logarithmic singularities in \cite{Caron-Huot:2018kta}.

We now construct a similar differential operator for the giant graviton two-point functions. The operator turns out to be fourth order 
\begin{equation}\label{Delta4}
   \widehat{\mathbf{\Delta}}^{(4)} =  \frac{x \bar x \beta \bar{\beta}}{(x-\bar x)(\beta-\bar{\beta})}(x\partial_x -\beta \partial_{\beta}) (x\partial_x -\bar{\beta} \partial_{\bar{\beta}})  (\bar{x}\partial_{\bar{x}} -\beta \partial_{\beta})(\bar{x}\partial_{\bar{x}} - \bar{\beta} \partial_{\bar{\beta}}) \frac{(x-\bar x)(\beta-\bar{\beta})}{x \bar x \beta \bar{\beta}}\;,
\end{equation}
where we have defined 
\begin{equation}
    x = 1-z\;,\quad \bar{x}=1-\bar{z}\;,\quad \beta = \frac{\alpha}{1-\alpha}\;, \quad \bar{\beta} = \frac{\bar{\alpha}}{1-\bar{\alpha}}\;.
\end{equation}
One can easily check that
the action of $\widehat{\mathbf{\Delta}}^{(4)}$ on the long superconformal block is diagonal
\begin{equation}
    \widehat{\mathbf{\Delta}}^{(4)} \left(\widehat{g}_{\widehat{\tau}+s+2,s}^+ \widehat{\mathcal{Q}}_{r_1,r_2+2}^+ \right) = M_t^{(2)} M_{t+s+1}^{(2)} \left( \widehat{g}_{\widehat{\tau}+s+2,s}^+ \widehat{\mathcal{Q}}_{r_1,r_2+2}^+ \right)\;,
\end{equation}
and the eigenvalue is exactly the numerator of the anomalous dimension \eqref{defectanomalousdim}.\footnote{More precisely, this is in the ranges $0<V<1$, $0<\frac{\sigma}{\tau}<1$ or $V>1$, $\frac{\sigma}{\tau}>1$ which correspond to the defect channel OPE.}

Using $\widehat{\mathbf{\Delta}}^{(4)}$ one can easily write down the leading $\log^{\kappa} V$ coefficients of the reduced correlator at any loop order $\kappa$ (similar to defect free four-point functions that discussed in Section IV in \cite{Aprile:2018efk} and eq.(5.42) of \cite{Caron-Huot:2018kta}). We have
\begin{equation}
\label{logVanyloop}
    \mathcal{H}^{(\kappa)}_{22} \big|_{\log^{\kappa} V} =  \left( \widehat{\mathbf{\Delta}}^{(4)} \right)^{\kappa-1} h^{(\kappa)}(z,\bar{z})\;,
\end{equation}
where $h^{(\kappa)}(z,\bar{z})$ has the 10d bosonic defect channel conformal block expansion
\begin{equation}
\label{pre-log}
    h^{(\kappa)}(z,\bar{z}) = c(\kappa) \sum_{s=0}^{\infty} \left[(s+1)_3\right]^{1-\kappa} \widehat{g}_{4+s,s}^{(4,6)}\;,
\end{equation}
and $c(\kappa)$ is a constant  depending on $\kappa$. Factoring out the powers of the $\widehat{\mathbf{\Delta}}^{(4)}$ operator leads to the natural expectation that the pre-log functions $h^{(\kappa)}(z,\bar{z})$ are much simpler than the full  $\log^\kappa V$ coefficients.

The advantage can be illustrated with a few examples. At tree level, i.e., $\kappa = 1$, the differential operator does not appear. It is obvious that the pre-log coincides with \eqref{treelog} and reads 
\begin{equation}
   h^{(1)}(z,\bar{z}) =  \mathbf{H}\big|_{\log V} = \frac{4 (1-z)^2(1- \bar z)^2}{z \bar{z}(z \bar{z} -z -\bar{z})^3}\;.
\end{equation}
At one-loop level with $\kappa=2$, we can perform the sum in \eqref{pre-log} and get a pre-log function containing at most logarithms
\begin{equation}
\begin{aligned}
    h^{(2)}(z,\bar{z}) = {}&\frac{(1-z)^2 (1-\bar{z})^2 \left( z^3 (\bar{z}-1)-\bar{z}^3(z-1)+3 z \bar{z}(z-\bar{z})+2z^2 \bar{z}^2 \left(  \log \bar{z}-  \log
   z\right)\right)}{ (z-\bar{z})^3 (z \bar{z}-z-\bar{z})^3}\;.
   \end{aligned}
\end{equation}
It is straightforward to obtain the $\log^2 V$ coefficient from \eqref{logVanyloop} and one can be easily convinced that the $\log^2 V$ coefficient is much lengthier. It is also not difficult to continue on to two loops and compute the pre-log function $ h^{(3)}(z,\bar{z}) $. The result is structurally similar but now also involves the higher weight polylogarithms $\operatorname{Li}_2(1-z)$, $\operatorname{Li}_2 (1-\bar z)$. The detailed expression however is a bit complicated and we will refrain from writing it down here. 

We expect that the differential operator $\widehat{\mathbf{\Delta}}^{(4)}$ will play an important role in studying the $1/N$ corrections to the giant graviton correlators. In particular, it will be useful for bootstrapping the loop corrections in position space, which is complementary to the Mellin space approach developed in \cite{Chen:2024orp}. We will not  pursue this extension further in this work, but leave it to a separate publication.

\section{Defect as natural description of heavy limit}\label{Sec:defectforLLHH}

\subsection{Kinematic emergence of defects}
In the previous sections, we have demonstrated that the giant graviton four-point functions $\langle\mathcal{O}_{k_1}\mathcal{O}_{k_2}\mathcal{D}\mathcal{D}\rangle$ should be most naturally viewed as defect two-point functions $\llangle\mathcal{O}_{k_1}\mathcal{O}_{k_2}\rrangle$. The intuition follows from the dual bulk picture where the heavy giant gravitons are represented by a D-brane. However, this defect treatment is in fact more widely applicable. We would like to show here that it applies to any Light-Light-Heavy-Heavy (LLHH) four-point functions. The analysis will show that this defect language is not only a natural description but also in a sense inevitable. 

Let us restrict our attention to the case where the dimensions of the two heavy operators are identical, i.e., $\Delta_3=\Delta_4=Q$. The four-point function 
\begin{equation}
    \langle\mathcal{O}_{\Delta_1}(x_1)\mathcal{O}_{\Delta_2}(x_2)\mathcal{O}_Q(x_3)\mathcal{O}_Q(x_4) \rangle=F^{\rm 4pt}(x_i)=\frac{1}{(x_{12}^2)^{\frac{\Delta_1+\Delta_2}{2}}(x_{34}^2)^Q}\left(\frac{x_{14}^2}{x_{24}^2}\right)^{\frac{\Delta_2-\Delta_1}{2}}\mathcal{G}^{\rm 4pt}(U,V)\;,
\end{equation}
admits the standard conformal block decomposition 
\begin{equation}\label{4ptcrossing}
\begin{split}
    \mathcal{G}^{\rm 4pt}(U,V)={}&\sum_{\Delta,\ell}C_{12\mathcal{O}}C_{34\mathcal{O}}g^{s,{\rm 4pt}}_{\Delta,\ell}(U,V)=\sum_{\Delta',\ell'}C_{14\mathcal{O}'}C_{23\mathcal{O}'}g^{t,{\rm 4pt}}_{\Delta',\ell'}(U,V)\\
    ={}&\sum_{\Delta',\ell'}C_{13\mathcal{O}'}C_{24\mathcal{O}'}g^{u,{\rm 4pt}}_{\Delta',\ell'}(U,V)\;,
\end{split}
\end{equation}
where $g^{s,t,u,{\rm 4pt}}_{\Delta,\ell}(U,V)$ are four-point function conformal blocks. We now write the four-point function as a defect two-point function in the same way as we did in Section \ref{Sec:corrkinematics}. We define the defect one-point functions for the light operators as 
\begin{equation}
    \llangle \mathcal{O}_{\Delta_i}(x_i)\rrangle=\frac{\langle\mathcal{O}_{\Delta_i}(x_i)\mathcal{O}_Q(x_3)\mathcal{O}_Q(x_4) \rangle}{\langle\mathcal{O}_Q(x_3)\mathcal{O}_Q(x_4) \rangle}=a_i\left(\frac{x_{34}^2}{x_{3i}^2x_{4i}^2}\right)^{\frac{\Delta_i}{2}}\;,
\end{equation}
where the defect one-point function coefficient is the same as the defect-free three-point function coefficient $a_i=C_{34i}$. Then we define the defect two-point function as
\begin{equation}
\begin{split}
    \llangle \mathcal{O}_{\Delta_1}(x_1)\mathcal{O}_{\Delta_2}(x_2)\rrangle={}&\frac{\langle\mathcal{O}_{\Delta_1}(x_1)\mathcal{O}_{\Delta_2}(x_2)\mathcal{O}_Q(x_3)\mathcal{O}_Q(x_4) \rangle}{\langle\mathcal{O}_Q(x_3)\mathcal{O}_Q(x_4) \rangle}\\
    ={}&\left(\frac{x_{34}^2}{x_{31}^2x_{41}^2}\right)^{\frac{\Delta_1}{2}}\left(\frac{x_{34}^2}{x_{32}^2x_{42}^2}\right)^{\frac{\Delta_2}{2}}\mathcal{G}(\xi,\chi)\;,
\end{split}
\end{equation}
where in the second line we have extracted a product of defect one-point functions (the kinematic part). Note that because the two heavy operators are the same, the four-point function is invariant under $3\leftrightarrow 4$. Using the defect cross ratios and the four-point cross ratios are the same, and there is no ambiguity discussed in Section \ref{Sec:corrkinematics}. For the same reason, when we decompose the correlator into defect conformal blocks we will only find parity even conformal blocks in the defect channel. However, for the moment let us not do that. Instead, we simply use the connection between two-point  and four-point functions to rewrite (\ref{4ptcrossing}). We get  
\begin{equation}\label{4ptas2ptcrossing}
    \mathcal{G}(\xi,\chi)=\sum_{\Delta,\ell} C_{12\mathcal{O}}a_{\mathcal{O}} \tilde{g}_{\Delta,\ell}(\xi,\chi)=\sum_{\Delta',\ell'} C_{14\mathcal{O}'}a_{\mathcal{O}'}\widehat{\tilde{g}}_{\Delta',\ell'}(\xi,\chi)=\sum_{\Delta',\ell'} C_{13\mathcal{O}'}a_{\mathcal{O}'}\widehat{\tilde{g}}'_{\Delta',\ell'}(\xi,\chi)\;,
\end{equation}
where 
\begin{equation}
a_{\mathcal{O}}=C_{34\mathcal{O}}\;,\quad a_{\mathcal{O}'}=C_{23\mathcal{O}'}=C_{24\mathcal{O}'}\;, 
\end{equation}
and we have defined
\begin{equation}
\begin{split}
    {}&\tilde{g}_{\Delta,\ell}(\xi,\chi)=U^{-\frac{\Delta_1+\Delta_2}{2}}V^{\frac{\Delta_2}{2}}g^{s,{\rm 4pt}}_{\Delta,\ell}(U,V)\;,\\
    {}&\widehat{\tilde{g}}_{\Delta',\ell'}(\xi,\chi)=U^{-\frac{\Delta_1+\Delta_2}{2}}V^{\frac{\Delta_2}{2}}g^{t,{\rm 4pt}}_{\Delta',\ell'}(U,V)\;,\\
    {}&\widehat{\tilde{g}}'_{\Delta',\ell'}(\xi,\chi)=U^{-\frac{\Delta_1+\Delta_2}{2}}V^{\frac{\Delta_2}{2}}g^{u,{\rm 4pt}}_{\Delta',\ell'}(U,V)\;.
\end{split}
\end{equation}
Note that the last two equalities in (\ref{4ptas2ptcrossing}) are physically the same because identical heavy operators mean u- and t-channels are the same by crossing. Although (\ref{4ptas2ptcrossing}) has the form of the crossing equation of a defect two-point function (taking the first two nontrivial equalities), at this point it is still just the four-point function decomposition (\ref{4ptcrossing}) in disguise. The ``defect'' conformal blocks with the tildes are really four-point function conformal blocks.
However, we will use the following remarkable fact: In the heavy limit $Q\to\infty$, we can remove the tildes and the four-point function conformal blocks coincide with the defect conformal blocks. The s-channel conformal block is identified with the bulk channel conformal block
\begin{equation}\label{gheavylimit}
g_{\Delta,\ell}=\tilde{g}_{\Delta,\ell}\big|_{Q\to \infty}\;.
\end{equation}
This is not surprising. In fact, in Section \ref{subsec:Superconformal blocks} we have pointed out that they are the same even before taking the large $Q$ limit. The nontrivial part is that the t-channel conformal block will reduce to the defect channel conformal block. More precisely, the defect channel conformal block is the {\it union} of t- and u-channel conformal blocks in the large $Q$ limit\footnote{Here only parity even defect conformal blocks shows up because the correlator is invariant under $\{U,V\}\to\{U/V,1/V\}$. One can trivially obtain the parity odd case by adding a minus sign in the second row. This will be relevant when the two heavy operators have the same dimension but differ by some global charges.}
\begin{equation}\label{ghatheavylimit}   \widehat{g}^+_{\widehat{\Delta},s}=\left\{\begin{array}{l}\widehat{\tilde{g}}_{\widehat{\Delta}+Q,s}\big|_{Q\to\infty}\;,\quad \text{when } 0<V<1\;, \\ \widehat{\tilde{g}}'_{\widehat{\Delta}+Q,s}\big|_{Q\to\infty}\;,\quad \text{when } V\geq 1\;.\end{array}\right.
\end{equation}
One can easily be convinced by checking a few explicit examples of conformal blocks. But we will also outline a proof for the general case in Appendix \ref{app:heavylimit}. A notable feature in this formula is the subtraction of the background, namely the defect operator dimension $\widehat{\Delta}$ is given by the dimension exchanged in the four-point function with the background dimension $Q$ removed. Therefore, we should think of the defect operator with dimension $\widehat{\Delta}$ as a small fluctuation above the defect background $Q$. These identities turn (\ref{4ptas2ptcrossing}) into a true crossing equation for a defect two-point function
\begin{equation}
    \mathcal{G}(\xi,\chi)=\sum_{\Delta,\ell} C_{12\mathcal{O}}a_{\mathcal{O}} \tilde{g}_{\Delta,\ell}(\xi,\chi)=\sum_{\widehat{\Delta},s} C_{14\mathcal{O}'}a_{\mathcal{O}'}\widehat{{g}}^+_{\widehat{\Delta},s}(\xi,\chi)\;,
\end{equation}
where in the second sum we should identify $s$ with $\ell'$ and subtract the background $Q$ to define $\widehat{\Delta}$. 

We hasten to add that directly taking the large $Q$ limit does not always make sense. A good counter example is the giant graviton correlators which we computed in this paper. If we start from the tree-level four-point functions obtained in \cite{Rastelli:2016nze,Rastelli:2017udc} for all light supergravitons and set the KK levels of two of the operators equal to $Q$, the infinite $Q$ limit does not match the result here. In fact, this is already expected to fail from the mismatched $1/N$ counting.\footnote{The leading connected giant graviton correlator is of order $1/N$ while the leading connected supergraviton four-point function gives only $1/N^2$. The reason for failure here is also clear: In taking the large $N$ limit the supergravitons are assumed to have $\Delta\ll N$ but the giant gravitons have $\Delta\sim N$.} This failure does not undermine our analysis in this subsection. On the contrary, the correct lesson to draw is the identities (\ref{gheavylimit}) and (\ref{ghatheavylimit}), which tell us the four-point conformal blocks will inevitably become  defect conformal blocks in the heavy limit. Note that conformal blocks are kinematic objects that make up any correlator. Therefore, the statement that LLHH correlators should be viewed as defect correlators holds nonperturbatively and regardless of supersymmetry.

\subsection{Heavy limit of four-point Witten diagrams}
\label{subsec:heavylimit}
In the previous subsection, we have established that defects are the most natural way to describe the kinematics of LLHH correlators. Here we provide another way to see the emergence of defects by considering the heavy limit of four-point Witten diagrams in empty AdS. We will see that defect Witten diagrams appear naturally in this limit, essentially as the saddle points of these integrals.

\begin{figure}
\centering  \includegraphics[width=0.7\linewidth]{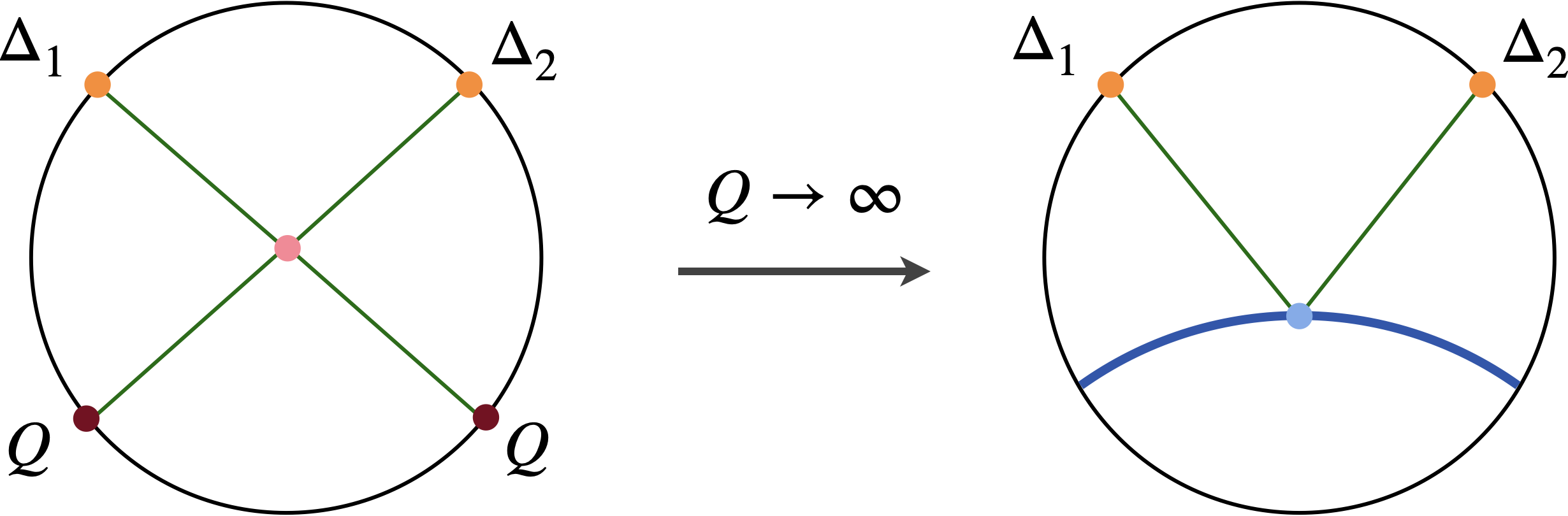}
  \caption{In the heavy limit, a four-point contact Witten diagram reduces to a defect two-point contact Witten diagram.}
  \label{Fig:LargeQcontact}
\end{figure}

The simplest and most instructive example to consider is the contact Witten diagram (Figure \ref{Fig:LargeQcontact}). In an empty AdS$_{d+1}$ space, a four-point contact Witten diagram is known in the literature as a $D$-function and is defined by the integral
\begin{equation}
\begin{split}
    D_{\Delta_1\Delta_2QQ}={}&\int \frac{d^{d+1}z}{z_0^{d+1}}\prod_{i=1}^4G_{B\partial}^{\Delta_i}(x_i,z)\\
    ={}& \int \frac{d^{d+1}z}{z_0^{d+1}} e^{-Q\log\left(\frac{(z_0^2+(\vec{z}-\vec{x}_3)^2)(z_0^2+(\vec{z}-\vec{x}_4)^2)}{z_0^2}\right)} \prod_{i=1}^2G_{B\partial}^{\Delta_i}(x_i,z)\;,
\end{split}
\end{equation}
where we have set $\Delta_3=\Delta_4=Q$ and explicitly expressed these two propagators as exponentials in the second line. It is clear from this expression that in the $Q\to\infty$ limit the dominant contribution of the integral will come from the locus defined by the condition
\begin{equation}
    \frac{(z_0^2+(\vec{z}-\vec{x}_3)^2)(z_0^2+(\vec{z}-\vec{x}_4)^2)}{z_0^2}=1\;,
\end{equation}
which is precisely the one dimensional geodesic line $\Gamma$ connecting $x_3$ and $x_4$ on the boundary. One can perform a more careful saddle point analysis which also takes into account the Gaussian fluctuations around the saddle. The analysis is straightforward and shows that the AdS$_{d+1}$ integrals completely localize to the geodesic
\begin{equation}
    \lim_{Q\to\infty} Q^{\frac{d}{2}} x_{34}^{2Q} D_{\Delta_1\Delta_2QQ} \sim \int_\Gamma d\tau G^{\Delta_1}_{B\partial}(x_1,z(\tau))G^{\Delta_2}_{B\partial}(x_2,z(\tau))\;,
\end{equation}
where the RHS is precisely the contact Witten diagram with no derivatives (\ref{defcontact}) for defect two-point functions. 

\begin{figure}
\centering  \includegraphics[width=0.7\linewidth]{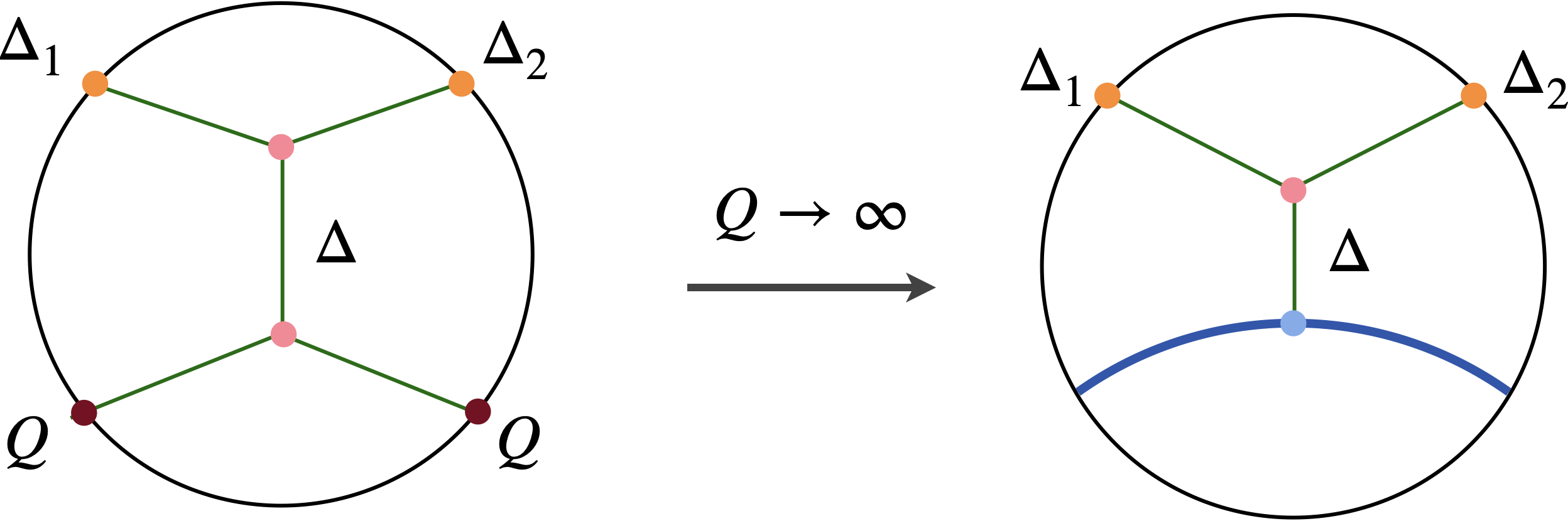}
  \caption{In the heavy limit, the s-channel four-point exchange Witten diagram reduces to a bulk channel exchange Witten diagram for a defect two-point function.}
  \label{Fig:LargeQbulk}
\end{figure}

The analysis can be straightforwardly generalized to the four-point exchange Witten diagram in the LLHH channel (Figure \ref{Fig:LargeQbulk}). For simplicity, let us focus on the case where the exchanged field is a scalar. The four-point exchange Witten diagram is defined by
\begin{equation}
    W^{(s),\Delta}_{\Delta_1\Delta_2QQ}=\int \frac{d^{d+1}z}{z_0^{d+1}}\frac{d^{d+1}w}{w_0^{d+1}} G^{\Delta_1}_{B\partial}(x_1,z)G^{\Delta_2}_{B\partial}(x_2,z)G^\Delta_{BB}(z,w)G^Q_{B\partial}(x_3,w)G^Q_{B\partial}(x_4,w)\;,
\end{equation}
where we need to integrate both $z$ and $w$ over AdS$_{d+1}$. The exchanged field is assumed to be light and has $\Delta\ll Q$. We can focus on the $w$ integral and the saddle point approximation at large $Q$ turns it into a geodesic integral. This gives us
\begin{equation}
    \lim_{Q\to\infty} Q^{\frac{d}{2}} x_{34}^{2Q} W^{(s),\Delta}_{\Delta_1\Delta_2QQ} \sim \int_\Gamma d\tau \int \frac{d^{d+1}z}{z_0^{d+1}} G^{\Delta_1}_{B\partial}(x_1,z)G^{\Delta_2}_{B\partial}(x_2,z)G^\Delta_{BB}(z,w'(\tau))\;,
\end{equation}
which shows the LLHH four-point exchange Witten diagram reduces in the heavy limit to a bulk channel defect two-point exchange Witten diagram.

\begin{figure}
\centering  \includegraphics[width=0.8\linewidth]{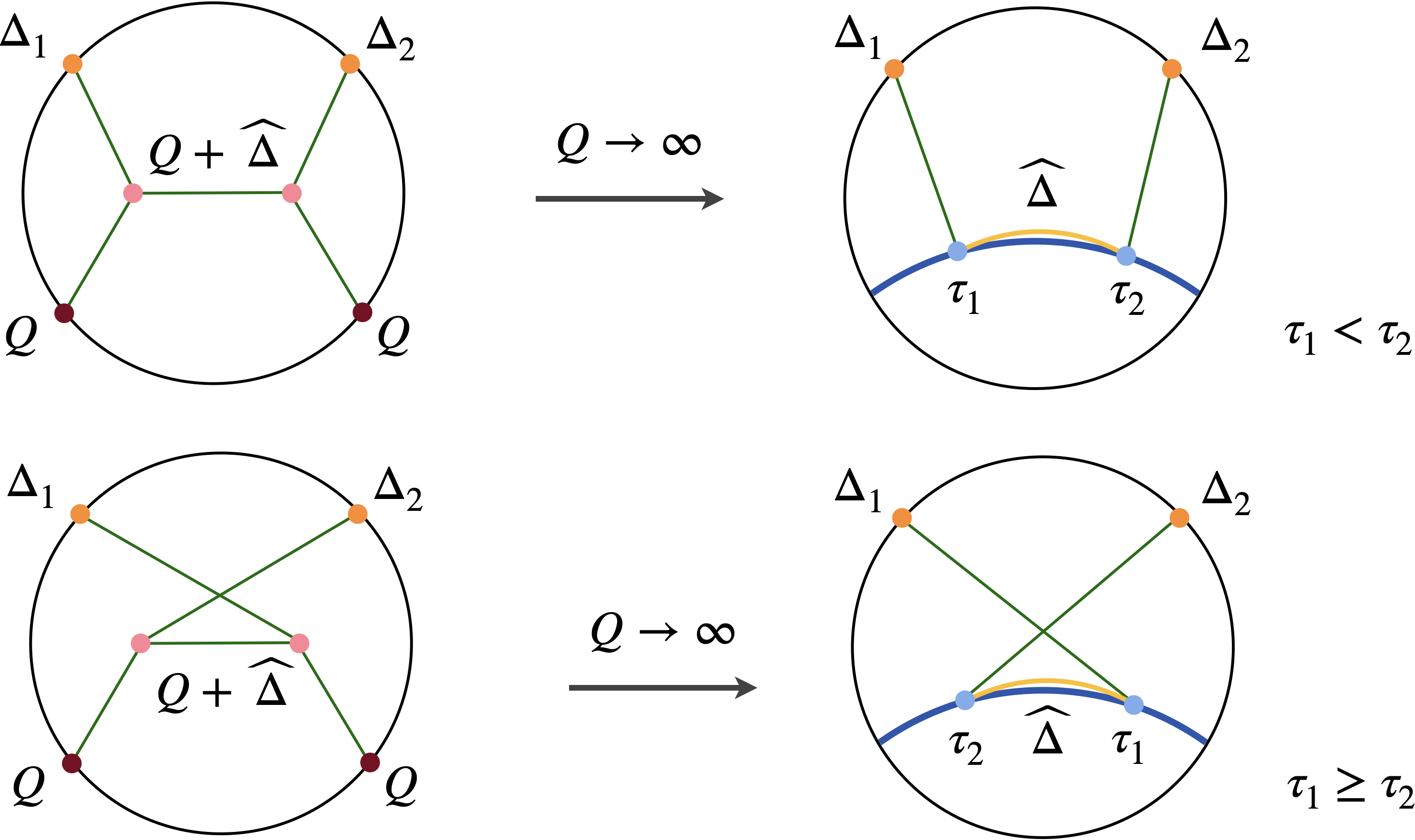}
  \caption{In the heavy limit, the t- and u-channel four-point exchange Witten diagrams reduce  to the two halves of the integral that defines a defect channel exchange Witten diagram for a defect two-point function.}
  \label{Fig:LargeQdefect}
\end{figure}

Finally, let us consider the heavy limit of four-point exchange Witten diagrams in the Light-Heavy-Light-Heavy (LHLH) limit (Figure \ref{Fig:LargeQdefect}). We define 
\begin{equation}
    W^{(u),\widehat{\Delta}+Q}_{\Delta_1\Delta_2QQ}=\int \frac{d^{d+1}z}{z_0^{d+1}}\frac{d^{d+1}w}{w_0^{d+1}} G^{\Delta_1}_{B\partial}(x_1,z)G^Q_{B\partial}(x_3,z)G^{\widehat{\Delta}+Q}_{BB}(z,w)G^{\Delta_2}_{B\partial}(x_2,w)G^Q_{B\partial}(x_4,w)\;,
\end{equation}
and similarly 
\begin{equation}
    W^{(t),\widehat{\Delta}+Q}_{\Delta_1\Delta_2QQ}=\int \frac{d^{d+1}z}{z_0^{d+1}}\frac{d^{d+1}w}{w_0^{d+1}} G^{\Delta_1}_{B\partial}(x_1,z)G^Q_{B\partial}(x_4,z)G^{\widehat{\Delta}+Q}_{BB}(z,w)G^{\Delta_2}_{B\partial}(x_2,w)G^Q_{B\partial}(x_3,w)\;.
\end{equation}
Here we have assumed that the exchanged intermediate particle is also heavy and has dimension $\widehat{\Delta}+Q$ with $\widehat{\Delta}\ll Q$. 
In the large $Q$ limit, one can also perform the saddle point analysis but it now involves solving saddle point equations with respect to both $z$ and $w$. Here we will not present the full detailed analysis (which is straightforward albeit slightly technical) but only demonstrate the key underlying physics. Because the bulk-to-bulk propagator is of the same magnitude heavy as the two bulk-to-boundary propagators from 3 and 4, it is not difficult to convince oneself that both the $z$ and $w$ integrals will localize to the geodesic connecting $x_3$ and $x_4$. We can then consider the approximation\footnote{Here we will stop keeping track of overall powers of $Q$.} 
\begin{equation}\label{Wuapprox}
\begin{split}
    W^{(u),\widehat{\Delta}+Q}_{\Delta_1\Delta_2QQ}\propto {}&\int_\Gamma d\tau_1 d\tau_2 G^{\Delta_1}_{B\partial}(x_1,z(\tau_1))G^Q_{B\partial}(x_3,z(\tau_1))G^{\widehat{\Delta}+Q}_{BB}(z(\tau_1),w(\tau_2))\\
    {}&\times G^{\Delta_2}_{B\partial}(x_2,w(\tau_2))G^Q_{B\partial}(x_4,w(\tau_2))\;.
\end{split}
\end{equation}
For large $Q$, we can further approximate the AdS$_{d+1}$ propagator by
\begin{equation}
    G^{\widehat{\Delta}+Q}_{BB}(z(\tau_1),w(\tau_2))\propto e^{-(\widehat{\Delta}+Q)|\log z(\tau_1)-\log w(\tau_2)|}\;,
\end{equation}
which is essentially the same as the scalar propagator on the geodesic (\ref{geoprop}) up to factors of $Q$ not explicitly written. It is convenient to go to the frame $\Gamma=\Gamma_0$ where the geodesic is a straight line. Then (\ref{Wuapprox}) can be explicitly written as
\begin{equation}
\begin{split}
   {}& \lim_{x_4\to\infty}x_4^{2Q}W^{(u),\widehat{\Delta}+Q}_{\Delta_1\Delta_2QQ} \propto\int_0^\infty d z_0 \int_0^\infty dw_0  \left(\frac{z_0}{z_0^2+x_1^2}\right)^{\Delta_1}\left(\frac{1}{z_0}\right)^Q e^{-(\widehat{\Delta}+Q)|\log(\frac{z_0}{w_0})|}w_0^Q\\
    {}&\quad\quad\quad\quad\quad\quad=\int_0^\infty \frac{dw_0}{w_0} \int_0^{w_0} \frac{dz_0}{z_0} \left(\frac{z_0}{z_0^2+x_1^2}\right)^{\Delta_1}\left(\frac{z_0}{w_0}\right)^{\widehat{\Delta}} \left(\frac{w_0}{w_0^2+x_2^2}\right)^{\Delta_2}\\
{}&\quad\quad\quad\quad\quad\quad\quad+\int_0^\infty \frac{dw_0}{w_0} \int_{w_0}^\infty \frac{dz_0}{z_0} \left(\frac{z_0}{z_0^2+x_1^2}\right)^{\Delta_1}\left(\frac{w_0}{z_0}\right)^{\widehat{\Delta}+2Q} \left(\frac{w_0}{w_0^2+x_2^2}\right)^{\Delta_2}\;,
\end{split}
\end{equation}
where the integral splits into two pieces depending on $z_0\geq w_0$ or $z_0<w_0$. It is easy to see that the first integral is the defect integral $\widehat{I}^{(u)}$ in (\ref{intIu}) with dimension $\widehat{\Delta}$ and the second integral is $\widehat{I}^{(t)}$ in (\ref{intIt}) with dimension $\widehat{\Delta}+Q$. It is important to notice that the second piece is exponentially suppressed for large $Q$ because it corresponds to a heavy particle doubling back in time and is penalized by the action. Therefore, we find that the LHLH four-point exchange Witten diagrams gives in the heavy limit 
\begin{equation}
    x_{34}^{2Q} W^{(u),\widehat{\Delta}+Q}_{\Delta_1\Delta_2QQ} \xrightarrow{Q\to\infty} \widehat{I}^{(u)}_{\widehat{\Delta}}\;,
\end{equation}
and similarly
\begin{equation}
    x_{34}^{2Q} W^{(t),\widehat{\Delta}+Q}_{\Delta_1\Delta_2QQ} \xrightarrow{Q\to\infty} \widehat{I}^{(t)}_{\widehat{\Delta}}\;.
\end{equation}
It is important to notice that individual t- or u-channel exchange Witten diagrams do not correspond to defect channel exchange Witten diagrams in the heavy limit. Instead, the defect channel exchange Witten diagram only arises from the sum of four-point exchange Witten diagrams in both channels
\begin{equation}
    x_{34}^{2Q}( W^{(u),\widehat{\Delta}+Q}_{\Delta_1\Delta_2QQ}+W^{(t),\widehat{\Delta}+Q}_{\Delta_1\Delta_2QQ}) \xrightarrow{Q\to\infty} \widehat{E}^{\widehat{\Delta},0}_{\Delta_1\Delta_2}\;.
\end{equation}
This result shows two satisfying features which agree with our intuition. The first is the removal of the background dimension $Q$, as we have already seen in the large $Q$ limit of conformal blocks. The second is the restoration of symmetry under $3\leftrightarrow 4$. Unlike in the case of conformal blocks, where symmetry is regained from taking the union, here the $3\leftrightarrow 4$ exchange symmetry is restored by taking the sum of t- and u-channel exchange Witten diagrams.

\subsection{Comments on analytic functional approach}\label{Subsec:analyticfunctionals}

Using the results from the previous two subsections, we now make a few comments about the analytic functional approach for LLHH correlators. Such an approach is nonperturbative in nature and will further strengthen our claim that the defect perspective is the right language for studying LLHH correlators. Rather than giving a rigorous derivation, we will outline the key points by drawing analogy with other systems where the analytic functional approach has already been established \cite{Mazac:2016qev,Mazac:2018mdx,Mazac:2018ycv,Zhou:2018sfz,Kaviraj:2018tfd,Mazac:2018biw,Mazac:2019shk,Giombi:2020xah,Ghosh:2021ruh,Antunes:2025vvl}. We plan to confirm and flesh out these ideas in another future work.

The analytic functional approach is intimately related to Polyakov's original proposal in the '70s \cite{Polyakov:1974gs}, where he proposed an alternative basis for expanding conformal correlators. These basis functions, with the hindsight of AdS/CFT, turn out to be just tree-level exchange Witten diagrams with improved Regge behavior. On the other hand, it follows from the conformal block decomposition of these Witten diagrams that there is yet another basis for decomposition. These are the double-trace\footnote{We used  ``double-trace'' here to follow the terminology in the analytic functional literature. But in this context it is synonymous to ``double-particle'' which we used earlier.} conformal blocks from two different OPE channels, and their duals define a complete basis of analytic functionals of which the actions on generic conformal blocks can be read off from the conformal block expansion coefficients of the Witten diagram. The action of this basis of functionals on the crossing equation then gives the complete sum rules needed to be satisfied by the CFT data. This summarizes the basic logic of the analytic functional approach. In the particular context of defects, this logic has been explicitly spelt out and checked in \cite{Mazac:2018biw,Giombi:2020xah} where the defects have dimension $d-1$ and $-1$ respectively. 

The logic is extremely similar here for the LLHH correlators. Viewed directly as defect correlators, the associated defect exchange Witten diagrams have the same conformal block decomposition structure. In the direct channel (e.g., a bulk channel Witten diagram decomposed in the bulk channel), the exchange Witten diagram decomposes into the sum of a single-trace conformal block and infinitely many double-trace conformal blocks. In the crossed the channel, the decomposition consists of only double-trace conformal blocks. Therefore, following similar reasoning, we are naturally led to propose that the double-trace conformal blocks in both the bulk and defect channels form a complete basis for decomposing defect two-point functions. More precisely, the proposed basis is 
\begin{equation}\label{basisLLHH}
\{g_{\Delta^{12}_{n,\ell},\ell}, \widehat{g}_{\Delta^1_{m,s}}, \widehat{g}_{\Delta^2_{m,s}}\}\;,
\end{equation}
where $\Delta^{12}_{n,\ell}=\Delta_1+\Delta_2+2n+\ell$ are the conformal dimensions of the double-trace operators in the bulk channel and  $\Delta^i_{m,s}=\Delta_i+2m+s$ are the dimensions of the double-trace operators in the defect channel. This proposal is also reasonable to expect from the perspective of four-point functions where the functional approach has been established \cite{Mazac:2019shk}. We notice that (\ref{basisLLHH}) are just the large $Q$ limit of the four-point double-trace conformal blocks. In particular, the double-trace operators formed out of 3 and 4 with dimensions $2Q+2n+\ell$ are infinitely massive in this limit and therefore decouple from the basis. From (\ref{basisLLHH}) we can now define a basis of analytic functionals as their dual
\begin{equation}
\{\omega_{n,\ell}, \widehat{\omega}^{(1)}_{m,s},  \widehat{\omega}^{(2)}_{m,s}\}\;,
\end{equation}
which satisfies the orthonormal relations
\begin{equation}
\begin{split}
{}&\omega_{n,l}[g_{\Delta^{12}_{n',\ell'},\ell'}]=\delta_{nn'}\delta_{\ell\ell'}\;,\quad \omega_{n,l}[\widehat{g}_{\Delta^1_{m,s}}]=0\;,\quad   \omega_{n,l} [\widehat{g}_{\Delta^2_{m,s}}]=0\;,\\
{}& \widehat{\omega}^{(1)}_{m,s}[g_{\Delta^{12}_{n,\ell},\ell}]=0\;,\quad  \widehat{\omega}^{(1)}_{m,s}[\widehat{g}_{\Delta^1_{m',s'}}]=\delta_{mm'}\delta_{ss'}\;,\quad  \widehat{\omega}^{(1)}_{m,s}[\widehat{g}_{\Delta^2_{m',s'}}]=0\;,\\
{}& \widehat{\omega}^{(2)}_{m,s}[g_{\Delta^{12}_{n,\ell},\ell}]=0\;,\quad  \widehat{\omega}^{(2)}_{m,s}[\widehat{g}_{\Delta^1_{m',s'}}]=0\;,\quad  \widehat{\omega}^{(2)}_{m,s}[\widehat{g}_{\Delta^2_{m',s'}}]=\delta_{mm'}\delta_{ss'}\;.
\end{split}
\end{equation}
The actions of these functionals on generic conformal blocks can also be read off from the decomposition coefficients of the Witten diagrams using the techniques from \cite{Zhou:2018sfz}. However, we will not discuss its details here. 

It should be emphasized that while such a functional approach can be developed by studying the particular Witten diagrams which show up in the computation of giant gravitons, the functionals are nonperturbatively valid and can be applied to general LLHH correlators. It would also be interesting to explore its other mathematically equivalent formulations, namely dispersion relations \cite{Caron-Huot:2017vep,Lemos:2017vnx,Liendo:2019jpu,Carmi:2019cub,Barrat:2022psm} and Polyakov-Mellin bootstrap \cite{Gopakumar:2016wkt,Gopakumar:2016cpb}.

\section{Discussions and outlook}\label{Sec:outlook}
In this paper, we developed a formalism of zero dimensional defects which is ideally suited for studying LLHH correlators. In particular, we focused on the case where the heavy operators are giant gravitons. Their correlators encode rich non-planar physics of $\mathcal{N}=4$ SYM. Combining the defect perspective with bootstrap techniques, we computed at strong coupling all four-point functions in which two heavy operators are maximal giant gravitons and two light operators are super gravitons of arbitrary KK levels. We demonstrated how the defect perspective allows us to uncover a partially broken hidden conformal symmetry in higher dimensions which unifies all these correlators. We also pointed out that the same structure exists at weak coupling which serves as a new organizing principle for perturbative calculations. In addition, we performed a systematic OPE analysis using the defect kinematics we developed to extract the dynamical information encoded in these giant graviton correlators. We obtained the complete spectrum of anomalous dimensions at leading order in $1/N$ for all defect channel double-particle operators. This sets a concrete benchmark for other nonperturbative methods, such as integrability, to reproduce in the future. The results of our paper open up many possibilities of future research. We mention some of them below. 

Starting from our results for giant graviton correlators, there are two directions of extensions which can be immediately pursued. The first is the supergravity correction to these correlators at order $\mathcal{O}(1/N^2)$, which corresponds to the leading loop level processes in AdS. These loop corrections can be systematically computed using the unitarity method for defect correlators \cite{Chen:2024orp}, which uses the tree-level result computed here as the input. We also expect that the fourth-order differential operator found in Section \ref{Subsec:Delta4} will significantly simplify the analysis and in particular help to develop a bootstrap approach directly in position space. The second extension is to compute the string theory corrections beyond the supergravity approximation. This can be facilitated by using the flat-space limit formulas recently found for defect correlators \cite{Alday:2024srr,Chen:2025cod}. The same analyses, starting from  tree-level supergravity, can also be repeated for giant gravitons in other backgrounds, such as AdS$_7\times$S$^4$ and AdS$_4\times$S$^7$.  It would be very interesting to find similar universality in correlator structures across dimensions as was discovered in the supergraviton four-point functions \cite{Alday:2020dtb}. 

Another important direction is to include more operators. On the one hand, we can add more light supergravitons in the correlators and this will be interesting from the perspective of scattering form factors in AdS. For example, we can consider five-point functions with two giant gravitons and three supergravitons. This target seems reasonably within reach by extending the position space method used in this paper, similar to the all light case \cite{Goncalves:2019znr}. On the other hand, we can explore qualitatively different physics by including more giant graviton correlators. The correlator of four giant gravitons and no light supergravitons is already a nontrivial example. By considering this correlator, we can extract information about the interactions between these zero dimensional defects \cite{Vescovi:2021fjf}.

In this work our focus was on correlators involving maximal giant gravitons. However, there are two intimately related cases, namely  non-maximal giant gravitons and dual giant gravitons, which are also of great interest. To compute their correlators with light supergravitons at strong coupling, the defect formalism and the bootstrap strategy will continue to play a prominent role. But new technical ingredients, such as new types of Witten diagrams, will be needed, in particular for the latter case. These correlators are also interesting to study at weak coupling, and progress can be made by extending the techniques of \cite{Wu:2025ott}. The results at different regimes of coupling can be crosschecked with the recent prediction of integrated correlators from supersymmetric localization \cite{Brown:2025huy}. A particularly interesting question to explore for correlators of non-maximal and dual giant gravitons is whether the higher dimensional hidden symmetry, perhaps with certain modifications, will continue to exist. Moreover, it is also worth mentioning that in the large charge limit of dual giant gravitons we can access correlators on the Coulomb branch of $\mathcal{N}=4$ SYM \cite{Ivanovskiy:2024vel,Coronado:2025xwk}. 

Finally, another independent task is to further develop the zero dimensional defect framework and turn it into a powerful and universal tool for general LLHH correlators. In particular, we can focus on the analytic functional approach outlined in Section \ref{Subsec:analyticfunctionals}, which  should be established more rigorously and fleshed out with more details. Once we have explicitly constructed these  functionals, they can be applied to study an array of interesting problems, e.g., in relation to black hole microstates and the Eigenstate Thermalization Hypothesis.   

\acknowledgments

We thank Robert de Mello Koch and Congkao Wen for helpful discussions. We would also like to thank the Institute of Modern Physics at Northwest University for their hospitality where part of this work was done. X.Z. gratefully acknowledges the hospitality provided by Indian Institute of Science, NITheCS at Stellenbosch University, and the University of the Witwatersrand during the completion of the manuscript. The work of J.C. and X.Z. is supported by the NSFC Grant No. 12275273, funds from Chinese Academy of Sciences, University of Chinese Academy of Sciences, and the Kavli Institute for Theoretical Sciences. The work of Y.J. is partly supported by Startup Funding no. 4007022326 of Southeast University and National Natural
Science Foundation of China through Grant No.12575073. This work is also supported by the NSFC Grant No. 12247103.

\appendix

\section{More about superconformal blocks}\label{App:scfblocks}
As we discussed in the main text, the bulk channel conformal blocks and R-symmetry blocks are equal to the ones of four-point functions. Consequently, the bulk channel superconformal blocks are also identical to the four-point superconformal blocks. Therefore, in this section we focus only on the defect channel superconformal blocks.

The $\frac{1}{2}$-BPS giant graviton breaks the $\mathcal{N}=4$ superconformal symmetry into $SU(2|2)_L\times SU(2|2)_R$ (see, e.g., \cite{Imamura:2021ytr}). In particular, the $SO(6)$ R-symmetry gets broken into $SO(4)\times SO(2)=SU(2)_M\times SU(2)_{\bar{M}}\times U(1)_{r_2}$ where we have labeled the subgroups using the quantum numbers of their representations. The conformal group $SO(5,1)$ becomes $SU(2)_J\times SU(2)_{\bar{J}}\times SO(1,1)_{\widehat{\Delta}}$. The factor $SU(2|2)_L$ contains the bosonic subgroups $SU(2)_J\times SU(2)_M\times U(1)_Z$ with $Z=\widehat{\Delta}-r_2$ being the central charge. Similarly, the $SU(2|2)_R$ factor contains $SU(2)_{\bar{J}}\times SU(2)_{\bar{M}}\times U(1)_Z$. For defect operators which can appear in the exchange of the defect two-point functions, the quantum numbers of the representation must satisfy $J=\bar{J}=\frac{s}{2}$ and $M=\bar{M}=\frac{r_1}{2}$. Therefore, we will label the defect operator using $\{\widehat{\Delta},s,r_1,r_2\}$. Note that the central charge $Z=\widehat{\Delta}-r_2$ is the same for the entire superconformal multiplet. 

There are three types of superconformal multiplets which are relevant to us and they are classified by how much of the superconformal symmetry is preserved. We will refer to them as short, semi-short and long superconformal blocks. To obtain their corresponding superconformal blocks, we will use the strategy where we write them as a sum of products of conformal blocks and R-symmetry blocks with unknown coefficients and then fix these coefficients using the superconformal Ward identities. More precisely, the overall positive parity requires only the combinations $\widehat{g}^+\widehat{\mathcal{Q}}^+$ and $\widehat{g}^-\widehat{\mathcal{Q}}^-$ to be present. Due to the piecewise nature of these blocks, we should impose the superconformal Ward identities on all four possible regions
\begin{equation}
\begin{split}
   {}& {\rm I:}\quad \{0<V<1,0<\sigma/\tau<1\}\;,\quad\quad {\rm II:}\quad \{0<V<1,\sigma/\tau\geq 1\}\;, \\
    {}& {\rm III:}\quad \{V\geq 1,0<\sigma/\tau<1\}\;,\quad\quad {\rm IV:}\quad \{V\geq 1,\sigma/\tau\geq 1\}\;.
\end{split}
\end{equation}
These conditions are also supplemented by the constraints $U,V,\sigma,\tau>0$ assumed in the definitions of the blocks. Note that because parities for $\widehat{g}$ and $\widehat{\mathcal{Q}}$ are the same in each term, I, II and III, IV are related by the $3\leftrightarrow 4$ exchange symmetry and we only need to restrict ourselves to the regions I and II. On the other hand, in the implementation of the superconformal Ward identities, one needs to impose the twisting of the cross ratios $\alpha=1/z$ (recall the change of variables for the cross ratios in (\ref{crossratiostozzbalphaalphabar})). This can always be achieved in these regions by restricting the independent cross ratios to the following ranges
\begin{equation}\label{twistrange}
\begin{split}
    {}&{\rm I:}\; 0<\bar{\alpha}<1\;,\; z>\frac{1}{1-\bar{\alpha}}\;,\; 1<\bar{z}<\frac{z}{z-1}\;,\\
    {}&{\rm II:}\;\left\{0<\bar{\alpha}<1\;,\; 1<z<\frac{1}{1-\bar{\alpha}}\;,\; 1<\bar{z}<\frac{z}{z-1}\right\}\;\\{}&\quad \quad {\rm or}\; \left\{\bar{\alpha}>1\;,0<z<1,0<\bar{z}<1\right\}\;.
\end{split}
\end{equation}

We label the superconformal blocks using the quantum numbers of the superconformal primaries and the results are as follows. It is convenient to define the combination
\begin{equation}
\label{deff}
    \mathbb{f}_{\widehat{\Delta},s,r_1,r_2}=\frac{1}{2}\left( \widehat{g}^{+}_{\widehat{\Delta},s}\widehat{\mathcal{Q}}^{+}_{r_1,r_2}+\widehat{g}^{-}_{\widehat{\Delta},s}\widehat{\mathcal{Q}}^{-}_{r_1,r_2}\right)\;.
\end{equation}
The short superconformal blocks contain the smallest number of components and read
\begin{equation}
     \widehat{\mathfrak{G}}^{\rm sh}_{s,r_1,r_2}=\mathbb{f}_{r_1+r_2,0,r_1,r_2}+\mathbb{f}_{r_1+r_2+1,1,r_1-1,r_2+1} + \mathbb{f}_{r_1+r_2+2,0,r_1-2,r_2+2}\;.
\end{equation}
The first term in this expression is the superconformal primary. Here and in what follows, whenever the quantum numbers $s$ or $r_1$ are negative, we define the component blocks to be zero (recall these are $SO(4)$ quantum numbers). The solution for the semi-short superblock is
\begin{equation}
    \begin{aligned}
        \widehat{\mathfrak{G}}^{\rm se}_{s,r_1,r_2}={}& \mathbb{f}_{r_1+r_2+s+2,s,r_1,r_2}+\mathbb{f}_{r_1+r_2+s+3,s-1,r_1-1,r_2+1}\\{}&+\mathbb{f}_{r_1+r_2+s+3,s+1,r_1-1,r_2+1}+\mathbb{f}_{r_1+r_2+s+3,s+1,r_1+1,r_2+1} \\{}&+ \mathbb{f}_{r_1+r_2+s+4,s,r_1-2,r_2+2}  + \eta_{r_1}\mathbb{f}_{r_1+r_2+s+4,s,r_1,r_2+2} \\{}&+   \mathbb{f}_{r_1+r_2+s+4,s+2,r_1,r_2+2}+\mathbb{f}_{r_1+r_2+s+5,s+1,r_1-1,r_2+3} \;,
    \end{aligned}
\end{equation}
where the superconformal primary is the first term and $\eta_{r_1}=1-\delta_{r_1,0}$. Note that only $\mathbb{f}_{\widehat{\Delta},s,r_1,r_2}$ show up in these two family of solutions. As a result, it is easy to check that these superconformal blocks vanish in the II, III regions and are nonzero only in the I, IV regions. We also find independent solutions involving only $\widehat{g}^{+}_{\widehat{\Delta},s}\widehat{\mathcal{Q}}^{+}_{r_1,r_2}-\widehat{g}^{-}_{\widehat{\Delta},s}\widehat{\mathcal{Q}}^{-}_{r_1,r_2}$ combinations, where the relative minus sign ensures they have zero overlap with the short and semi-short superconformal blocks. These additional solutions are needed for reproducing correlators from resummation in II, III regions. However, they identically vanish in I, IV regions which correspond to the defect OPE, and therefore are not needed in the OPE analysis.

Finally, we consider the long superconformal blocks. The solution can be most conveniently written in the form
\begin{equation}
\label{longsuperblock}
\widehat{\mathfrak{G}}_{\widehat{\Delta},s,r_1,r_2}^{\text{long}}=R(V\sigma\tau)^{-1}\widehat{g}_{\widehat{\Delta}+2, s}^{+} \widehat{\mathcal{Q}}_{r_1,r_2+2}^{+}\;.
\end{equation}
We can rewrite it in terms of individual components using the following prescription. We note that the factor in \eqref{longsuperblock} can be expanded as 
\begin{equation}
\label{weightshiftR}
    R(V\sigma\tau)^{-1} = \frac{1}{2}\left( W^{+}_2\widetilde{W}^{+}_2+W^{-}_2\widetilde{W}^{-}_2\right)+  \frac{1}{2}\mathcal{C}_1\widetilde{\mathcal{C}}_1\left( W^{+}_1\widetilde{W}^{+}_1+W^{-}_1\widetilde{W}^{-}_1\right)+\mathcal{C}_2 + \widetilde{\mathcal{C}}_2\;,
\end{equation}
where the building blocks are
\begin{equation}
    W_{\delta}^{\pm} = V^{\frac{\delta}{2}}\pm V^{-\frac{\delta}{2}} \;,\quad   \widetilde{W}_{\delta}^{\pm} = (\sigma/\tau)^{\frac{\delta}{2}}\pm (\sigma/\tau)^{-\frac{\delta}{2}} \;,
\end{equation}
and 
\begin{equation}
    \mathcal{C}_{\delta} = C_{\delta}^{(1)}(\chi/2)\;,\quad \widetilde{\mathcal{C}}_{\delta} = C_{\delta}^{(1)}\left((\Sigma-\bar{\Sigma})/2\right)\;.
\end{equation}
Here $W_{\delta}^{\pm} $ and $\widetilde{W}_{\delta}^{\pm} $ can be viewed as  weight-shifting operators for the conformal blocks and R-symmetry blocks changing the quantum numbers $\widehat{\Delta}$ and $r_2$
\begin{equation}
\begin{aligned}
    W_{\delta}^{+} \left(\widehat{g}_{\widehat{\Delta}, s}^{\pm}  \right) = {}&  \widehat{g}_{\widehat{\Delta}+\delta, s}^{\pm}  + \widehat{g}_{\widehat{\Delta}-\delta, s}^{\pm} \;, \quad  W_{\delta}^{-} \left(\widehat{g}_{\widehat{\Delta}, s}^{\pm}  \right) =  \widehat{g}_{\widehat{\Delta}+\delta, s}^{\mp}  - \widehat{g}_{\widehat{\Delta}-\delta, s}^{\mp} \;,\\ \widetilde{W}_{\delta}^{+} \left(\widehat{\mathcal{Q}}_{r_1,r_2}^{\pm}  \right) = {}&  \widehat{\mathcal{Q}}_{r_1,r_2+\delta}^{\pm}  + \widehat{\mathcal{Q}}_{r_1,r_2-\delta}^{\pm} \;, \quad  \widetilde{W}_{\delta}^{-} \left(\widehat{\mathcal{Q}}_{r_1,r_2}^{\pm}  \right) =  \widehat{\mathcal{Q}}_{r_1,r_2+\delta}^{\mp}  - \widehat{\mathcal{Q}}_{r_1,r_2-\delta}^{\mp}\;,
    \end{aligned}
\end{equation}
and $\mathcal{C}_i$, $\widetilde{\mathcal{C}}_i$ shift the quantum numbers $s$ and $r_1$
\begin{equation}
\begin{aligned}
    \mathcal{C}_1  \left(\widehat{g}_{\widehat{\Delta}, s}^{\pm} \right) = {}& \widehat{g}_{\widehat{\Delta}, s+1}^{\pm} + \widehat{g}_{\widehat{\Delta}, s-1}^{\pm}\;,\quad   \widetilde{\mathcal{C}}_1  \left(\widehat{\mathcal{Q}}_{r_1,r_2}^{\pm} \right) = \widehat{\mathcal{Q}}_{r_1+1,r_2}^{\pm}+\widehat{\mathcal{Q}}_{r_1-1,r_2}^{\pm}\;,\\      \mathcal{C}_2  \left(\widehat{g}_{\widehat{\Delta}, s}^{\pm} \right) = {}& \widehat{g}_{\widehat{\Delta}, s+2}^{\pm} +\widehat{g}_{\widehat{\Delta}, s}^{\pm}+ \widehat{g}_{\widehat{\Delta}, s-2}^{\pm}\;,\quad \widetilde{\mathcal{C}}_2  \left(  \widehat{\mathcal{Q}}_{r_1,r_2}^{\pm} \right) = \widehat{\mathcal{Q}}_{r_1+2,r_2}^{\pm} +\widehat{\mathcal{Q}}_{r_1,r_2}^{\pm}+\widehat{\mathcal{Q}}_{r_1-2,r_2}^{\pm} \;.
    \end{aligned}
\end{equation}
Following this procedure, the long superconformal block can be expanded in terms of $\widehat{g}^+\widehat{\mathcal{Q}}^+$ and $\widehat{g}^-\widehat{\mathcal{Q}}^-$. But the explicit expression is lengthy and we do not include it here. We only record the superconformal primary 
\begin{equation}
\widehat{\mathfrak{G}}_{\widehat{\Delta},s,r_1,r_2}^{\text{long}}\Big|_{\rm superprimary} = \mathbb{f}_{\widehat{\Delta},s,r_1,r_2}\;.
\end{equation}
As a side comment, we also point out that in the I, IV regions, where the OPE analysis is performed, we have the following decomposition of long superconformal blocks into short and semi-short superconformal blocks at the unitarity bound
\begin{equation}
    \begin{aligned}
     \widehat{\mathfrak{G}}_{\widehat{\Delta},s,r_1,r_2}^{\text{long}} \Big|_{\widehat{\Delta} \to r_1+r_2+s+2} & = \widehat{\mathfrak{G}}_{s,r_1,r_2}^{\rm se} + \widehat{\mathfrak{G}}_{s-1,r_1+1,r_2+1}^{\rm se}\;, \\ \widehat{\mathfrak{G}}_{\widehat{\Delta},0,r_1,r_2}^{\text{long}} \Big|_{\widehat{\Delta} \to r_1+r_2+2} & = \widehat{\mathfrak{G}}_{0,r_1,r_2}^{\rm se} + \widehat{\mathfrak{G}}_{r_1+2,r_2+2}^{\rm sh}\;.
    \end{aligned}
\end{equation}

\section{More about tree-level Witten diagrams}\label{App:moreWittendiag}

Here we outline the proofs for the relations (\ref{Ire1}-\ref{Ire3}) of the defect integral. To prove (\ref{Ire1}), we start from (\ref{Iu2F1}) and use the  identity
\begin{equation}
\begin{split}
{}_2F_1(a,b;c;z)={}&\frac{\Gamma(a+b-c)\Gamma(c)}{\Gamma(a)\Gamma(b)}(1-z)^{c-a-b}{}_2F_1(c-a,c-b;c-a-b+1;1-z)\\
{}&+\frac{\Gamma(c-a-b)\Gamma(c)}{\Gamma(c-a)\Gamma(c-b)}{}_2F_1(a,b;a+b-c+1;1-z)\;,
\end{split}
\end{equation}
to rewrite the ${}_2F_1$ function. Then we can use
\begin{equation}
{}_3F_2(a_1,a_2,a_3;b_1,b_2;z)=\frac{\Gamma(b_2)}{\Gamma(a_3)\Gamma(b_2-a_3)}\int_0^1dt\, t^{a_3-1}(1-t)^{-a_3+b_2-1}{}_2F_1(a_1,a_2;b_1;tz)\;,
\end{equation}
to perform the $\kappa$ integral and get
\begin{equation}\label{IuinH}
\widehat{\mathcal{I}}^{(u)}_{\widehat{\Delta}}=\frac{\pi}{4\sin(\frac{\pi(\Delta_2-\Delta_1)}{2})\Gamma(\Delta_1)\Gamma(\Delta_1)}\left(r^{\Delta_1}H(\Delta_1,\Delta_2,\widehat{\Delta},r^2)-r^{\Delta_2}H(\Delta_2,\Delta_1,\widehat{\Delta},r^2)\right)\;,
\end{equation}
where
\begin{equation}
H(\Delta_1,\Delta_2,\widehat{\Delta},x)=\frac{\Gamma(\frac{\widehat{\Delta}+\Delta_1}{2})\Gamma(\Delta_1)\Gamma(\frac{\Delta_1+\Delta_2}{2}){}_3F_2(\frac{\Delta+\Delta_1}{2},\Delta_1,\frac{\Delta_1+\Delta_2}{2};\frac{2+\widehat{\Delta}+\Delta_1}{2},\frac{2+\Delta_1-\Delta_2}{2};x)}{\Gamma(\frac{2+\widehat{\Delta}+\Delta_2}{2})\Gamma(\frac{2+\Delta_1-\Delta_2}{2})}\;.
\end{equation}
It is then clear that $\widehat{\mathcal{I}}^{(u)}_{\widehat{\Delta}}$ is symmetric in $\Delta_1$,  $\Delta_2$ and similarly for $\widehat{\mathcal{I}}^{(t)}_{\widehat{\Delta}}$. Moreover, the ${}_3F_2$ function satisfies 
\begin{equation}
   \left(\vartheta (\vartheta+b_1-1)(\vartheta+b_2-1)-x(\vartheta+a_3)(\vartheta+a_2)(\vartheta+a_1)\right){}_3F_2(a_1,a_2,a_3;b_1,b_2;x)=0\;,
\end{equation}
where we have defined $\vartheta=x d/dx$. It is not difficult to see using (\ref{IuinH}) that this gives rise to the relations (\ref{Ire2}). Let us also comment that (\ref{intIu}) might appear singular for $\Delta_1-\Delta_2\in 2\mathbb{Z}$. However, it is not difficult to convince oneself that the singularities are spurious using basic properties of ${}_3F_2$.

Finally, to prove (\ref{Ire3}) we start from the definition (\ref{intIu}) and use the symmetry property (\ref{Ire1}). We can see that sum of $\widehat{\mathcal{I}}^{(u)}_{\widehat{\Delta}}$ and its shadow $\widehat{\mathcal{I}}^{(u)}_{-\widehat{\Delta}}$ is equivalent to 
\begin{equation}
\widehat{\mathcal{I}}^{(u)}_{\widehat{\Delta}}(r)+\widehat{\mathcal{I}}^{(u)}_{-\widehat{\Delta}}(1/r)=(x_1^2)^{\frac{\Delta_1}{2}}(x_2^2)^{\frac{\Delta_2}{2}}\int_0^\infty \frac{dw_0}{w_0} \int_0^\infty \frac{dz_0}{z_0} \left(\frac{z_0}{z_0^2+x_1^2}\right)^{\Delta_1}\left(\frac{z_0}{w_0}\right)^{\widehat{\Delta}} \left(\frac{w_0}{w_0^2+x_2^2}\right)^{\Delta_2}\,.
\end{equation}
After rescaling $z_0\to z_0 \sqrt{x_1^2}$, $w_0\to w_0 \sqrt{x_2^2}$ we get
\begin{equation}
\widehat{\mathcal{I}}^{(u)}_{\widehat{\Delta}}(r)+\widehat{\mathcal{I}}^{(u)}_{-\widehat{\Delta}}(1/r)=r^{-\widehat{\Delta}}\int_0^\infty \frac{dw_0}{w_0} \int_0^\infty \frac{dz_0}{z_0} \left(\frac{z_0}{z_0^2+1}\right)^{\Delta_1}\left(\frac{z_0}{w_0}\right)^{\widehat{\Delta}} \left(\frac{w_0}{w_0^2+1}\right)^{\Delta_2}\;,
\end{equation}
which can then be evaluated the RHS of (\ref{Ire3}) by using 
\begin{equation}
\int_0^\infty dt\, t^{a-1}(1+t)^{-b}=\frac{\Gamma(a)\Gamma(b-a)}{\Gamma(b)}\;.
\end{equation}

\section{Heavy limit of conformal blocks} \label{app:heavylimit}

In this appendix, we present an explicit proof of \eqref{ghatheavylimit}. Thanks to the $3 \leftrightarrow 4$ symmetry, it is sufficient to focus only on the $0<V<1$ region where the defect channel conformal block is related to the four-point function conformal block in the t-channel.

The t-channel conformal block $\widehat{\tilde{g}}_{\Delta',\ell'}$  admits a standard radial expansion \cite{Hogervorst:2013sma}. With $\Delta'=\widehat{\Delta}+Q$, $\ell'=s$, the radial expansion can be written as
\begin{equation}
\label{radialexpant}
\widehat{\tilde{g}}_{\widehat{\Delta}+Q,s}=\sum_{n=0}^{\infty} V^{\frac{\widehat{\Delta}+n}{2}}   \sum_{m=-n}^{n} \mathcal{A}_{n,m} \widehat{C}_{s+m}^{\frac{d}{2}-1}(\chi/2)\;,
\end{equation}
where the coefficients $\mathcal{A}_{n,m}$ depend on $\widehat{\Delta}$, $s$, $d$, $\Delta_1$, $\Delta_2$ (with the normalization $\mathcal{A}_{0,0}=2^{s}$), and 
\begin{equation}
    \widehat{C}_{j}^{\nu}(x)= \frac{j!}{2^j (\nu)_j} C^{\nu}_j(x)\;,
\end{equation}
is the normalized Gegenbauer polynomial. From \eqref{radialexpant} we recognize that the defect channel conformal block \eqref{gdefectpiecewise} is nothing but the $n=m=0$ term. Therefore, we only need to prove that in the $Q\to \infty$ limit all the other coefficients with $n>0$ are suppressed. This can be most straightforwardly seen from the Casimir equation satisfied by the t-channel conformal block
\begin{equation}
\begin{aligned}
\label{casimireqfort}
   & \left(r^2 (1-r\phi )\partial_{r}^2- (d-2 Q+\left(\Delta _1+\Delta _2+1\right)r \phi  -1) r \partial_r \right.\\&\left.-2 r^2(\phi ^2-1)  \partial_{r} \partial_{\phi}+(\phi ^2-1) (1-r\phi  ) \partial_{\phi}^2 \right.\\&\left. + \left((d-1) \phi +r (\Delta
   _1+\Delta _2-d+2-\left(\Delta _1+\Delta _2+1\right) \phi ^2 )\right)\partial_{\phi}\right.\\&\left. - \widehat{\Delta}  (\widehat{\Delta}-d +2 Q) -\Delta _1 \Delta _2 r\phi - s (s +d-2)  \right) \widehat{\tilde{g}}_{\widehat{\Delta}+Q,s}=0\;,
   \end{aligned}
\end{equation}
where $V=r^2$ and $\phi$ was defined in \eqref{defrhophi}. After plugging the radial expansion \eqref{radialexpant} into \eqref{casimireqfort} and using the following Gegenbauer identities
\begin{equation}
    \begin{aligned}
     \widehat{C}_{j}^{\nu}(x)' ={}& \frac{x (2 \nu +j)}{1-x^2} \widehat{C}_{j}^{\nu}(x)-\frac{2 (\nu +j)}{1-x^2} \widehat{C}_{j+1}^{\nu}(x) \;,\\ \widehat{C}_{j}^{\nu}(x)''={}& \frac{(2 \nu +j) (j (x^2-1)+(2 \nu +1) x^2)}{(1-x^2)^2} \widehat{C}_{j}^{\nu}(x)-\frac{2 (2 \nu +1) x (\nu +j)}{(1-x^2)^2} \widehat{C}_{j+1}^{\nu}(x)\;,\\ x \,\widehat{C}_{j}^{\nu}(x)={}&\frac{j (2 \nu +j-1)}{4 (\nu +j-1) (\nu +j)} \widehat{C}_{j-1}^{\nu}(x) +\widehat{C}_{j+1}^{\nu}(x)\;,
     \end{aligned}
\end{equation}
we can obtain the following recursion relation for $\mathcal{A}_{n,m}$
\begin{equation}
    \begin{aligned}
    \label{Anmrecureq}
   &  \big(m^2+m (d+2 s-2)+n (2 \widehat{\Delta }+n+2 Q-d)\big) \mathcal{A}_{n,m} =\\&\frac{(m+s+1) (m+d+s-2)  (\widehat{\Delta }+\Delta _1-m+n-d-s) (\widehat{\Delta }+\Delta
   _2-m+n-d-s)}{(2 m+d+2 s-2) (2 m+d+2 s)} \\&\times \mathcal{A}_{n-1,m+1}  +(\widehat{\Delta }+\Delta _1+m+n+s-2) (\widehat{\Delta }+\Delta _2+m+n+s-2)\,\mathcal{A}_{n-1,m-1}\;.
    \end{aligned}
\end{equation}
It shows that $\mathcal{A}_{n,m}$ is completely determined from $\mathcal{A}_{n-1,m}$ and the $Q$-dependence comes only from the coefficient before $\mathcal{A}_{n,m}$. This immediately implies that $\mathcal{A}_{n,m}$ is more suppressed than $\mathcal{A}_{n-1,m}$ by $1/Q$ and therefore proves the $n=m=0$ term is the leading contribution of (\ref{radialexpant}) in the large $Q$ limit. Let us mention that a relation analogous to \eqref{ghatheavylimit} also exists for the R-symmetry blocks and can be proven in similar ways.

\bibliography{refs} 
\bibliographystyle{utphys}
\end{document}